\documentclass[usenatbib]{mn2e}
\usepackage{graphicx}
\newcommand{\text}[1]{\quad\mbox{#1}\quad}

\newcommand{\vgrad}[1]{\nabla{#1}}

\newcommand{\pder}[2]{\frac{\partial #1}{\partial #2}}
\newcommand{\Pd}[1]{\partial_{#1}}

\newcommand{\massint}{k}
\newcommand{\vvartheta}{\theta_{\rm v}}
\newcommand{\thetag}{\theta_{\rm m}}
\newcommand{\alphap}{{\alpha}}

\newcommand{\apj}{ApJ}

\newcommand{\mnras}{MNRAS}

\newcommand{\aap}{A\&A}

\newcommand{\apss}{Ap\&SS}
\title{Magnetic acceleration of ultra-relativistic jets in 
gamma-ray burst sources} 
\author[S.~S. Komissarov et al.]{Serguei
S.~Komissarov,$^{1}$\thanks{E-Mail: serguei@maths.leeds.ac.uk~(SSK);
vlahakis@phys.uoa.gr~(NV); arieh@jets.uchicago.edu~(AK);
bmv@maths.leeds.ac.uk~(MVB) 
} 
Nektarios Vlahakis,$^{2}$\footnotemark[1]
Arieh K\"onigl$^{3}$\footnotemark[1] and 
Maxim V.~Barkov,$^{1,4}$\footnotemark[1]\\
$^{1}$Department of Applied Mathematics, The University of Leeds,
Leeds, LS2 9GT\\
$^{2}$Section of Astrophysics, Astronomy and Mechanics, 
Physics Department, University of Athens, 15784 Zografos, Athens, Greece\\
$^{3}$Department of Astronomy and Astrophysics and
Enrico Fermi Institute, University of Chicago, 5640 South Ellis
Avenue,\\ \hskip 1cm Chicago, IL 60637, USA\\
$^{4}$Space Research Institute, 84/32 Profsoyuznaya Street, Moscow
117997, Russia}
\begin{document}
\date{Received/Accepted}
\maketitle

%%%%%%%%%%%%%%%%%%%%%%%%%%%%%%%%%%%%%%%%%%%%%%%%%%
\begin{abstract}
%%%%%%%%%%%%%%%%%%%%%%%%%%%%%%%%%%%%%%%%%%%%%%%%%
We present numerical simulations of axisymmetric, magnetically driven
outflows that reproduce the inferred properties of ultra-relativistic
gamma-ray burst (GRB) jets. These results extend our previous
simulations (Komissarov et al. 2007) of outflows accelerated to
moderately relativistic speeds, which we applied to
jets of active galactic nuclei.  In contrast to several recent
investigations, which have employed the magnetodynamics
approximation, our numerical scheme solves the full set of equations of
special-relativistic, ideal MHD, which enables us to
explicitly calculate the jet velocity and magnetic-to-kinetic energy
conversion efficiency --- key parameters of interest for astrophysical
applications. We confirm that the magnetic acceleration scheme remains
robust into the ultra-relativistic regime, as previously indicated by
semi-analytic self-similar solutions. We find that all current-carrying
outflows exhibit self-collimation and consequent acceleration near the
rotation axis, but that unconfined outflows lose causal connectivity
across the jet and therefore do not collimate or accelerate efficiently
in their outer regions.  We show that magnetically accelerated jets
confined by an external pressure that varies as $z^{-\alphap}$ 
($0<\alphap \le 2$) assume a paraboloidal shape $z \propto r^{a}$
(where $r\,,\, z$ are cylindrical coordinates and $a>1$), and we obtain 
analytic expressions for the one-to-one correspondence between 
the pressure distribution and the asymptotic jet shape.
We demonstrate that the acceleration
efficiency of jets with paraboloidal streamlines is $\ga 50\%$, with the
numerical value being higher the lower the initial magnetization.
We derive asymptotic analytic expressions for the acceleration of
initially cold outflows along paraboloidal streamlines
and verify that they provide good descriptions of the simulated flows. 
Our modelled jets (corresponding to $3/2<a<3$) attain Lorentz
factors $\Gamma \ga 10^2$ on scales $\sim 10^{10}-10^{12}\; {\rm cm}$,
consistent with the possibility that long/soft GRB jets are accelerated
within envelopes of collapsing massive stars, and $\Gamma \ga 30$ on
scales $\sim 9\times 10^{8}-3\times 10^{10}\; {\rm cm}$, consistent with
the possibility that short/hard GRB jets are accelerated on scales where
they can be confined by moderately relativistic winds from accretion
discs. We also find that 
$\Gamma \vvartheta \sim 1$ for magnetically accelerated jets,
where $\vvartheta$ is the half-opening angle of the poloidal
streamlines, which 
implies that the $\gamma$-ray emitting components of GRB outflows are very
narrow, with $\vvartheta\la 1^\circ$ in regions where $\Gamma \ga 100$,
and that the afterglow light curves of these components would
either exhibit a very early jet break or show no jet break at all.

\end{abstract}
                                                              

%%%%%%%%%%%%%%%%%%%%%%%%%%%%%%%%%%%%%%%%%%%%%%%%%%
\section{Introduction}
%%%%%%%%%%%%%%%%%%%%%%%%%%%%%%%%%%%%%%%%%%%%%%%%%%
\label{introduction}

In the ``standard'' model of long-duration, soft-spectrum gamma-ray
bursts (GRBs; e.g.  \citealt{Pir05}), the prompt high-energy emission
arises in ultra-relativistic (bulk Lorentz factor $\Gamma \ga 10^2$),
highly collimated (opening half-angle of a few degrees) jets. The high
Lorentz factors are inferred from the requirement of a sufficiently low
opacity to photon-photon annihilation or to scattering by photon
annihilation-produced electron-positron pairs \citep[e.g.][]{LS01},
whereas the jet opening angle is deduced from the detection of a
panchromatic break in the light curve of the lower-energy afterglow
emission \citep[e.g.][]{Rho99,Sari99}. Recent observations by the
{\it Swift}\/ satellite have indicated that various aspects of this
model may need to be modified \citep[e.g.][]{Mes06,Pan07,Lia08}, but the
basic picture of a collimated $\Gamma \ga 10^2$ outflow is still the
accepted paradigm.

Observations of long/soft GRBs and their afterglows have revealed that
these events typically involve the release of a few times $10^{51}\;
{\rm erg}$, although the fraction of this energy that corresponds to the
$\gamma$-ray emitting outflow component may vary from source to source
\citep[e.g.][]{BKF03,Fra05}. The outflows in these GRBs have been argued
to originate either in a magnetar or in a rapidly accreting stellar-mass
black hole, formed in the collapse of a massive star. The jets could tap
into the rotational energy of the neutron star, black hole or accretion
disc through the agency of an ordered magnetic field that threads the
source
\citep[e.g.][]{U92,T94,MR97,K97,KR98,VK01,VK03a,VK03b,Bla02,DS02,VPK03,P03,M06,Lyu06b,Lev06,KB07,B08,BK08}. For
typical burst energies and durations the field amplitudes should be
$\sim 10^{14}-10^{15}\;$G. Early models have postulated that GRB
outflows are driven purely thermally via annihilation of neutrinos
emitted by the accretion disc. Although this model remains very popular,
some recent studies have indicated that the neutrino heating may not be
as efficient as previously thought \citep[e.g.][]{DPN02}.  At present,
both the magnetic and the thermal mechanisms seem equally possible and
it may well be that in many cases they operate simultaneously. In
particular, neutrino heating may play an important role in the initial
acceleration of magnetized outflows \citep[e.g.][]{VK03a} and in
determining their mass load \citep[e.g.][]{Lev06, BL08}.

While short/hard GRBs
evidently have different progenitors (quite possibly merging neutron
stars or neutron star/black hole pairs) and on average involve a smaller
energy release, a lower Lorentz factor, and weaker collimation than
long/soft GRBs, they may well represent the same basic phenomenon and
arise in relativistic outflows that are driven in a similar way
\citep[e.g.][]{Nak07}.

The magnetic acceleration and collimation of GRB outflows needs to be
studied within the framework of relativistic magnetohydrodynamics
(MHD). Although general-relativistic effects may influence the
conditions near the base of the flow, most of the action takes place
sufficiently far away from the central mass that the simpler equations
of special-relativistic MHD can be employed. Since our focus in this
paper is on the global structure of GRB jets, we henceforth consider
only the special-relativistic theory. However, even in this case there
are qualitatively new effects in comparison with Newtonian MHD. These
include the fact that, when the bulk Lorentz factor becomes large, the
electric force can no longer be neglected relative to the magnetic force
and, in fact, becomes comparable to it in magnitude. Correspondingly,
one needs to retain the displacement current and the electric charge
density in Maxwell's equations. Another consequence of relativistic
motion (which also affects unmagnetized flows) is the coupling between
different spatial components of the momentum conservation equation
brought about by the appearance of the Lorentz factor (which is
calculated from the total velocity) in each of the component equations.
Furthermore, in cases where the temperature (i.e. the characteristic
velocity of internal motions) is relativistic, one needs to take into
account the enthalpy contribution to the inertia of the flow. On account
of these various factors, relativistic MHD does not naturally yield
simple generalizations of results obtained in Newtonian MHD. To simplify
the treatment, various authors have adopted the force-free
electrodynamics (also termed ``magnetodynamics'') approximation, in
which the matter inertia is neglected altogether. While this approach
has led to useful insights and interesting exact solutions, it is
inherently limited in that one cannot explicitly calculate the fluid
velocity and hence the efficiency of transforming electromagnetic energy
into kinetic form, which are key parameters of interest for
astrophysical applications.

In a pioneering work, \citeauthor{LCB92} (\citeyear{LCB92}; see also
\citealt{Con94}) derived exact semi-analytic MHD solutions of steady,
axisymmetric, ``cold'' relativistic flows patterned after the Newtonian
radially self-similar outflow solutions of \citet{BP82}. In contrast
with the Newtonian solutions, one cannot match the flow in the
relativistic case to a given power-law radial distribution of the
rotation velocity of the source (e.g. the $\propto r^{-1/2}$ rotation
law of a Keplerian accretion disc) because the relativistic equations
already contain the (constant) speed of light $c$. However, this
constraint only affects the base of the flow (and, as shown by
\citealt{VK03a}, it is possible to approximate a Keplerian disc even in
this case by judiciously parametrizing the disc height above the origin
of the coordinate system), and one can proceed to obtain the global
structure of the outflow as in the Newtonian case. \citet{LCB92}
identified as a key property of the relativistic outflow solutions the
spatially extended nature of the acceleration region, which continues
well beyond the classical fast-magnetosonic surface. These results were
further generalized to initially ``hot'' outflows by
\citet{VK03a,VK03b}, who went on to apply the relativistic self-similar
solutions to GRB outflows (see also \citealt{VK01} and \citealt{VPK03})
and to the lower-$\Gamma$ jets imaged in active galactic nuclei
\citep[AGNs;][]{VK04}. The solutions obtained in these papers confirmed
that spatially extended acceleration is a generic property of MHD
outflows that distinguishes it from purely hydrodynamic, thermally
driven winds. \citet{VK01,VK03a} noted that this property can be
understood from the fact that the magnetic acceleration is determined
from the joint solution of the Bernoulli equation (derived from the
momentum conservation equation along the poloidal magnetic field) and
the trans-field equation (which describes the force balance in the
transverse direction). The effective singular surface (the ``event
horizon'' for the propagation of fast-magnetosonic waves) is the
so-called modified fast magnetosonic surface, which can lie well beyond
the corresponding classical surface. (The classical fast-magnetosonic
surface is singular only when one solves the Bernoulli equation alone,
assuming that the shape of the field lines is given; in
Section~\ref{theory} we further elaborate on the strong connection
between acceleration and poloidal field-line shape in magnetically
driven flows.)

The semi-analytic solutions have also established the collimation
properties of MHD outflows, demonstrating that they converge
asymptotically to cylinders for flows that are Poynting flux-dominated
at the source and to cones when the enthalpy flux is initially dominant
\citep[e.g.][]{VK03a,VK03b}. These solutions are, however, limited by
the self-similarity assumption, which, besides restricting the angular
velocity distribution at their base, also requires the magnetic flux
distribution to be a power law in radius and only enables one current
regime (current-carrying or return-current, but not a global current
circuit) to be modelled by any given solution. To validate the
applicability of these results under more realistic circumstances and
to ascertain their dynamical stability, one needs to resort to numerical
simulations. However, the large spatial extent of the acceleration
region (which, according to the semi-analytic solutions, typically
covers several decades in spherical radius) has posed a strong challenge
for such calculations: in fact, early attempts to simulate such flows
were limited by numerical dissipation to maximum Lorentz factors that
were only a small fraction (less than $1\%$) of the potentially
achievable terminal value.

\citeauthor{KBVK07} (\citeyear{KBVK07}; hereafter Paper~I) have taken a
major step toward overcoming this challenge by employing a
special-relativistic, ideal-MHD numerical scheme that was specifically
designed to optimize accuracy and resolution and to minimize numerical
dissipation. A key element of their approach was the implementation of a
grid-extension method that made it possible to follow the flow up to six
decades in spatial scale while reducing the computation time by up to
three orders of magnitude. They were able to model cold flows that
converted nearly $80\%$ of the initial Poynting flux into kinetic energy
of $\Gamma_\infty \ga 10$ baryons and demonstrated that the results were
consistent with the available data on the acceleration of relativistic
jets in AGNs. They found that the numerical solutions assumed a
quasi-static configuration that was qualitatively in accord with the
self-similar AGN jet models of \citet{VK04}. The simulations were,
however, able to examine various aspects of the flow that could not be
studied within the framework of a self-similar model (including the
structure of outflows in which both the current and the return current
flow within the jet and the dependence of the collimation properties on
the shape of the jet boundary) and uncovered new features (such as the
formation of a cylindrical core around the jet axis) that were
inherently non--self-similar.

In this paper we further extend the scheme presented in Paper~I to cover
the regime of GRB outflows. In particular, we present simulations of
outflows that attain terminal Lorentz factors $\Gamma_\infty \ga 10^2$,
following them over up to eight decades in axial scale. Besides cold
jets, we also consider the case of an outflow in which the enthalpy flux
is a significant fraction of the injected energy flux. Owing to the
larger range in $\Gamma$ in comparison with the solutions presented in
Paper~I, the magnetic acceleration region can now be better isolated,
which enables us to more accurately compare its behaviour with that of
the self-similar solutions and to analyse it using the asymptotic forms
of the Bernoulli and trans-field equations. We begin by reviewing the
relativistic MHD formalism (Section~\ref{equations}) and our numerical
scheme (Section~\ref{simulations}). We present key simulation results in
Section~\ref{results} and discuss them in the context of the theory of
magnetic acceleration in Section~\ref{theory}. Section~\ref{application}
deals with applications of our results to GRBs. Our conclusions are
given in Section~\ref{conclusion}.

%%%%%%%%%%%%%%%%%%%%%%%%%%%%%%%%%%%%%%%%%%%%%%%%%%
\section{Basic equations}
%%%%%%%%%%%%%%%%%%%%%%%%%%%%%%%%%%%%%%%%%%%%%%%%%%
\label{equations}

Since most of the acceleration takes place far away from the source,
we assume that the space-time is flat. In an inertial frame at rest
relative to the source, the relativistic ideal-MHD equations that
describe the flow take the following form: {\it
continuity equation}
\begin{equation}
(1/c)\Pd{t}(\sqrt{-g}\rho u^t)+ \Pd{i}(\sqrt{-g}\rho u^i)=0\, ,
\label{cont1}
\end{equation}
where $\rho$ is the rest mass density of matter, $u^\nu$ is its
4-velocity, and $g$ is the determinant of the metric tensor; {\it
energy-momentum equations}
\begin{equation} 
(1/c)\Pd{t}(\sqrt{-g}T^t_{\ \nu})+ \Pd{i}(\sqrt{-g}T^i_{\ \nu})=
\frac{\sqrt{-g}}{2} \Pd{\nu}(g_{\alpha\beta}) T^{\alpha\beta}\, ,
\label{en-mom1} 
\end{equation} 
where $T^{\kappa\nu}$ is the total stress-energy-momentum tensor;
{\it induction equation}
\begin{equation} (1/c)\Pd{t}(B^i)+e^{ijk}\Pd{j}(E_k) =0\, , \label{ind1}
\end{equation} 
where $e_{ijk} = \sqrt{\gamma} \epsilon_{ijk} $ is the Levi-Civita
tensor of the absolute space ($\epsilon_{123}=1$ for right-handed
systems and $\epsilon_{123}=-1$ for left-handed ones) and $\gamma$ is
the determinant of the spatial part of the metric tensor
($\gamma_{ij}=g_{ij}$); the {\it solenoidal condition}
\begin{equation}
\Pd{i}(\sqrt{\gamma} B^i) =0\, .
\end{equation}

The total stress-energy-momentum tensor, $T^{\kappa\nu}$, is a sum of
the stress-energy momentum tensor of matter,
\begin{equation}
T_{(m)}^{\kappa\nu} = wu^\kappa u^\nu /c^2 + p g^{\kappa\nu}\, ,
\end{equation}
where $p$ is the thermodynamic pressure and $w$ is the enthalpy per
unit volume, and the stress-energy momentum tensor of the
electromagnetic field,
\begin{equation}
   T_{(e)}^{\kappa\nu} = \frac{1}{4\pi}\left[F^{\kappa\alpha} F^\nu_{\
   \alpha} - \frac{1}{4}(F^{\alpha\beta}F_{\alpha\beta})g^{\kappa\nu}
   \right]\, ,
\end{equation}
where $F^{\nu\kappa}$ is the Maxwell tensor. The electric and magnetic
fields are defined as measured by an observer stationary relative to the
spatial grid, which gives 
\begin{equation}
     B^i= \frac{1}{2}e^{ijk} F_{jk}
\label{B-def}
\end{equation}
and
\begin{equation}
   E_i = F_{it}\, .
\label{E-def}
\end{equation}
In the limit of ideal MHD
\begin{equation}
  E_i=-e_{ijk}v^jB^k/c\, ,
\label{perf-cond}
\end{equation}
where $v^i=u^i/u^t$ is the usual 3-velocity of the plasma.
 
In all of our simulations we use an isentropic equation of state
\begin{equation}
  p=Q\rho^s\, ,
\label{eos}
\end{equation}
where $Q=$const and $s=4/3$. 
This relation enables us to exclude the
energy equation from the integrated 
system.  However, the momentum equation remains intact, including the
non-linear advection term.  Therefore, if the conditions for shock
formation were to arise, our calculation would capture that
shock.\footnote{Since entropy is fixed, the compression of our shocks
would be the same as for continuous compression waves. This would give a higher
jump in density for the same jump in pressure than in a proper
(dissipative) shock. Fortunately, we do not need to contend with this
issue in practice as shocks do not form in our simulations.}
The enthalpy per unit volume is 
\begin{equation}
w=\rho c^2 + \frac{s}{s-1} p \,.
\label{w-def}
\end{equation}

%%%%%%%%%%%%%%%%%%%%%%%%%%%%%%%%%%%%%%%%%%%%%%%%%
\subsection{Field-line constants}
%%%%%%%%%%%%%%%%%%%%%%%%%%%%%%%%%%%%%%%%%%%%%%%%%
\label{section_integrals}

The poloidal magnetic field is fully described by the azimuthal
component of the vector potential,
\begin{equation}
  B^i = \frac{1}{\sqrt{\gamma}} \epsilon^{ij\phi} \pder{A_\phi}{x^j}\,
  .
\end{equation}
For axisymmetric solutions $A_\phi=\Psi/2\pi$, where $\Psi(x^i)$, the
so-called magnetic flux function, is the total magnetic flux enclosed
by the circle $x^i=$const ($x^i$ being the coordinates of the
meridional plane). Stationary and axisymmetric ideal MHD flows have five
quantities that propagate unchanged along the magnetic field lines and
thus are functions of $\Psi$ alone. These are 
$\massint$, the mass flux per unit magnetic flux; 
$\Omega$, the angular velocity of magnetic field lines; 
$l$, the total angular momentum flux per unit rest-mass flux; 
$\mu$, the total energy flux per unit rest-mass energy flux; 
and $Q$, the entropy per particle:
\begin{equation}
  \massint = \frac{\rho u_p}{B_p}\, ,
\label{kappa}
\end{equation}
\begin{equation}
  \Omega=\frac{v^{\hat{\phi}}}{r}-\frac{v_p}{r}
  \frac{B^{\hat{\phi}}}{B_p} \, ,
\label{omega-def}
\end{equation}
\begin{equation}
    l = -\frac{I}{2\pi\massint c}+r \frac{w}{\rho c^2} \Gamma v^{\hat{\phi}} \,,
\label{angm-def}
\end{equation}
\begin{equation}
    \mu = \mu_h + \mu_m,
\label{kap-def}
\end{equation}
and
\begin{equation}
    Q = P/\rho^s,
\end{equation}
where $u_p=\Gamma v_p$ is the magnitude of the poloidal component of
the 4-velocity, $B_p$ is the magnitude of the poloidal component of
the magnetic field, $r$ is the cylindrical radius,
\begin{equation}
   I = \frac{c}{2} r B^{\hat\phi}
\label{I}
\end{equation}
is the total electric current flowing through a loop of radius $r$
around the rotation axis,
\begin{equation}
 \mu_h= \frac{w}{\rho c^2}\Gamma
\label{mu_h-def}
\end{equation}
is the total hydrodynamic energy (rest mass plus thermal plus kinetic)
flux per unit rest-mass energy flux,
\begin{equation}
 \mu_m=\mu_h \sigma=-\frac{\Omega I}{2\pi\massint c^3}
\label{mu_m-def}
\end{equation}
is the Poynting flux per unit rest-mass energy flux, and
$\sigma$ is the ratio of the Poynting flux to the hydrodynamic
(rest-mass plus thermal plus kinetic) energy flux.
For cold flows $Q=0$, $w=\rho c^2$.  (Here and in the
rest of the paper we use a hat symbol over vector indices to indicate
their components in a normalized coordinate basis.)  From
equation~(\ref{kap-def}) it follows that the Lorentz factor $\Gamma$
cannot exceed $\mu$.

%%%%%%%%%%%%%%%%%%%%%%%%%%%%%%%%%%%%%%%%%%%%%%%%%%
\section{Numerical Simulations}
%%%%%%%%%%%%%%%%%%%%%%%%%%%%%%%%%%%%%%%%%%%%%%%%%%
\label{simulations}

To maintain a firm control over the jet's confinement and to prevent
complications related to numerical diffusion of the denser
plasma from the jet's surroundings, we study outflows
that propagate inside a solid funnel of a prescribed
shape.\footnote{As was already noted in Paper~I, in real astrophysical
systems the shape of the boundary is determined by the spatial
distribution of the pressure or the density of the confining ambient medium.
The effective ambient pressure
distributions implied by the adopted funnel shapes are considered in
Section~\ref{pressure}.}  Specifically, we consider axisymmetric
funnels
\begin{eqnarray}
\nonumber z \propto r^a\, ,
\end{eqnarray}
where $z$ and $r$ are the cylindrical coordinates of the funnel wall 
and $a=2/3$, $1$, $3/2$, $2$ and $3$. 
We employ elliptical coordinates $\{\xi,\eta,\phi\}$, where
\begin{equation}
 \xi=rz^{-1/a}
\label{xi}
\end{equation}
and
\begin{equation}
 \eta^2=\frac{r^2}{a} + z^2
\label{eta}
\end{equation}
(see Paper~I for details).

We use a Godunov-type numerical code based on the scheme described in
\citet{K99}.  To reduce numerical diffusion we apply parabolic
reconstruction instead of the linear one of the original code. Our
procedure, in brief, is to calculate minmod-averaged first and second
derivatives and use the first three terms of the Taylor expansion for spatial
reconstruction.
This simple procedure results in a noticeable
improvement in the solution accuracy even though the new scheme is still
not 3rd-order accurate because of the non-uniformity of the grid.

The grid is uniform in the $\xi$ direction (the polar angle direction
when we use spherical coordinates), where in most runs it has a total
of 60 cells.  To check the convergence, some runs were repeated with a
doubled resolution. The cells are elongated in the $\eta$ direction
(the radial direction when we use spherical coordinates), reflecting
the elongation of the funnel. Very elongated cells lead to a
numerical instability, so we imposed an upper limit of 40 on the
length/width ratio.

To speed up the simulations, we implement a sectioning of the
computational grid as described in \citet{KL04}. In each section,
which is shaped as a ring, the numerical solution is advanced using a
time step based on the local Courant condition. It is twice as large
as the time step of the adjacent inner ring and twice as small as the
time step of the adjacent outer ring.  This approach is particularly
effective for conical flows but less so for highly collimated, almost
cylindrical configurations.

The equations are dimensionalized in the following manner. The unit of length, $L$, is
such that $\eta_i=1$, where the subscript $i$ refers to the inlet
boundary. The unit of time is $T=L/c$. The unit of mass is $M=L^3
B_0^2/4\pi c^2$, where $B_0$ is the dimensional magnitude of the $\eta$
component of magnetic field at the inlet (so the dimensionless magnitude
of $B^{\hat\eta}$ at the inlet is $\sqrt{4\pi}$).  In applications, $L$
is the typical length-scale of the launch region, $T$ is the light
crossing time of that region and $B_0$ is the typical strength of the
poloidal magnetic field at the origin.  Notice that $L$ does not have to be 
the size of the rotating object at the base of the jet and in particular it   
cannot be identified with the radius of the black hole event horizon which 
allows only inflows. When dimensional estimates are required we use the
expected magnitude of the light-cylinder radius, $r_{\rm lc} \equiv c/\Omega$. 
The mass scale $M$ does not represent the mass of the central object
but rather the rest-mass equivalent of the magnetic energy within the
magnetosphere.

%----------------------------------- 
\subsection{Boundary conditions}
%----------------------------------- 

\subsubsection{Inlet boundary}
\label{inlet_section}

We treat the inlet boundary, $\eta_i=1$, as a surface of a perfectly
conducting rotator with either a uniform angular velocity $\Omega =
\Omega_0$ or with 
\begin{equation}
  \Omega = \Omega_0(1+a_2(\xi/\xi_j)^2+a_3(\xi/\xi_j)^3),
\label{omega}
\end{equation}
where the subscript $j$ refers to the jet boundary (funnel wall).
In this paper we set $a_2=0.778$ and $a_3=-1.778$. The angular
velocity profile is directly related to the distribution of the 
electric current in the jet, which for $r\gg r_{\rm lc}$ is given by
\begin{equation}\label{current}
    I\approx -\frac{1}{2}\Omega B_p r^2
\label{I1}
\end{equation}
(see Paper~I, or equation~\ref{I-Bp} in Section~\ref{power-law}). 
In fact, the current is driven
by the electric field associated with the rotating poloidal field, and
the electric charge conservation requires the circuit to eventually close.  In
the case of a constant $\Omega$ the return current flows over the jet
boundary, whereas in the case of differential rotation with
$\Omega(\xi_j)=0$ it flows mainly inside the jet (within $0.75<\xi/\xi_j <1$
for the $\Omega$ distribution given by equation~\ref{omega}). 
The solid-body rotation law provides a very good
description of the behaviour of magnetic field lines that thread the horizon
of a black hole or the surface of a magnetized star. This choice is therefore 
entirely appropriate for the black-hole or magnetar theory of GRB jets. 
On the other hand, differential rotation is a natural choice for jets
that are launched from an accretion disc, and although the
distribution~(\ref{omega}) does not correspond to a realistic disc
model, it should nevertheless capture the qualitative aspects of such a
system.\footnote{Note in this connection that \citet{TMN08} simulated a
force-free black-hole/disc outflow in which current flowed out along
field lines that threaded the uniformly rotating hole and returned along
field lines attached to the differentially rotating disc.}

The condition of perfect conductivity allows us to fix the azimuthal
component of the electric field and the $\eta$ component of the
magnetic field:
\begin{equation}
  E_\phi=0, \quad B^{\hat\eta}=B_0 \text{at} \eta=\eta_i\, .
\end{equation}
{}From the first of these conditions we derive
\begin{equation}
v^{\hat\xi} = \frac{v^{\hat\eta}}{B^{\hat\eta}} B^{\hat\xi}
\label{v_xi_b}
\end{equation}
and (using equation \ref{omega-def})
\begin{equation}
v^{\hat{\phi}} = r \Omega + \frac{v^{\hat\eta}}{B^{\hat\eta}}
B^{\hat{\phi}}\, .
\label{v_phi_b}
\end{equation}
The adopted uniform distribution of $B^{\hat{\eta}}$ is consistent with
transverse mechanical equilibrium at the inlet. 
We have also experimented with nonuniform distributions of
the magnetic field, in particular with $B^{\hat\eta}$ decreasing with
$\xi$. The results were not significantly different as the
field distribution downstream of the inlet underwent a rapid
rearrangement that restored the transverse force balance.

To have control over the mass flux, the flow
at the inlet boundary is set to be super--slow-magnetosonic. This
means that both the density and the radial component of the velocity
can be prescribed some fixed values:
\begin{eqnarray}
\nonumber \rho=\rho_0\, , \quad v^{\hat\eta}=v_{p_{0}}\, .
\end{eqnarray}
In the simulations we use $v_{p_{0}}=0.5\, c$ or $0.7\, c$, which is a
choice of convenience. On the one hand, this value is sufficiently small  
to insure that the flow at $\eta_i=1$ is sub-Alfv\'enic and hence that
the Alfv\'en and fast-magnetosonic critical surfaces are located
downstream of the inlet boundary. On the other hand, it is large enough 
to promote the rapid establishment of a steady state (in which the
outflow speed remains constant along the symmetry axis).
Because of the sub-Alfv\'enic nature of the inlet flow, we cannot fix the
other components of the magnetic field and the velocity --- they are to
be found as part of the global solution.  Following the standard
approach we extrapolate $B^{\hat\phi}$ and $B^{\hat\xi}$ from the
domain into the inlet boundary cells.  We then compute $v^{\hat\phi}$
and $v^{\hat\xi}$ from equations~(\ref{v_xi_b}) and~(\ref{v_phi_b}).

The magnitude of the angular velocity is chosen in such a way that 
the Alfv\'en surface is encountered close to the source. Specifically,
in the case of solid-body rotation the light cylinder radius, $r_{\rm
lc}$, is $\simeq 50\%$ larger than the initial jet radius.
In the differential rotation case, the 
closest point of the Alfv\'en surface is located at a distance
of $\simeq 1$ initial jet radius from the inlet surface. 

The inlet density varies from model to model in order to cover a
wide range of initial magnetizations. Table I gives the key parameters 
of all the jet models constructed in this study. Most of the models,
denoted by the letter B, correspond to the wall shape $z\propto r^{3/2}$
and differ only  by the value of the magnetization parameter: $\mu$
varies from the relatively small value of 39, which is more suitable to
AGN jets (see Paper~I), all the way up to 620. Model B2H is included to
study the effects of a high temperature at the source. The initial
effective thermal Lorentz factor in this model is 
$\Gamma_{t0}=w_0/\rho_0c^2=55$.
Models A and AW have a wall of conical 
shape. In model AW the half-opening angle of the cone is $90^\circ$,
which allows us to model the case of an unconfined outflow (which could be
relevant to pulsar winds).         
The remaining models help to explore the effects of 
differential rotation (model D), of various other paraboloidal wall shapes
($z \propto r^{2}$ in model C, $z \propto r^{3}$ in model F)
and of a wall shape whose opening angle increases with distance (model E).

%ttttttttttttttttttttttttttttttttttttttttttttttttttttttttttttttttttttttttt
\begin{table}
\caption{Parameters of simulation models.}
   \begin{tabular}{|c|l|l|l|l|l|}
   \hline
    Model & a & rotation & $w_0/\rho_0c^2$ & $\xi_j$ or $\theta_j$ &
$\mu_{\rm max}$\\
   \hline
   \hline
    A & 1 &  uniform & 1.0 & $\theta_j=0.2$ & 560\\  
   \hline
    AW & 1 &  uniform & 1.0 & $\theta_j=\pi/2$ & 560\\
   \hline 
    B1 & 3/2 & uniform & 1.0 & $\xi_j=2.0$ & 620\\
   \hline
    B2 & 3/2 & uniform & 1.0 & $\xi_j=2.0$ & 310\\
   \hline
    B2H & 3/2 & uniform & 55 & $\xi_j=2.0$ & 370\\
   \hline
    B3 & 3/2 & uniform & 1.0 & $\xi_j=2.0$ & 155 \\
   \hline
    B4 & 3/2 & uniform & 1.0 & $\xi_j=2.0$ & 78\\
   \hline
    B5 & 3/2 & uniform & 1.0 & $\xi_j=2.0$ & 39\\
   \hline
    C & 2 &  uniform & 1.0 & $\xi_j=2.0$ & 620\\
   \hline
    D & 3/2 & differential & 1.0 & $\xi_j=2.0$ & 600\\
   \hline
    E & 2/3 & uniform & 1.0 & $\xi_j=0.1$ & 300\\
   \hline
    F & 3.0 & uniform & 1.0 & $\xi_j=2.0$ & 540\\
   \hline
   \end{tabular}
\end{table}
%ttttttttttttttttttttttttttttttttttttttttttttttttttttttttttttttttttttttttt

%fffffffffffffffffffffffffffffffffffffffffffffffffffffffffffffffffff
\begin{figure*}
\includegraphics[width=55mm]{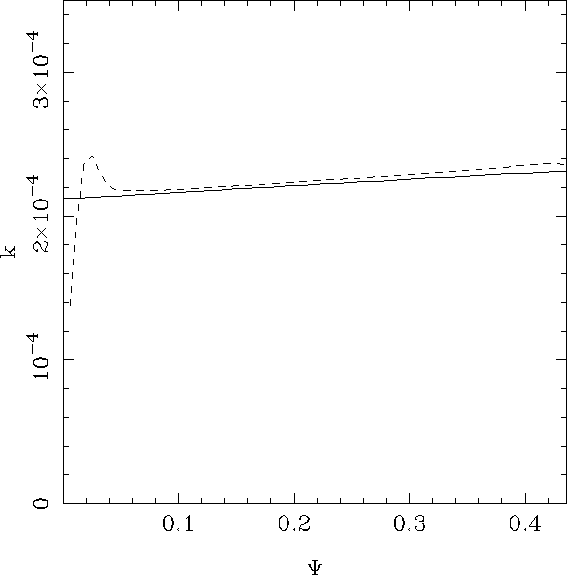}
\includegraphics[width=55mm]{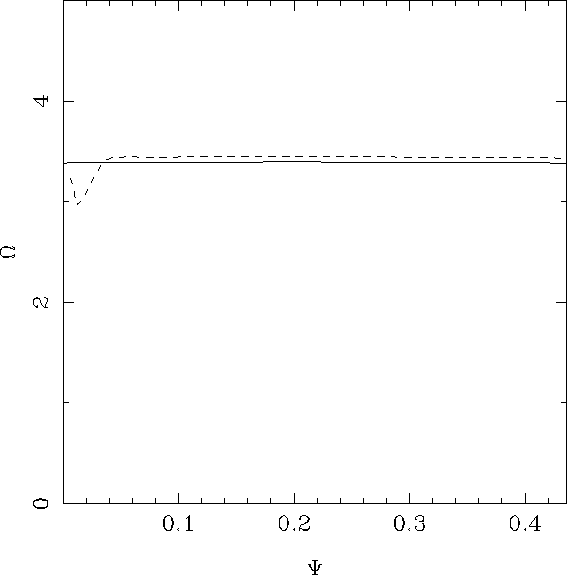}
\includegraphics[width=55mm]{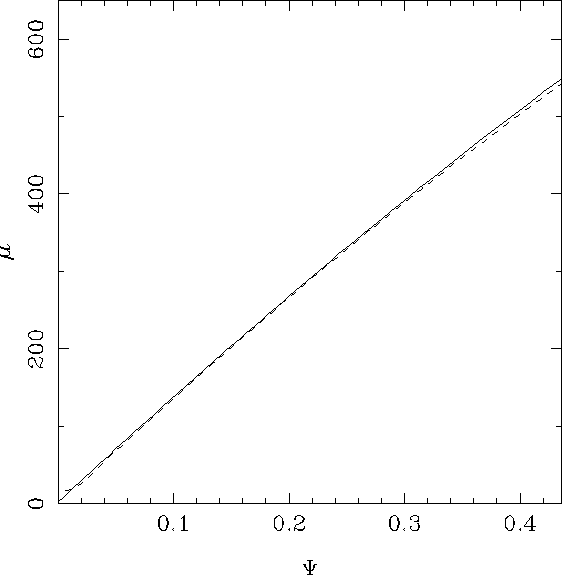}
\includegraphics[width=55mm]{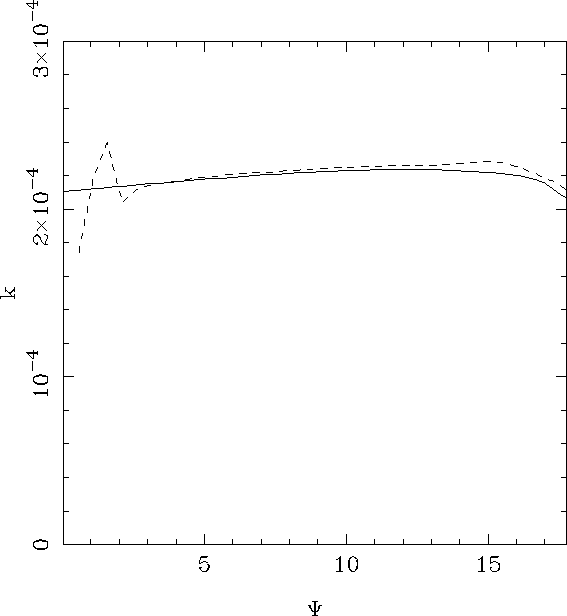}
\includegraphics[width=55mm]{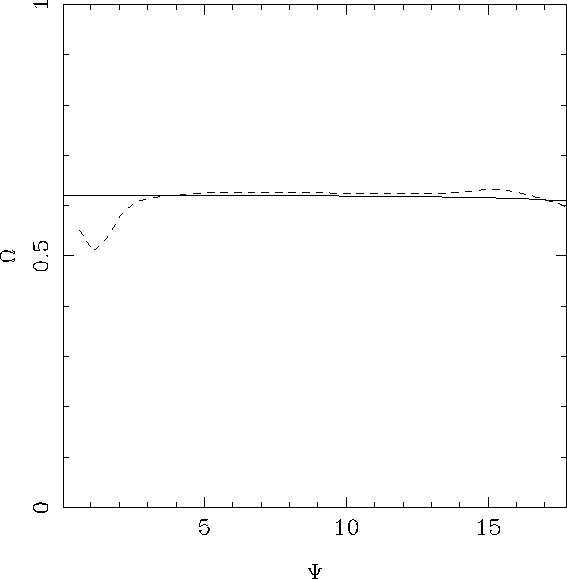}
\includegraphics[width=55mm]{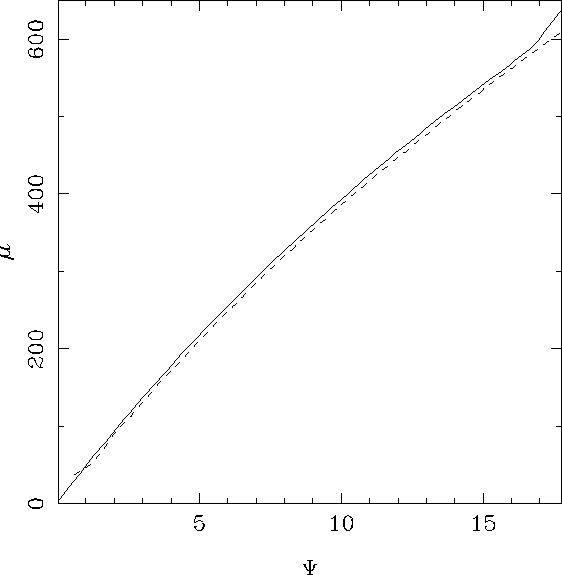}
\caption{Computational errors for models A (top row) and B1 (bottom row). 
The plots show the flow parameters $\massint(\Psi)$, $\Omega(\Psi)$ and
$\mu(\Psi)$ at the inlet (solid lines) and at $\eta=1\times 10^5$ 
for model A and $\eta=5\times 10^7$ for model B1 (dashed lines).
}
\label{constants}
\end{figure*}
%fffffffffffffffffffffffffffffffffffffffffffffffffffffffffffffffffff

\subsubsection{Other boundaries}

The computational domain is always chosen to be long enough for the
jet to be super--fast-magnetosonic when it approaches the outlet
boundary $\eta=\eta_o$. This justifies the use of radiative boundary
conditions at this boundary (i.e. we determine the state variables of
the boundary cells via extrapolation of the domain solution).

At the polar axis, $\xi=0$, we impose symmetry boundary conditions for
the dependent variables that are expected to pass through zero there,
\begin{eqnarray}
\nonumber f(-\xi)=-f(\xi)\, .
\end{eqnarray}
These variables include $B^{\hat\xi}$, $B^{\hat\phi}$, $u^{\hat\xi}$
and $u^{\hat\phi}$.  For other variables we impose a ``zero second
derivative'' condition,
\begin{eqnarray}
\nonumber \partial^2{f}/\partial{\xi^2} = 0\, ,
\end{eqnarray}
which means that we use linear interpolation to calculate the values of these
variables in the boundary cells.

We do this in order to improve the numerical representation of a
narrow core that develops in all cases as a result of the magnetic
hoop stress. Within this core the gradients in the $\xi$ direction are
very large and the usual zero-gradient condition, $f(-\xi)=f(\xi)$,
results in increased numerical diffusion in this region. We have
checked that this has a noticeable effect only on the axial region and
that the global solution does not depend on which of these two
conditions is used.

At the wall boundary, $\xi=\xi_j$, we use a reflection condition,
\begin{eqnarray}
\nonumber f(\xi_j+\Delta \xi)=-f(\xi_j-\Delta\xi) \, ,
\end{eqnarray}
for $B^{\hat\xi}$ and $u^{\hat\xi}$ and a zero-gradient condition for
all other variables.

%fffffffffffffffffffffffffffffffffffffffffffffffffffffffffffffffffff
\begin{figure*}
\includegraphics[width=70mm]{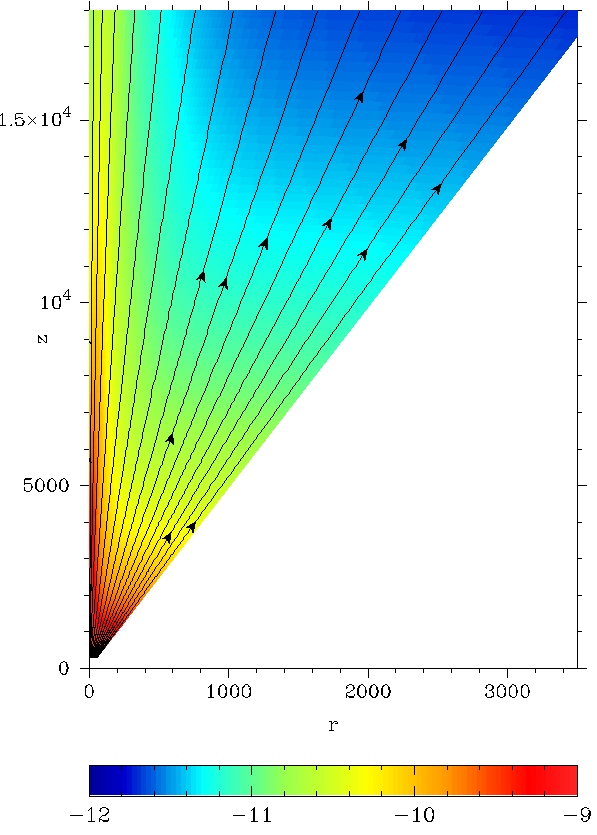}
\includegraphics[width=70mm]{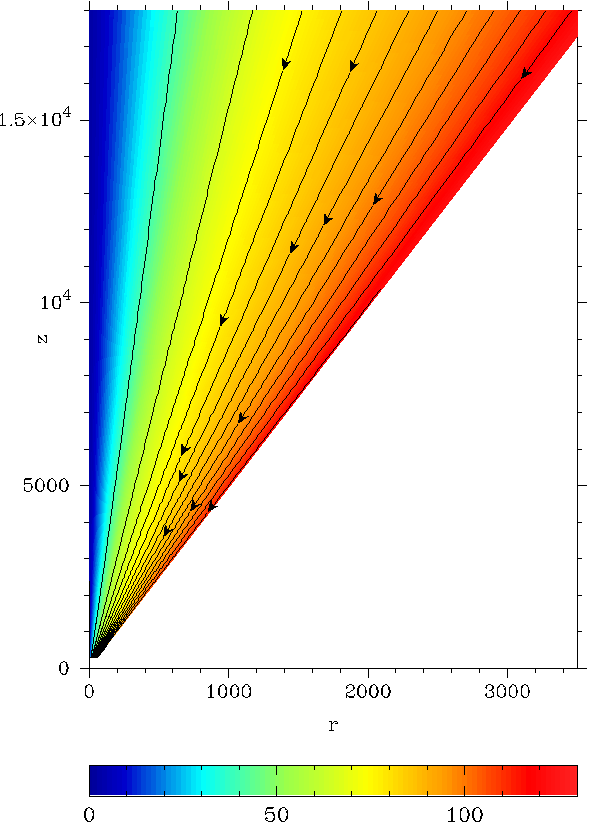}
\caption{Model A. Left panel shows $\log_{10}\rho'$
(colour), where $\rho'=\Gamma\rho$ is the jet density as measured in
the frame of jet source, and the magnetic field lines. Right panel
shows the Lorentz factor (colour) and the current lines. 
The light cylinder radius is $r_{\rm lc}=0.29$. 
}
\label{model-a}
\end{figure*}
%fffffffffffffffffffffffffffffffffffffffffffffffffffffffffffffffffff

%fffffffffffffffffffffffffffffffffffffffffffffffffffffffffffffffffff
\begin{figure*}
\includegraphics[width=70mm]{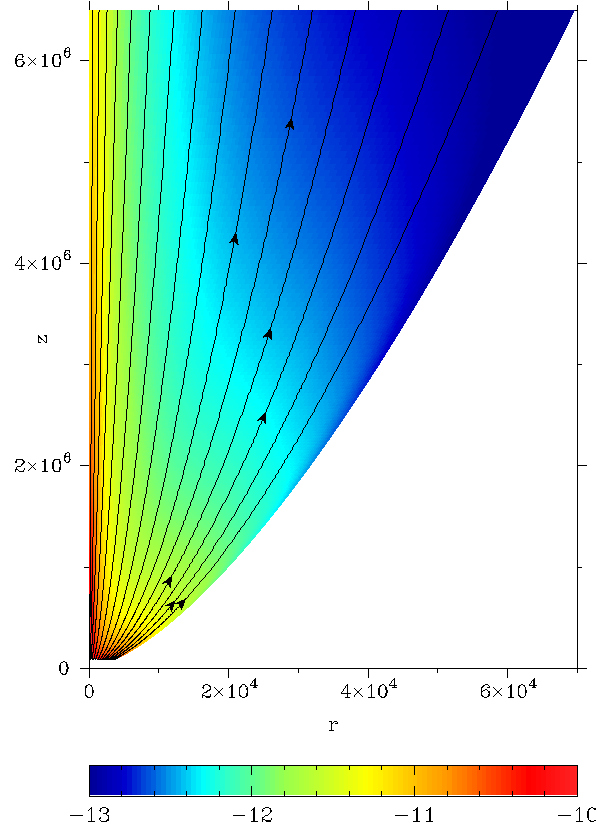}
\includegraphics[width=70mm]{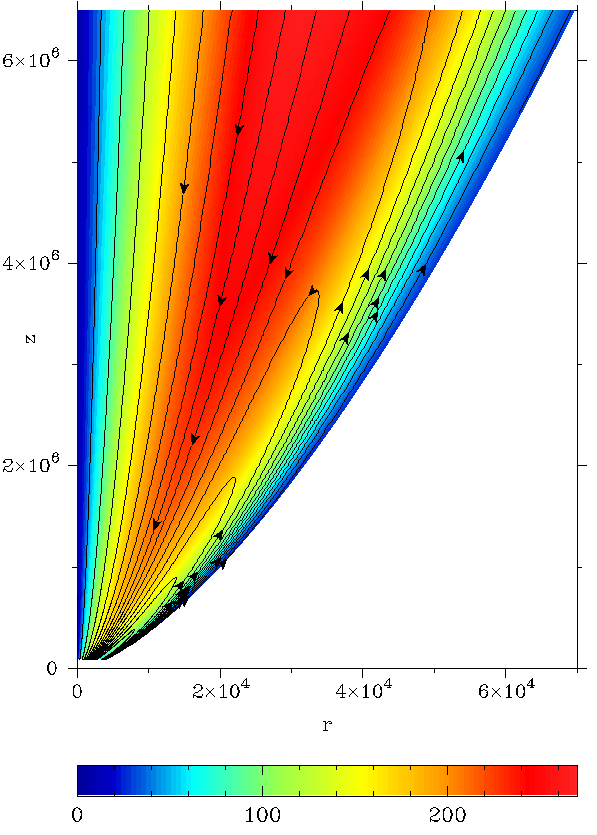}
\caption{Same as in Fig.~\ref{model-a}, but for model D. 
The closest to the inlet point of the Alfv\'en surface has the radius  
$r_{\rm lc}=1.3$. 
}
\label{model-d}
\end{figure*}
%fffffffffffffffffffffffffffffffffffffffffffffffffffffffffffffffffff
%fffffffffffffffffffffffffffffffffffffffffffffffffffffffffffffffffff
\begin{figure*}
\includegraphics[width=70mm]{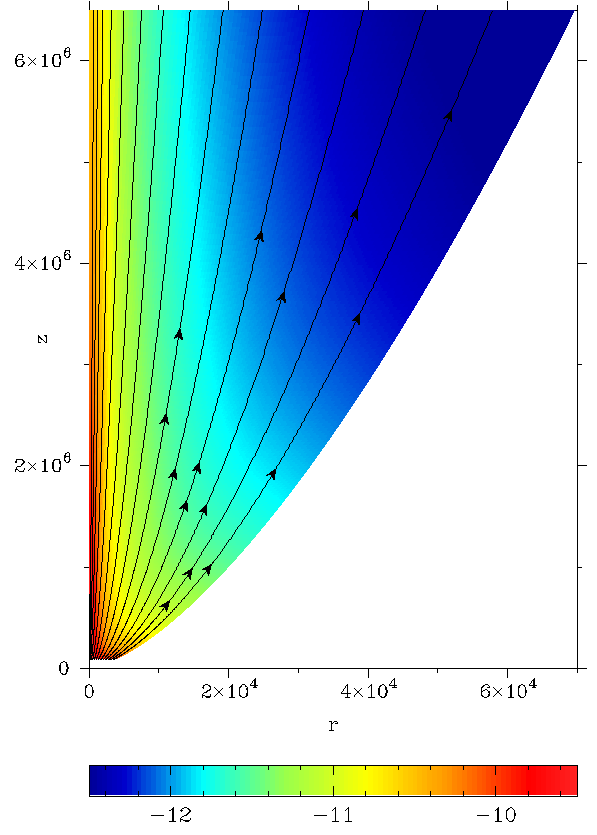}
\includegraphics[width=70mm]{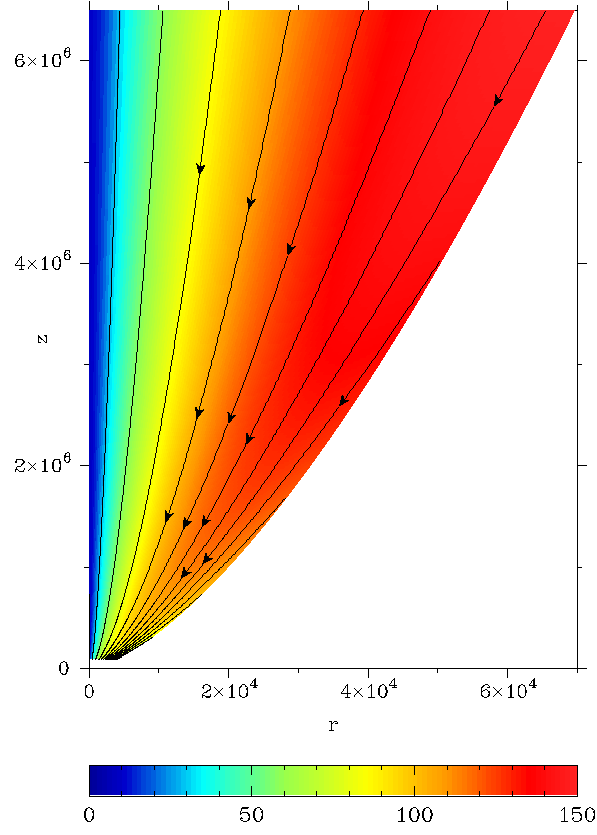}
\caption{Same as in Fig.~\ref{model-a}, but for model B2.
The light cylinder radius is $r_{\rm lc}=1.6$. 
}
\label{model-b2}
\end{figure*}
%fffffffffffffffffffffffffffffffffffffffffffffffffffffffffffffffffff
%fffffffffffffffffffffffffffffffffffffffffffffffffffffffffffffffffff
\begin{figure*}
\includegraphics[width=70mm]{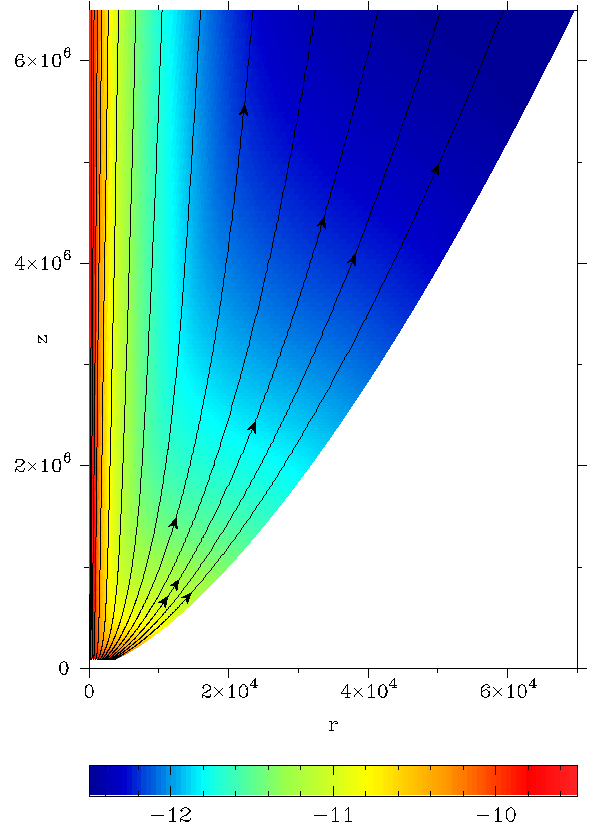}
\includegraphics[width=70mm]{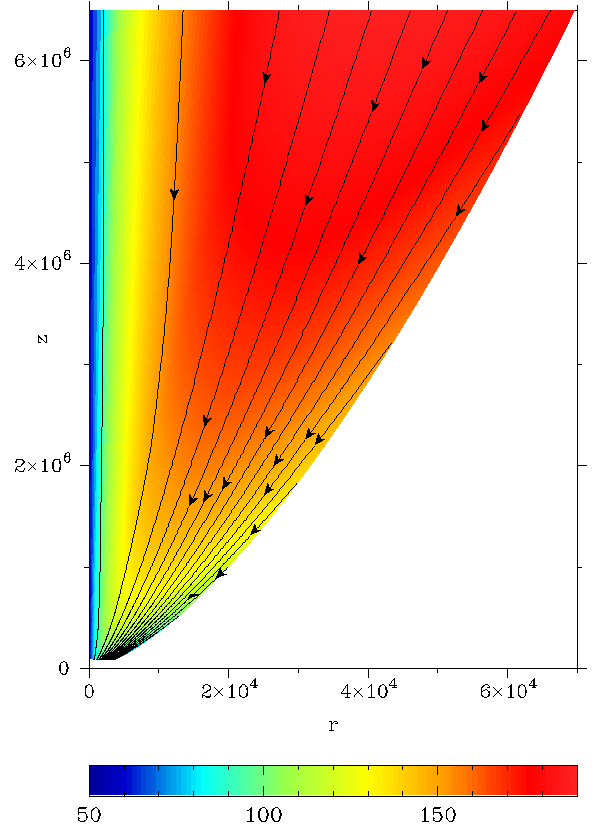}
\caption{Same as in Fig.~\ref{model-a}, but for model B2H. 
The light cylinder radius is $r_{\rm lc}=1.6$. 
}
\label{model-b2h}
\end{figure*}
%ffffffffffffffffffffffffffffffffffffffffffffffffffffffffffffffffff

%----------------------------------- 
\subsection{Initial setup}
%----------------------------------- 
\label{setup}

The initial configuration corresponds to a non-rotating, purely poloidal
magnetic field with approximately constant magnetic pressure across the
funnel. The plasma density within the funnel is set to a small value so
that the outflow generated at the inlet boundary can easily sweep it
away. In order to speed this process up the $\eta$ component of velocity
inside the funnel is set equal to $0.7\, c$, whereas the $\xi$ component
is set equal to zero.

%------------------------------------------
\subsection{Grid extensions}
%-------------------------------------------
 
The inner rings of the grid, where the grid cells are small and,
therefore, so is also
the time step, are the computationally most intensive regions of the
simulation domain. If we kept computing these inner rings during the
whole run then we would not be able to advance very far from the jet
origin. Fortunately, the transonic nature of the jet flow allows us
to cease computations in the inner region once the solution there
settles to a steady state.  To be more precise, we cut the funnel
along the $\xi$-coordinate surfaces into overlapping sectors with the
intention of computing only within one sector at any given time,
starting with the sector closest to the inlet boundary.  Once the
solution in the ``active'' sector settles to a steady state we switch
to the subsequent sector, located further away from the inlet. During
the switch the solution in the outermost cells of the active sector is
copied into the corresponding inner boundary cells of the subsequent
sector.  During the computation within the latter sector these inner
boundary cells are not updated.  This procedure is justified
only when the flow in a given sector cannot communicate with the flow
in the preceding sector through hyperbolic waves, and thus we 
ensure that the Mach cone of the fast-magnetosonic waves points
outward at the sector interfaces (see Paper~I). 

In these simulations we used up to 7 sectors, with each
additional sector being ten times longer than the preceding one. This
technique has enabled us to reduce the computation time by more than 
three orders of magnitude.  Although
the grid extension can in principle be continued indefinitely, there
are other factors that limit how far along the jet one can advance in
practice. Firstly, once the paraboloidal jets become highly collimated
the required number of grid cells along the jet axis increases, and
each successive sector becomes more computationally expensive than the
previous one. Secondly, errors due to numerical diffusion gradually
accumulate in the downstream region of the flow and the solution
becomes progressively less accurate (see Fig.~\ref{constants}).

%%%%%%%%%%%%%%%%%%%%%%%%%%%%%%%%%%%%%%%%%%%%%%%%%%%%%%%%%%%%%%%%%
\section{Results}
%%%%%%%%%%%%%%%%%%%%%%%%%%%%%%%%%%%%%%%%%%%%%%%%%%%%%%%%%%%%%%%%%
\label{results}

As is generally the case in numerical simulations, our computations are
subject to numerical errors, mainly the truncation errors of our RMHD
scheme.  The field-line constants described in
Section~\ref{section_integrals} can be used for a straightforward
evaluation of the absolute error.  Fig.~\ref{constants} shows the
ideal-MHD constants $\massint,\Omega$ and $\mu$ as functions of magnetic
flux at the inlet and near the outer boundary of the computational
domain for models A and B1. If the curves do not exactly coincide, this
is indicative of computational errors.  Although the plots exhibit
noticeable deviations, they remain relatively small, and we
conclude that the results are trustworthy.

%fffffffffffffffffffffffffffffffffffffffffffffffffffffffffffffffffff
\begin{figure}
\includegraphics[width=77mm]{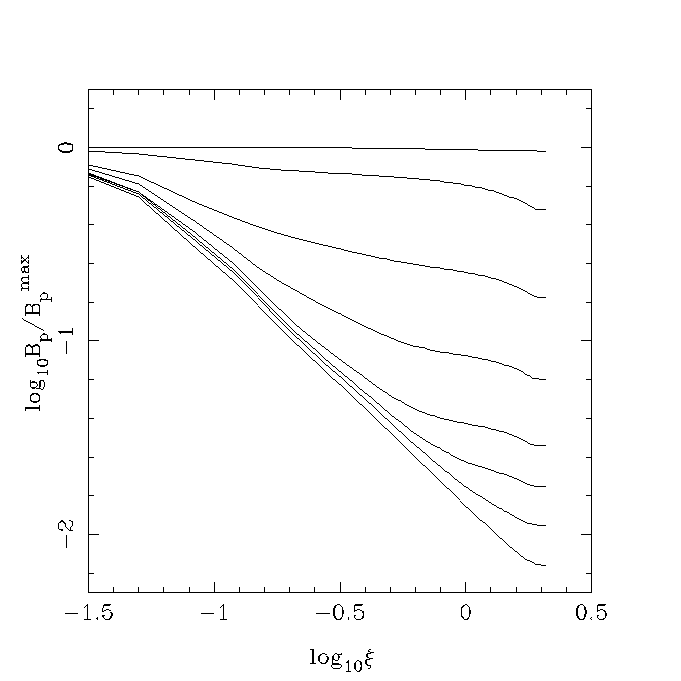}
\caption{Distribution of the poloidal magnetic field across the
jet of model B1, showing the development of an axial core as the
distance from the origin increases. From top to bottom, the curves
correspond to $\eta = 1$, $50$, $5\times 10^2$, $5 \times 10^3$, $5 \times
10^4$, $5\times 10^5$, $5\times 10^6$ and $5\times 10^7$, respectively.
}
\label{bp}
\end{figure}
%fffffffffffffffffffffffffffffffffffffffffffffffffffffffffffffffffff

%fffffffffffffffffffffffffffffffffffffffffffffffffffffffffffffffffff
\begin{figure*}
\includegraphics[width=77mm]{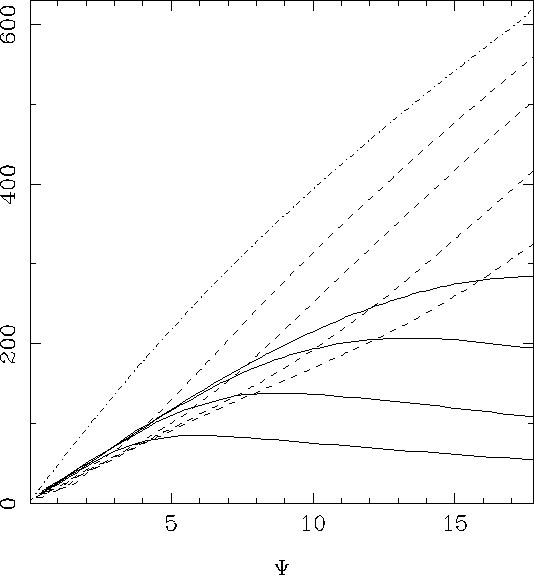}
\includegraphics[width=77mm]{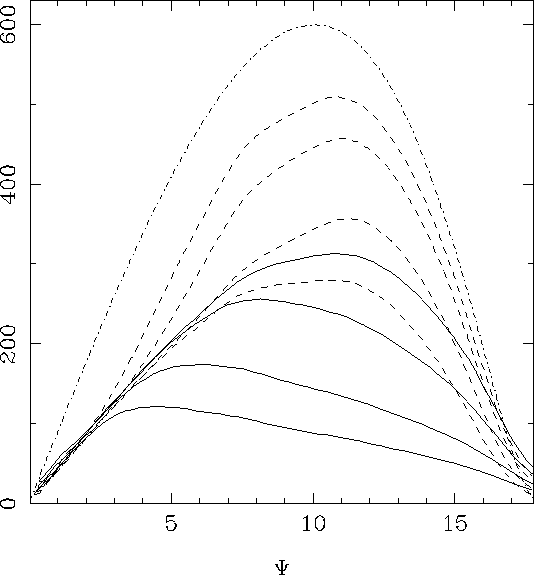}
\caption{Distribution of $\Gamma$ and $\mu_m=\mu_h\sigma$ across the jet in 
models B1 (left panel) and D (right panel).  Solid lines show $\Gamma$ at 
$\eta=5\times10^4,5\times10^5,5\times10^6,5\times10^7$ (increasing upward),
dashed lines show $\mu_h\sigma$ at the same locations (increasing downward), 
and the dash-dotted line shows $\mu$.  
}
\label{cross2}
\end{figure*}
%fffffffffffffffffffffffffffffffffffffffffffffffffffffffffffffffffff
                                                                               
Figs.~\ref{model-a}--\ref{model-b2h} show the general 2D structure of
the derived jet solutions for models A, D, B2 and B2H midway from the
inlet surface. We selected these particular cases since they represent the most
significant variations in the model parameters, namely the
transition (i) from conical to paraboloidal shape of the confining wall
(A and B2), (ii) from uniform to differential rotation at the base (A
and D)\footnote{Note that, when displaying results for model D, we
define the fiducial light-cylinder radius in terms of the angular
velocity $\Omega_0$ of the innermost field line.}
and (iii) from cold to initially hot flows (B2 and B2H).  In
general, the structure of the simulated ultra-relativistic jets is very
similar to that of the moderately-relativistic conical jets studied in
Paper~I. All models show the development of a central core where the
source-frame mass density $\rho'=\Gamma\rho$ peaks. 
The mass concentration is accompanied by a bunching-up of the poloidal
magnetic field lines near the axis, as further illustrated in
Fig.~\ref{bp}. The development of an axial core is a generic property of
axisymmetric MHD outflows from a rotating source \citep{Bog95} and was
also a feature of the jets simulated in Paper~I. The distribution
of the Lorentz factor across the jet varies, however, from case to case.
In model A $\Gamma$ has 
its maximum value at the jet boundary (Fig.~\ref{model-a}). In model D
the maximum is located approximately midway between the symmetry
axis and the boundary (Fig.~\ref{model-d}). This reflects the fact that
the angular velocity of magnetic field lines, and hence the
electromagnetic energy flux (equation~\ref{mu_m-def}), vanishes at the
boundary in this model, resulting in $\mu \approx \mu_h \approx 1$ near
the wall (see Fig.~\ref{cross2}).
The Lorentz factor of the initially cold jet in model B2 at first peaks
near the axis, with its value decreasing slightly on the way to the jet
boundary. However, further downstream the maximum shifts towards the
boundary and eventually disappears. In the initially hot jet of model
B2H the Lorentz factor at first peaks right on the symmetry axis, where
the acceleration is due to the by the gas pressure. However, further
downstream its evolution is similar to that of model B2.

\begin{figure*}
\includegraphics[width=70mm]{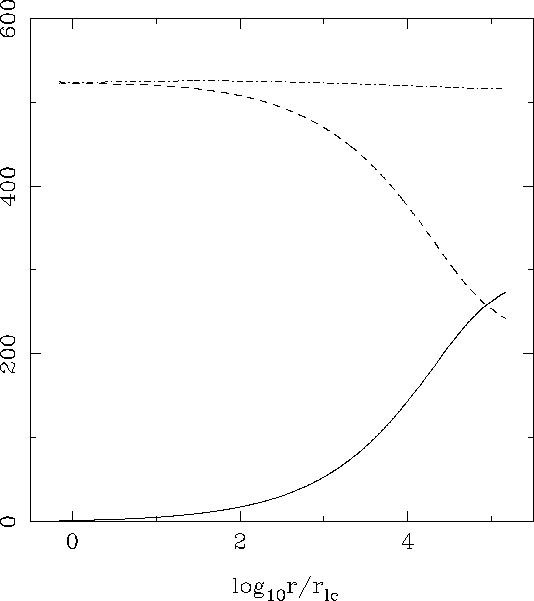}
\includegraphics[width=70mm]{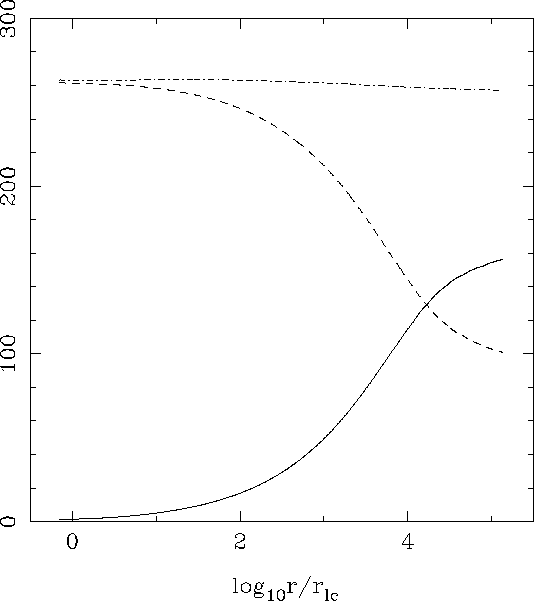}
\includegraphics[width=70mm]{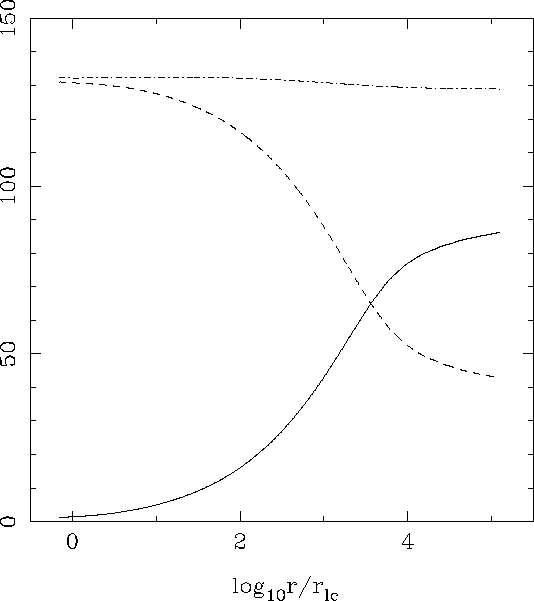}
\includegraphics[width=70mm]{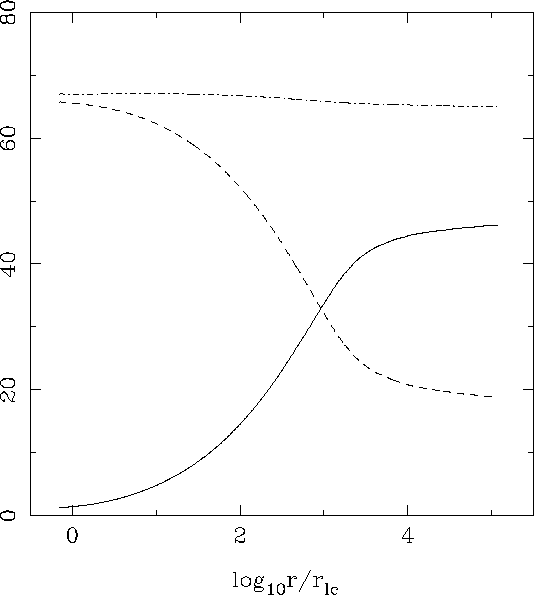}
\caption{$\Gamma$  (solid line), $\mu_m=\mu_h\sigma$  (dashed line) and
$\mu$ (dash-dotted line) along the magnetic field line with
$\Psi=0.8\Psi_{\rm max}$
as a function of cylindrical radius for models 
B1 (top left panel), 
B2 (top right panel), 
B3 (bottom left panel) and
B4 (bottom right panel).
}
\label{sigma}
\end{figure*}

\begin{figure*}
\includegraphics[width=55mm]{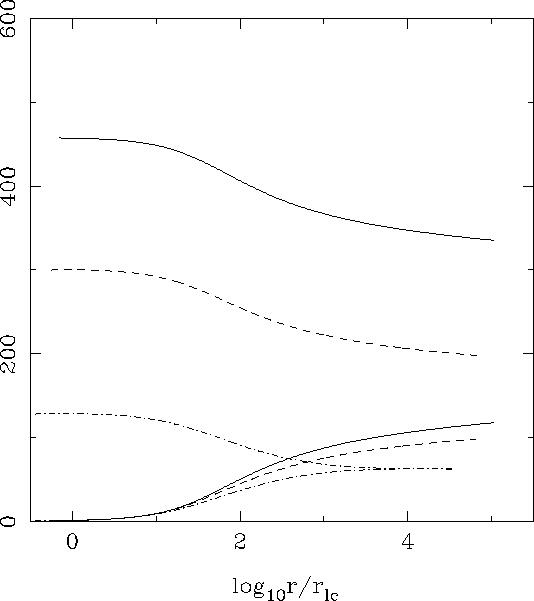}
\includegraphics[width=55mm]{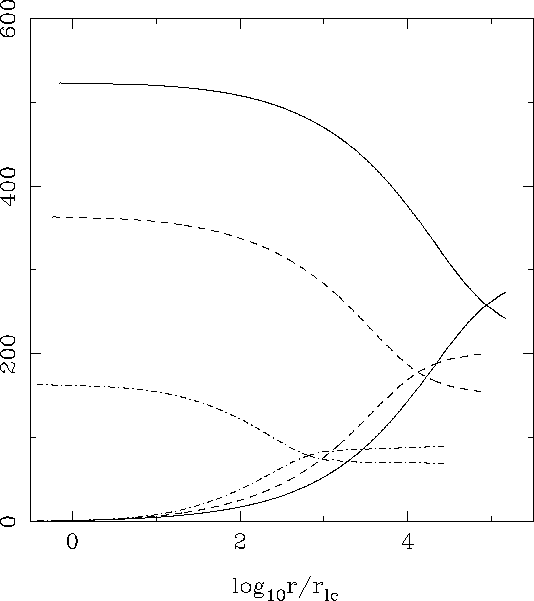}
\includegraphics[width=55mm]{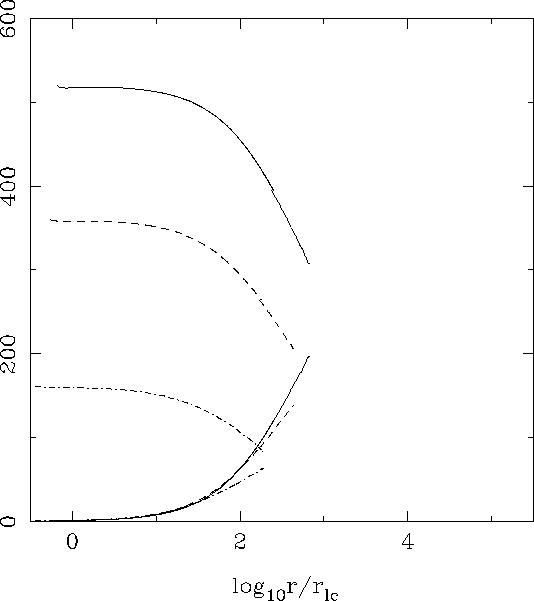}
\caption{$\Gamma$ (increasing functions of r) and 
$\mu_m=\mu_h\sigma$ (decreasing functions of r) 
along the magnetic field lines $\Psi=0.8\Psi_{\rm max}$ (solid lines), 
$\Psi=0.5\Psi_{\rm max}$ (dashed lines) and $\Psi=0.2\Psi_{\rm max}$
(dash-dotted lines) in models A (left panel), B1 (middle panel) and C
(right panel).
}
\label{geom-effect}
\end{figure*}

\begin{figure*}
\includegraphics[width=77mm]{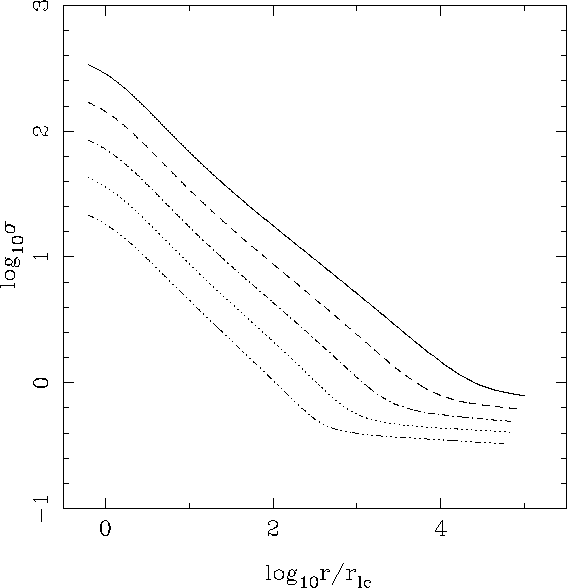}
\includegraphics[width=77mm]{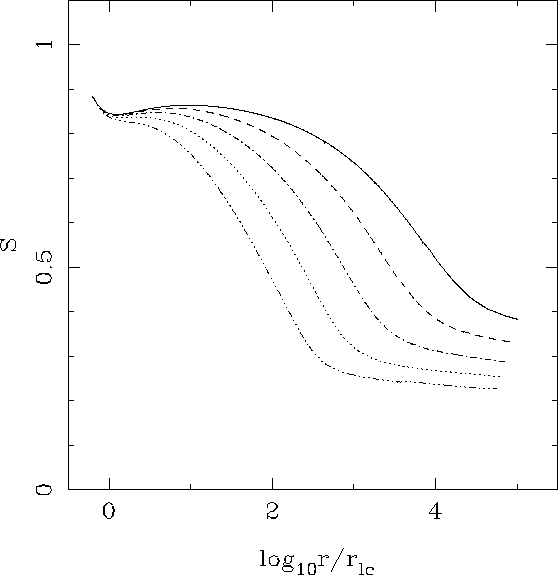}
\caption{Left panel: Evolution of $\sigma$ along the magnetic field line 
$\Psi=0.8\Psi_{\rm max}$ in models B1 (solid line), B2 (dashed line), 
B3 (dash-dotted line), B4 (dotted line) and B5 (dash-triple-dotted line). 
Right panel: Evolution of the bunching function ${\cal S}=\pi B_p r^2/\Psi$
for the same models along the same magnetic field line.   
}
\label{sigma-ev}
\end{figure*}

Fig.~\ref{sigma} shows the efficiency of plasma acceleration 
along the magnetic surface $\Psi=0.8\Psi_{\rm max}$ (located near the
jet boundary) for models B1--B4, which differ only by the strength of
the initial magnetization. One can see that in all four cases the
kinetic energy flux, $\simeq \mu_h \rho u_p c^2 \simeq \Gamma\rho u_p c^2$, 
eventually exceeds the Poynting flux, $\mu_m \rho u_p c^2$. 
This magnetic surface is not exceptional 
and a similar behaviour is exhibited along other flux surfaces. This is 
illustrated by Figs.~\ref{cross2}~and~\ref{geom-effect}. These figures also 
show that soon after reaching equipartition the plasma acceleration 
slows down significantly:
this is consistent with the relation $\mu \approx \Gamma (1+\sigma)$
obtained from equations~(\ref{kap-def}), (\ref{mu_m-def})
and~(\ref{mu_h-def}), in which crossing the equipartition point
corresponds to the magnetization parameter $\sigma$ dropping below 1.
Fig.~\ref{sigma} further indicates that the 
efficiency of magnetic acceleration is higher the lower the initial
magnetization. This is reflected in the behaviour of $\sigma$, the ratio
of the Poynting flux to the matter energy flux  (see
Section~\ref{section_integrals}). The left panel of Fig.~\ref{sigma-ev}
shows that the fast initial decrease of $\sigma$ slows down at
a higher value of $\sigma$ when the initial magnetization is larger. If
this behaviour in fact extends to values of $\mu_{m0}\approx \mu$ that are
low enough for the maximum attainable speed to remain nonrelativistic then the
indicated inverse correlation is consistent with the very high
acceleration efficiency exhibited by MHD outflow solutions in the
Newtonian regime \citep[e.g.][]{V00}.

%fffffffffffffffffffffffffffffffffffffffffffffffffffffffffffffffffff
%fffffffffffffffffffffffffffffffffffffffffffffffffffffffffffffffffff

%fffffffffffffffffffffffffffffffffffffffffffffffffffffffffffffffffff
%fffffffffffffffffffffffffffffffffffffffffffffffffffffffffffffffffff

%fffffffffffffffffffffffffffffffffffffffffffffffffffffffffffffffffff
%fffffffffffffffffffffffffffffffffffffffffffffffffffffffffffffffffff

The high efficiency of magnetic acceleration is not unique to models in
which the magnetic field lines rotate uniformly. Fig.~\ref{cross2}, in
which the results for model B1 are compared with those for model D,
shows that equally effective acceleration is achievable in the case of a
differentially rotating source.

\begin{figure}
\includegraphics[width=77mm]{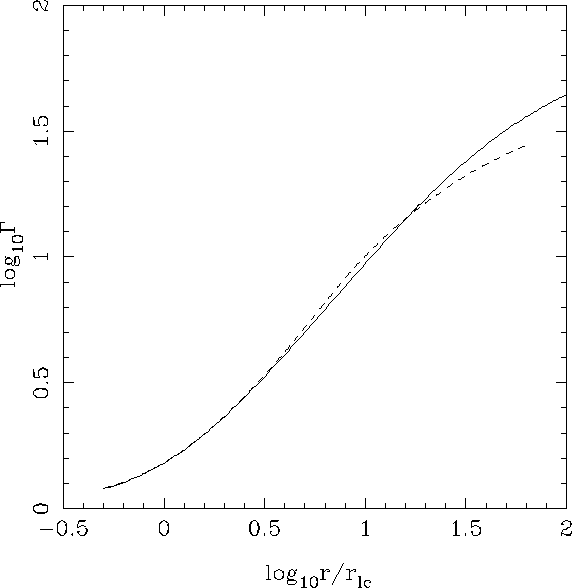}
\caption{
Lorentz factor along the magnetic field line with 
$\Psi=0.5\Psi_{\rm max}$ for model A (solid line) and model E (dashed line). 
}
\label{geu2}
\end{figure}

The geometry of the bounding wall has a pronounced effect on the
acceleration efficiency, as demonstrated by Fig.~\ref{geom-effect}. A
larger value of the power-law index $a$ in the shape function $z\propto
r^a$ corresponds to a more rapidly rising function $\Gamma(r/r_{\rm
lc})$ along a given magnetic flux surface $\Psi=$const. Whereas in the
model B1 ($a=3/2$) the acceleration slows down only after the
equipartition point, in model A ($a=1$) this occurs much earlier and, as
a result, equipartition between magnetic and kinetic energy is reached
only near the jet axis.  Equipartition is not reached in model C ($a=2$)
either (see Fig.~\ref{geom-effect}), but for a different reason. Due to
the higher degree of external collimation, this jet eventually becomes
very thin. This makes our simulation increasingly expensive and we are
forced to terminate it before reaching sufficiently large jet radii.
(Moreover, the computational errors are accumulated over a longer path
along the jet and would become rather high if we continued.)  However,
Fig.~\ref{geom-effect} shows that in this model the Lorentz factor is a
faster growing function of cylindrical radius compared to model B1.
Finally, in model E ($a=2/3$) we consider a jet propagating in a channel
with a progressively diverging wall, which in practice may correspond to
the polar funnel of a thick accretion disc \citep[e.g.][]{PW80}. In this
case the jet eventually becomes detached from the wall and then expands
as a conical outflow (Fig.~\ref{geu1}). The acceleration rate is similar
to that of model A (see Fig.~\ref{geu2}).

The initially hot jet, model B2H, is subject to both magnetic and
thermal acceleration, so, as expected, the Lorentz factor in this case
grows faster compared to the corresponding cold jet (see the right panel
of Fig.~\ref{gbut}). But a closer inspection reveals that the
acceleration process exhibits a new mode of behaviour in this case (one
that was, however, found before in semi-analytic self-similar solutions;
see \citealt{VK03b}). It is seen that a significant fraction of thermal
energy is at first converted into Poynting flux. The middle panel of
Fig.~\ref{gbut} shows that the Poynting-to-mass flux ratio $\mu_m c^2$
grows until $r\simeq 10^2r_{\rm lc}$ and only then starts to
decline. However, this decrease is quite fast and the terminal value of
$\mu_m$ for the chosen magnetic flux surface ($\Psi=0.5\Psi_{\rm max}$)
is, in fact, lower than in the corresponding cold jet (model B2) shown
in the left panel of this figure, with a correspondingly higher
asymptotic Lorentz factor.

The distribution of the terminal bulk Lorentz factor across these two
jet models is shown in right panel of Fig.~\ref{gbut}. One can see that
on the axis the Lorentz factor of the hot jet is higher than that of the
cold jet by approximately the value of the initial thermal Lorentz factor,
$\Gamma_{t0}=55$.  This is as expected given that magnetic acceleration
does not operate along the axis. However, at the wall the difference is
only half as large and in the middle of the jet it is higher than
40. These traits are evidently a consequence of the thermal-to-Poynting
energy conversion and its effect on the poloidal magnetic field
distribution, as discussed in Section~\ref{hot}.

Although the case of an unconfined wind may not be directly relevant to
GRB flows, which are inferred to undergo a fairly efficient collimation
(see Section~\ref{introduction}), it is certainly of interest to the
pulsar community. Furthermore, it is worth investigating from a purely
theoretical point of view. The acceleration details for this case (model
AW) are presented in Fig.~\ref{gpw}. The lower efficiency of magnetic
acceleration noted in the conical-wall case (model A), particularly near
the jet boundary, is even more pronounced in this instance. As can be
seen in the right panel of Fig.~\ref{gpw}, only $\simeq 5\%$ of the
Poynting flux injected at $\simeq 12^\circ$ to the equatorial direction has been
converted into kinetic energy by the time the cylindrical radius grew to
$r=10^6r_{\rm lc}$.  
Although, as shown in the left panel of Fig.~\ref{gpw}, the
efficiency is higher near the symmetry axis, the terminal Lorentz factor
there remains comparatively low because of the reduced effectiveness of
magnetic acceleration as the polar angle approaches zero.

%%%%%%%%%%%%%%%%%%%%%%%%%%%%%%%%%%%%%%%%%%%%%%%%%%%%%
\section{Analysis of the Results}
%%%%%%%%%%%%%%%%%%%%%%%%%%%%%%%%%%%%%%%%%%%%%%%%%%%%%
\label{theory}

%sssssssssssssssssssssssssssssssssssssssssssssss
\subsection{Efficiency of magnetic acceleration}
%sssssssssssssssssssssssssssssssssssssssssssssss
\label{efficiency}

The steady-state structure of a magnetized relativistic outflow can be
understood by analysing the momentum equation.  After the partial
integration described in Section~\ref{section_integrals}, two more
equations remain to be considered, corresponding to the two components
of the momentum equation in the poloidal plane.  Since the main part of
the acceleration occurs in the super-Alfv\'enic region of the flow, it
is sufficient to examine only this regime.  We further simplify the
discussion by taking the flow to be cold.  Thermal effects, when
present, in any case only affect the initial acceleration region of the
flow; we consider them in Section~\ref{hot}. We now proceed to extend
the discussion in Paper~I by taking the $\Gamma\gg 1$ of the constituent
equations, appropriate for the ultrarelativistic flows simulated in the
present work, which enables us to derive analytic scalings.

\begin{figure}
\includegraphics[width=77mm]{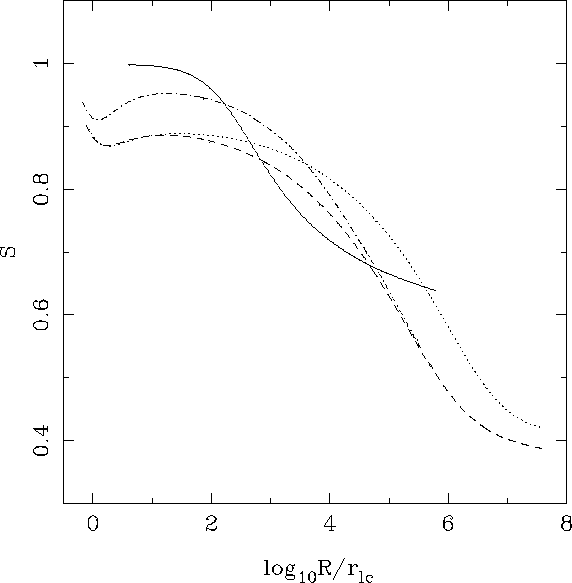}
\caption{Evolution of the function ${\cal S} =\pi B_p r^2/\Psi$ along 
the magnetic field line with $\Psi=0.5\Psi_{\rm max}$ in models A 
(solid line), B1 (dashed line), C (dash-dotted line) and D(dotted line).
}
\label{S-ev}
\end{figure}

For cold flows $\mu_h\approx \Gamma$ (equation~\ref{mu_h-def}), and from
equation~(\ref{kap-def}) one finds that $\Gamma\approx \mu-\mu_m$. 
Substituting the electric current from equation~(\ref{I1}) into
equation~(\ref{mu_m-def}), we get
\begin{equation}
  \mu_m \approx \frac{\Psi \Omega^2}{4 \pi^2 \massint c^3} \, {\cal S} \,,
\label{S}
\end{equation}
where 
\begin{equation}
   {\cal S} 
= \frac{\pi r^2 B_p}{\int \bmath{B}_p \! \cdot \! d \bmath{S}} 
= \frac{\pi r^2 B_p}{\Psi}
= \frac{r |\vgrad{\Psi} |}{2\Psi}\,.
\label{calS}
\end{equation}
Thus, the flow Lorentz factor can be written as 
\begin{equation}
\Gamma\approx \mu- \frac{\Psi \Omega^2}{4 \pi^2 \massint c^3} \, {\cal S} \,.
\label{moment1}
\end{equation}
All the quantities except for ${\cal S}$ on the right-hand side of this
equation are field-line constants, so an increase in
$\Gamma$ along a field line necessarily requires ${\cal S}$ to decrease.  
The function ${\cal S}$
is a measure of how bunched the poloidal field lines are --- indeed,
it is equal to the ratio of $B_p$ at some cylindrical radius $r$ along the
field line to the
mean magnetic field within that radius, $\Psi/\pi r^2$. 
For example, for a flow confined within a sufficiently small 
angle that satisfies $B_p\propto r^\lambda$, $\Psi\propto r^{\lambda+2}$ and
\begin{eqnarray}
{\cal S} = \frac{\lambda+2}{2}\,. 
\nonumber
\end{eqnarray}
For a uniform distribution of $B_p$ this yields ${\cal S}=1$, whereas
one has ${\cal S}>1$ if $B_p$ increases with $r$ and ${\cal S}<1$ 
if it decreases. This shows that magnetic acceleration requires 
a gradual concentration of magnetic flux in the central part of 
the flow. In the case of a collimating flow this can be achieved through
a faster collimation of the inner magnetic flux surfaces than of the
outer ones, and  in the case of a decollimating flow a faster
decollimation of the outer flux surfaces is required.       
Fig.~\ref{bp} illustrates the concentration of magnetic flux toward the axis
in one of our simulations. In this case, at large distances 
the poloidal magnetic field scales roughly as $B_p \propto r^{-1.2}$, 
corresponding to ${\cal S}_\infty \sim 0.4$.
This is indeed the asymptotic value of ${\cal S}$, as shown in Fig.~\ref{S-ev}.

%fffffffffffffffffffffffffffffffffffffffffffffffffffffffffffffffffff
\begin{figure}
\includegraphics[width=77mm]{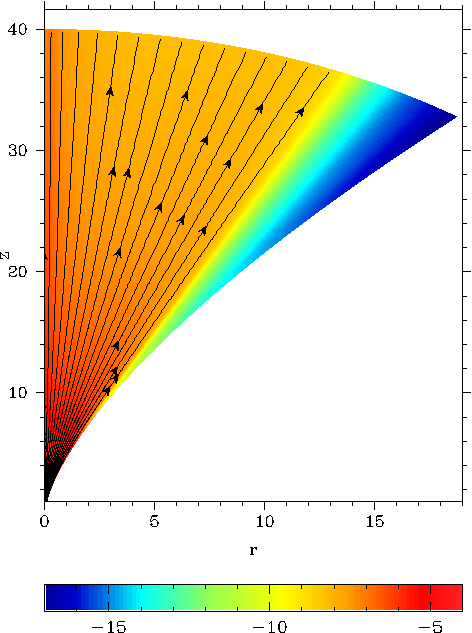}
\caption{Colour image shows
$\log_{10}p_{\rm tot}$ (with the total pressure given by $p_{\rm tot} =
p + B_{\rm co}^2/8\pi$, where $B_{\rm co}$ is the comoving magnetic
field) and the contours show the magnetic field lines for model E.
In this model the light cylinder radius is $r_{\rm lc}=0.29$. 
}
\label{geu1}
\end{figure}
%fffffffffffffffffffffffffffffffffffffffffffffffffffffffffffffffffff

%fffffffffffffffffffffffffffffffffffffffffffffffffffffffffffffffffff
%fffffffffffffffffffffffffffffffffffffffffffffffffffffffffffffffffff

%fffffffffffffffffffffffffffffffffffffffffffffffffffffffffffffffffff
\begin{figure*}
\includegraphics[width=55mm]{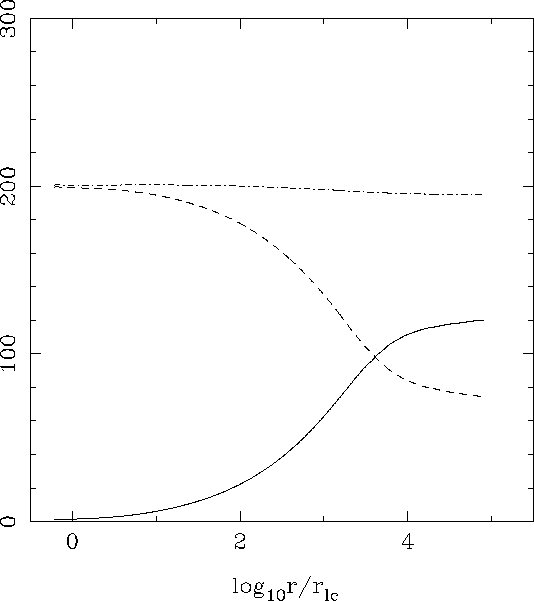}
\includegraphics[width=55mm]{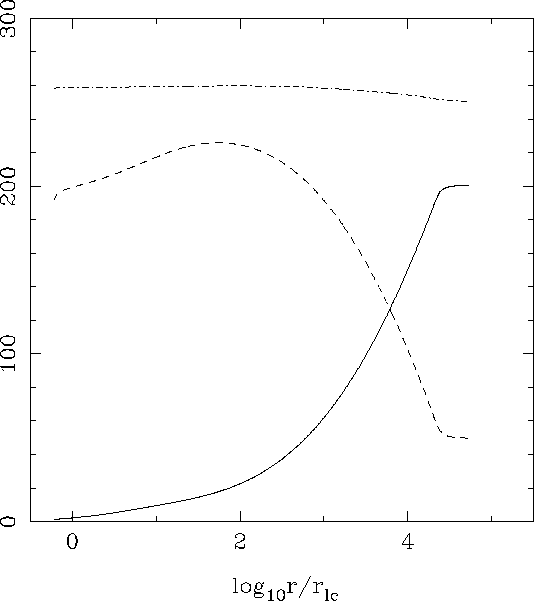}
\includegraphics[width=59mm]{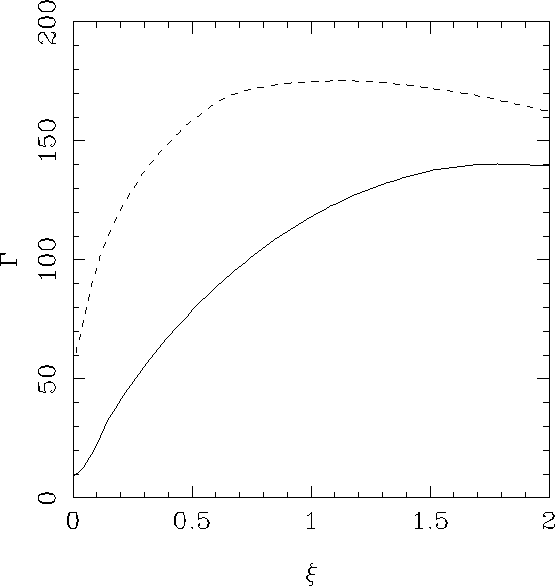}
\caption{Effects of thermal acceleration. Left panel: cold jet of model B2.
Middle panel: hot jet of model B2H (with $w_0/\rho_0 c^2=55$).
The lines show  $\Gamma$  (solid line), $\mu$  (dash-dotted line),
$\mu_m=\mu_h\sigma$ (dashed line) and $(w/\rho c^2-1)\Gamma$ (dotted
line) along the magnetic 
field line with $\Psi=0.5\Psi_{\rm max}$ as a function of cylindrical radius.
Right panel: Lorentz factor across the jet at $\eta=4\times10^6r_{\rm lc}$ 
for the cold jet of model B2 (solid line) and the hot jet of model B2H
(dashed line).}
\label{gbut}
\end{figure*}
%fffffffffffffffffffffffffffffffffffffffffffffffffffffffffffffffffff

%fffffffffffffffffffffffffffffffffffffffffffffffffffffffffffffffffff
\begin{figure*}
\includegraphics[width=80mm]{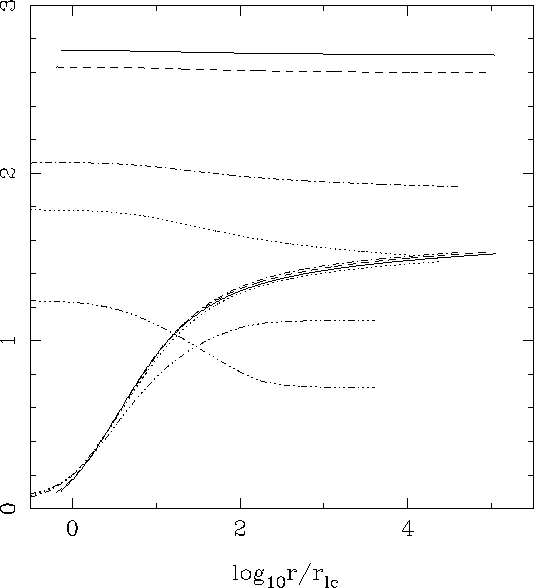}
\includegraphics[width=80mm]{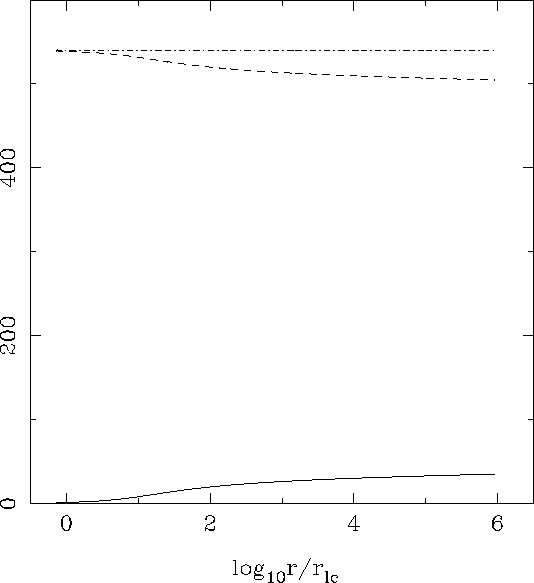}
\caption{Unconfined wind solution (model AW). Left panel:
Lorentz factor (increasing function) and
$\mu_h\sigma$ (decreasing function) along five different
magnetic field lines:
$\Psi=0.8\Psi_{\rm max}$ (solid line),
$\Psi=0.5\Psi_{\rm max}$ (dashed line),
$\Psi=0.2\Psi_{\rm max}$ (dash-dotted line),
$\Psi=0.1\Psi_{\rm max}$ (dotted line),
$\Psi=0.027\Psi_{\rm max}$ (dash-triple-dotted line),
the last line originating
from the same point at the inlet as the $\Psi=0.8\Psi_{\rm max}$ line of
model A.
Right panel:
$\Gamma$  (solid line), $\mu_h\sigma$  (dashed line) and
$\mu$ (dash-dotted line) along the magnetic field line with
$\Psi=0.8\Psi_{\rm max}$
as a function of cylindrical radius.
}
\label{gpw}
\end{figure*}
%fffffffffffffffffffffffffffffffffffffffffffffffffffffffffffffffffff

Equation~(\ref{moment1}) is a consequence of the momentum equation along
the flow. It shows how $\Gamma$ increases by the action of the $(1/c)
\bmath{J}_p \! \times \! \bmath{B}_\phi$ force when the function ${\cal
S}$ decreases along the flow, thereby demonstrating the intimate
connection between the acceleration efficiency and the evolution of the
poloidal shape of the flow. 
In evaluating this efficiency we can use ${\cal S}_{\rm f}$, the value of
${\cal S}$ at the fast surface, as a convenient proxy for the initial
value of ${\cal S}$. This is because, for $\mu\gg 1$, $\Gamma$ remains
$\ll \mu$ on this surface \citep[e.g.][]{K04}. In this case the two
terms on the right-hand side of equation~(\ref{moment1}) are comparable,
and we obtain
\begin{eqnarray}
{\cal S}_{\rm f}=\frac{4 \pi^2 \massint \mu c^3}{\Psi \Omega^2} \,.
\nonumber
\end{eqnarray}
We can legitimately use equation~(\ref{moment1}) since the fast surface
lies well outside the light cylinder and hence is in the
super-Alfv\'enic domain for most of the simulated field lines. We now
utilize this equation to write the asymptotic Lorentz factor in the form
\begin{equation}
\Gamma_\infty \approx \mu (1-{\cal S}_\infty/{\cal S}_{\rm f})\, .
\label{G_infty}
\end{equation}
In our simulations ${\cal S}_{\rm f}\approx 0.9$ (see Figs.~\ref{sigma-ev} 
and~\ref{S-ev}). This value reflects the adopted uniform distribution of
$B^{\hat{\eta}}$ at the inlet.\footnote{As we already in
Section~\ref{inlet_section}, we have experimented with other
distributions that put more flux near the axis and observed a quick
``uniformization'' of magnetic flux in the immediate vicinity of the
inlet under the action of magnetic pressure.} Beyond the Alfv\'en
surface the azimuthal magnetic field component becomes dominant, and its
hoop stress causes the inner flux surfaces to collimate faster than the
outer ones. As a result ${\cal S}$ decreases, attaining asymptotic values 
${\cal S}_\infty \approx 0.25 - 0.4$ for paraboloidal jets 
(see Figs.~\ref{sigma-ev} and~\ref{S-ev}).    
%
%fffffffffffffffffffffffffffffffffffffffffffffffffffffffffffffffffff
%fffffffffffffffffffffffffffffffffffffffffffffffffffffffffffffffffff
%
The implied asymptotic Lorentz factors thus satisfy
\begin{eqnarray}
\Gamma_\infty/\mu  \approx 0.55 - 0.72\, ,
\nonumber
\end{eqnarray}
which are indeed the values reached by our simulated flows
(see Figs.~\ref{cross2}--\ref{geom-effect}).
This result indicates that $\ga 50\%$ of the initial Poynting flux 
is converted into kinetic energy of bulk motion (see also
\citealp{V04dogl}). 
The significantly lower efficiency found in our simulations of flows
inside conical and diverging funnels, down to $25\%$ near the boundary
(models A and E), is most likely due to the loss of causal 
connection across the flow (see Section~\ref{causality}).

%sssssssssssssssssssssssssssssssssssssssssssssss
\subsection{Power-law acceleration phase}
%sssssssssssssssssssssssssssssssssssssssssssssss
\label{power-law}

Next we analyse the trans-field component of the momentum equation. 
The asymptotic form of the trans-field
equation in the highly relativistic limit is
\begin{equation}\label{transf}
\frac{\Gamma^2 r}{{\cal R}} \approx 
\frac{ {\displaystyle 
\left(\frac{2I}{\Omega B_p r^2} \right)^2
r \vgrad{\ln\left|\frac{I}{\Gamma}\right|} 
\!\cdot\! \frac{\vgrad{\Psi}}{|\vgrad{\Psi}|}
} }{ {\displaystyle 1+ \frac{w}{\rho c^2} \frac{4
\pi \rho u_p^2}{B_p^2} \frac{r_{\rm lc}^2}{r^2}} }
-\Gamma^2\frac{r_{\rm lc}^2}{r^2}
\frac{\vgrad{r} \!\cdot\!\vgrad{\Psi}}{|\vgrad{\Psi}|}
\end{equation}
where ${\cal R}$ is the curvature radius of poloidal field lines 
(see equation~16 and related discussion in \citealp{V04}).
The three terms of this equation are 
the poloidal curvature term (left-hand side), the electromagnetic term
(first on the right-hand side), which is of order 1, 
and the centrifugal term (second on the right-hand side).
This important equation, with the centrifugal term omitted,
was derived by \citet{CLB91}, \citet{2001ApJ...562..494L}, 
and \citet{2002ApJ...573L..31O}, while 
\citet{Bog95}, \citet{2000AstL...26..208B}, 
and \citet{2003ApJ...592..321T}
derived the same equation with the centrifugal
term included but the poloidal curvature term omitted.

Well outside the light cylinder, where 
$r\Omega \gg v^{\hat{\phi}}$ and $v\simeq c$, equations~(\ref{v_xi_b})
and~(\ref{v_phi_b}) imply
\begin{equation}
r B^{\hat{\phi}}=-\frac{1}{c} \Omega B_p r^2\, .
\end{equation}
\noindent
{}From this equation and equation~(\ref{I}) one finds that
\begin{equation}\label{I-Bp}
I=-\frac{1}{2} \Omega B_p r^2\, ,
\end{equation}
\noindent
where $B_p$ is the magnitude of the poloidal magnetic field.
Substituting this result into equation~(\ref{mu_m-def}) one also finds that 
\begin{equation}\label{mu_new}
\mu_m =\frac{1}{4\pi}\frac{r^2}{r^2_{\rm lc}} \frac{B_p^2\Gamma}{\rho u_p^2}. 
\end{equation}
Thus, in this regime one can rewrite equation~(\ref{transf}) as 
\begin{equation}\label{transf1}
\frac{\Gamma^2 r}{{\cal R}} \approx
\frac{ {\displaystyle
r \vgrad{\ln\left|\frac{I}{\Gamma}\right|}
\!\cdot\! \frac{\vgrad{\Psi}}{|\vgrad{\Psi}|} } }
{ {\displaystyle 1+ \frac{\mu_h}{\mu_m} }}
-\Gamma^2\frac{r_{\rm lc}^2}{r^2}
\frac{\vgrad{r} \!\cdot\!\vgrad{\Psi}}{|\vgrad{\Psi}|} \,.
\end{equation}  
In the magnetically dominated case, where $\mu_m\gg\mu_h$, order-of-magnitude 
evaluation of the last two terms in this equation gives the useful result 
\begin{equation}\label{transf2}
\frac{\Gamma^2 r}{{\cal R}} \approx 1
-\Gamma^2\frac{r_{\rm lc}^2}{r^2}\, .
\end{equation}

Depending on which term in equation~(\ref{transf}) can be neglected, we
can isolate the following three cases (ordered by increasing importance):

(i) If the electromagnetic part is negligible then the shape of the flow is 
determined by the centrifugal term, resulting in a hyperbolic line shape,
a characteristic of ballistic motion 
(see equation~20 and related discussion in \citealp{V04};
see also Sections~\ref{alphap>2}~and~\ref{A1.3}). 
None of the end-states of our simulations has this property.

\begin{figure*}
\includegraphics[width=65mm]{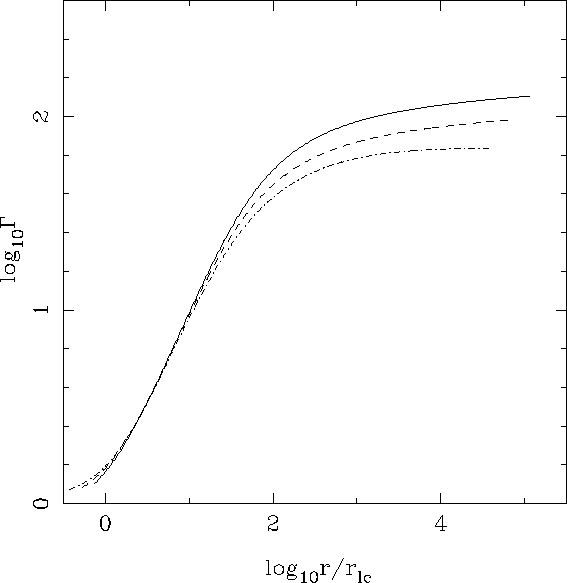}
\includegraphics[width=65mm]{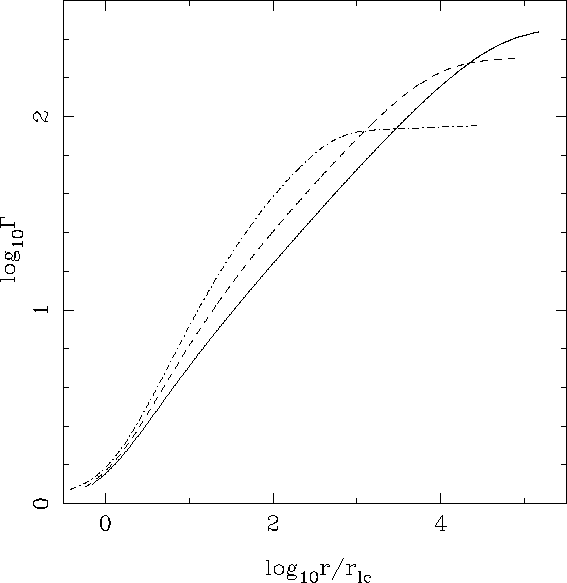}
\includegraphics[width=65mm]{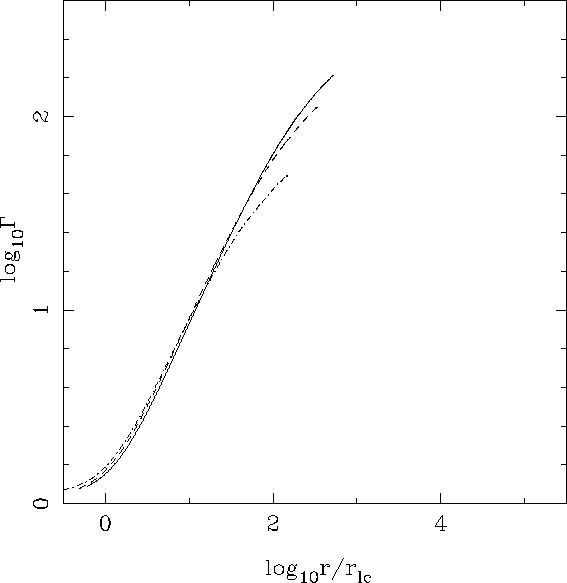}
\includegraphics[width=65mm]{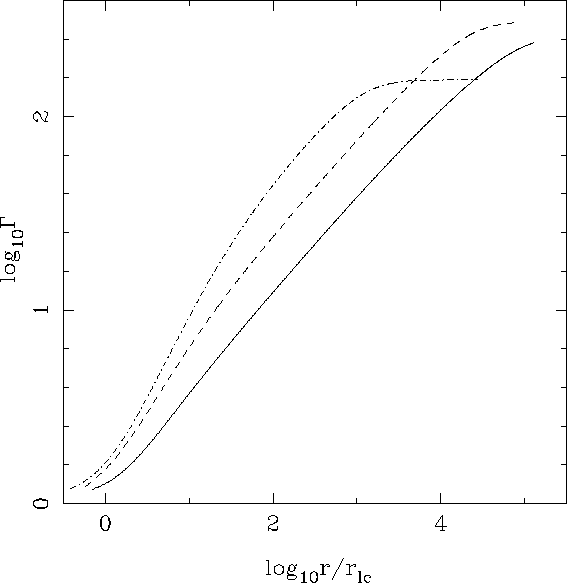}
\includegraphics[width=65mm]{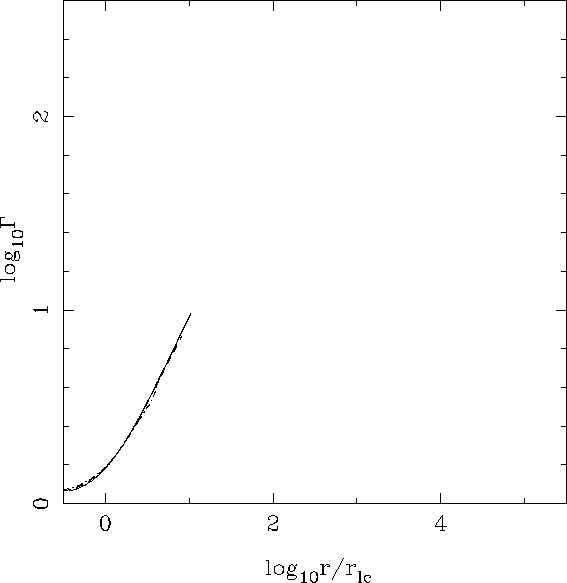}
\includegraphics[width=65mm]{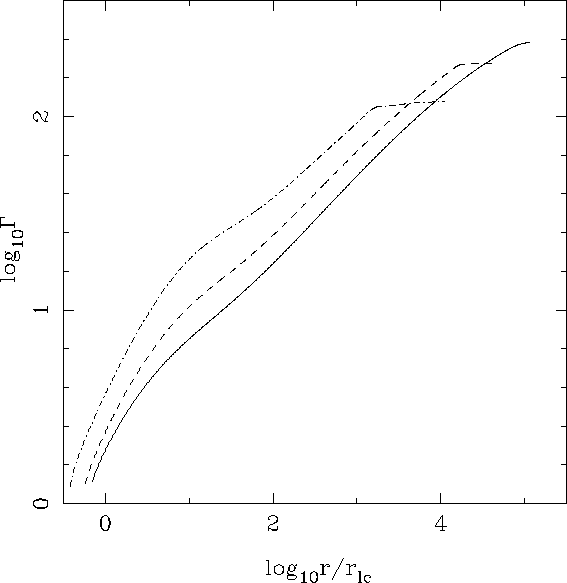}
\caption{Lorentz factor along three different magnetic field lines of 
models A (top left panel), B1 (top right panel), C (middle left panel),
D (middle right panel), F (bottom left panel), and B2H (bottom right panel) 
as a function of the cylindrical radius $r$.
Solid line: $\Psi=0.8\Psi_{\rm max}$;
dashed line: $\Psi=0.5\Psi_{\rm max}$;
dash-dotted line: $\Psi=0.2\Psi_{\rm max}$. 
}
\label{lor-r}
\end{figure*}

(ii) If the poloidal curvature term is negligible, the electromagnetic
and centrifugal terms balance each other.  This is the case very
close to the rotation axis (inside the cylindrical core) as well as for a
quasi-conical flow like our model A and for paraboloidal flows with $a >
2$ as in our model F (see Section~\ref{pressure}). In this case
equation~(\ref{transf2}) gives 
\begin{equation}
\Gamma \simeq \frac{r}{r_{\rm lc}}\,.
\label{Gamma2}
\end{equation}
Following different methods, this ``linear acceleration case'' was found by
\citet{2002ApJ...566..336C}, who analysed radial force-free flows beyond
the light cylinder (and hence their analysis holds in the regime between
the Alfv\'en and the fast-magnetosonic surfaces), and by \citet{BKR98},
who perturbed a quasi-conical flow (and found that $\Gamma \approx
r/r_{\rm lc}$ applies in the sub--fast-magnetosonic regime).  Our results
for models A and F agree with the scaling $\Gamma \approx r/r_{\rm lc}$;
see the top left panel of Fig.~\ref{lor-r}. 

\begin{figure*}
\includegraphics[width=55mm]{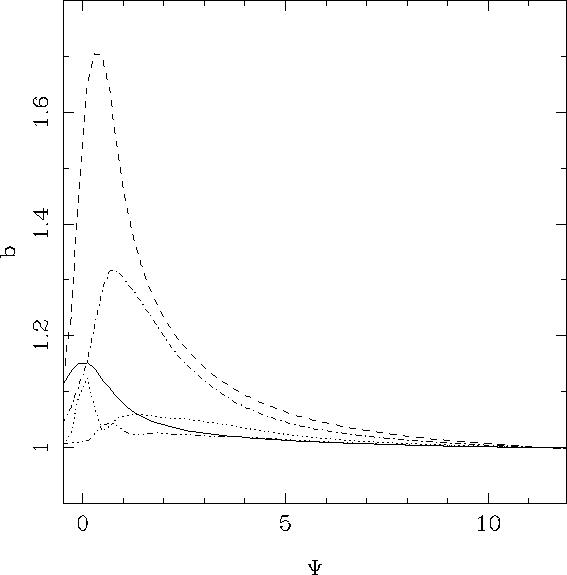}
\includegraphics[width=55mm]{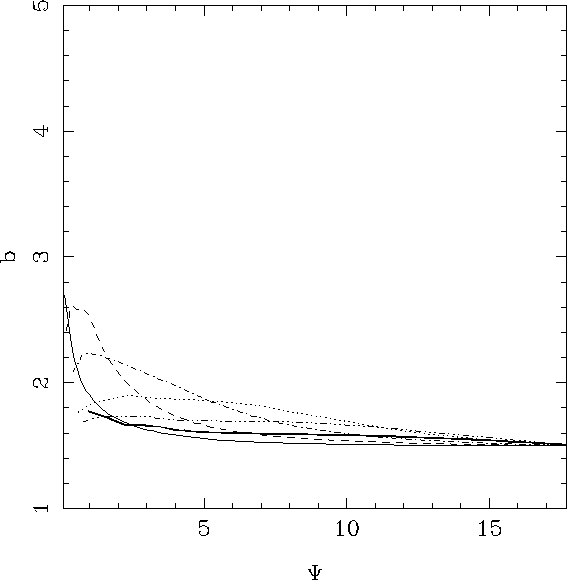}
\includegraphics[width=55mm]{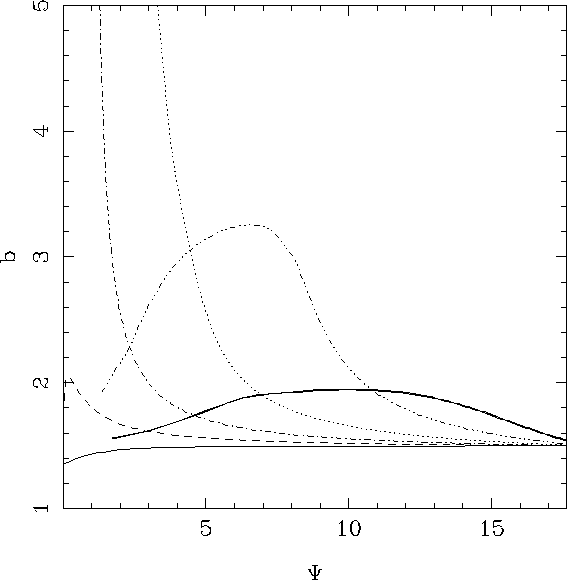}
\caption{The exponent $b$ of the poloidal shape function $z\propto r^b$
for models A (left panel), B2 (middle panel) and B2H (right panel) across the
jet. For model A the depicted cross sections are at
$R=10$ (solid line),
$R=10^2$ (dashed line),
$R=10^3$ (dash-dotted line),
$R=10^4$ (dotted line) and
$R=10^5$ (dash-triple-dotted line).
For models B2 and B2H the plotted cross sections are at
$\eta=5\times10^2$ (thin solid line),
$\eta=5\times10^3$ (dashed line),
$\eta=5\times10^4$ (dash-dotted line),
$\eta=5\times10^5$ (dotted line),
$\eta=5\times10^6$ (dash-triple-dotted line) and
$\eta=5\times10^7$ (thick solid line).
}
\label{par-b}
\end{figure*}

\begin{figure*}
\includegraphics[width=65mm]{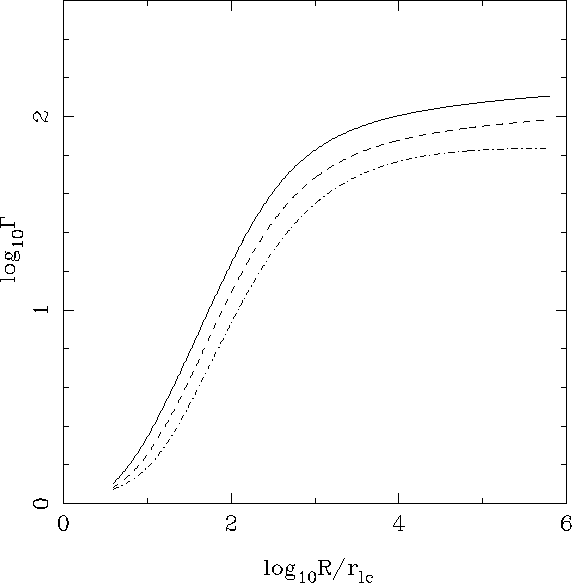}
\includegraphics[width=65mm]{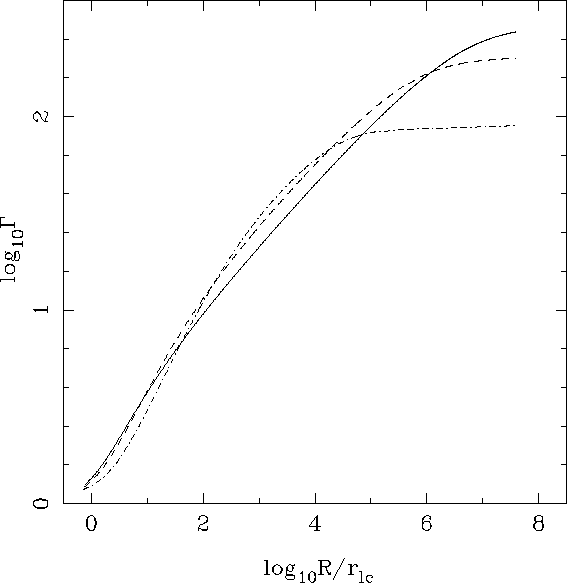}
\includegraphics[width=65mm]{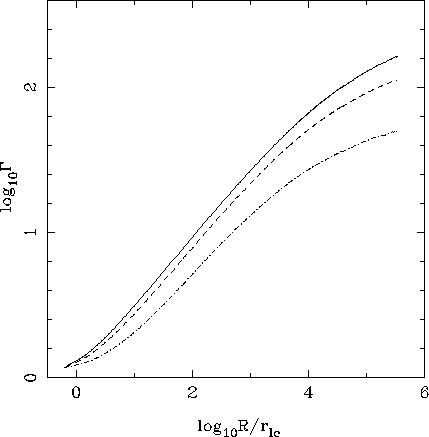}
\includegraphics[width=65mm]{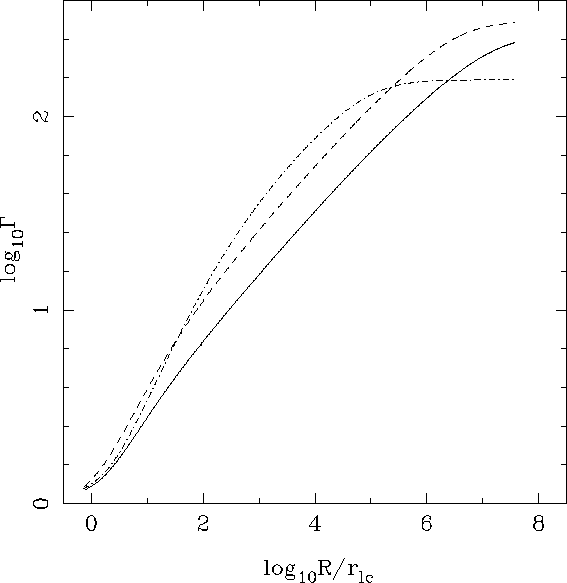}
\includegraphics[width=65mm]{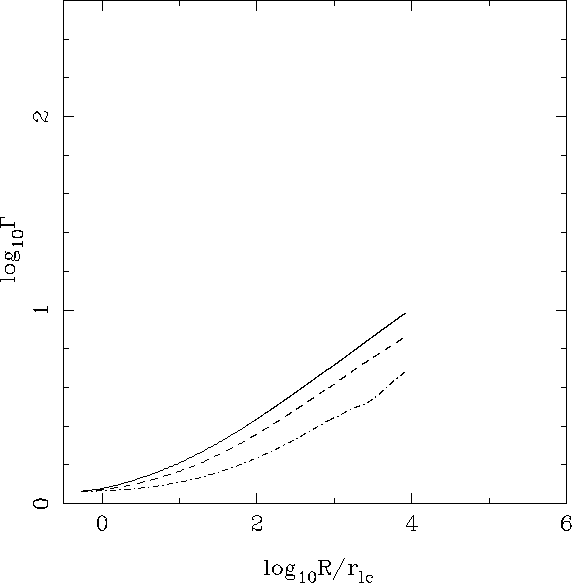}
\includegraphics[width=65mm]{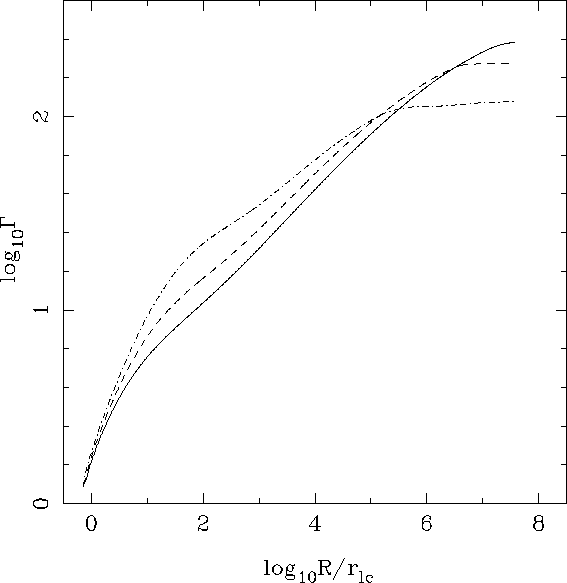}
\caption{Lorentz factor along three different magnetic field lines of
models A (top left panel), B1 (top right panel), C (middle left panel), 
D (middle right panel),  F (bottom left panel), and B2H (bottom right panel)
as a function of the spherical radius $R$.
Solid line: $\Psi=0.8\Psi_{\rm max}$;
dashed line: $\Psi=0.5\Psi_{\rm max}$;
dash-dotted line: $\Psi=0.2\Psi_{\rm max}$. 
}
\label{lor-R}
\end{figure*}

(iii) If the centrifugal term is negligible then the
shape of the flow is determined by the electromagnetic force.
This regime applies to the case of paraboloidal wall with $a \le 2$ (see
Section~\ref{pressure}). 
Equation~(\ref{transf2}) implies that in this case the radius of curvature 
of poloidal field lines is 
\begin{equation}
{\cal R} \approx \Gamma^2 r\, .
\label{R-curv2}
\end{equation}
Now, consider a field line of the shape, $z\propto r^b$. 
(In what follows we use the superscript $b$ to 
indicate the power-law index that describes the shape of given magnetic
field lines, whereas the superscript
$a$ is reserved for the power-law index that gives the shape of the funnel
wall in our numerical models. Note that the interior field lines in
these models have $b$ that is slightly larger than $a$, although $b
\rightarrow a$ as the wall is approached; see Fig.~\ref{par-b}.)
The curvature radius of such a line satisfies
\begin{eqnarray}
\frac{r}{\cal R}= - r \left(\frac{B_z}{B_p}\right)^3
 \frac{ \partial^2 r }{ \partial z^2 }
\approx \frac{ b-1 } {b^2} \left(\frac{r}{z}\right)^2 \,,
\label{R-curv}
\end{eqnarray}
where the final form is valid when $B_p \approx B_z$. Combining this
with equation~(\ref{R-curv2}) we get 
\begin{equation}
\Gamma \sim \frac{b}{\sqrt{b-1}}\frac{z}{r} \propto r^{b-1} \propto
z^{(b-1)/b}
\label{G_z_r}
\end{equation}
(see also \citealp{VK03b}), which applies when the power-law index lies
in the range $1<b\le 2$ and shows that the spatial growth of the Lorentz
factor is also a power law in this case (in either $r$ or $z$). 
Assuming that the flow is not too collimated within the light cylinder,
so that $z_{\rm lc}\simeq r_{\rm lc}$ for most of the field lines 
(an assumption that is well satisfied in our numerical models), we can
write the above result in the following useful forms:
\begin{equation}
  \Gamma \simeq (r/r_{\rm lc})^{b-1} \text{or} \Gamma \simeq  (R/r_{\rm
  lc})^{(b-1)/b}.
\label{G-scaling}
\end{equation}
This acceleration regime operates in our $1<a\le 2$ numerical models
before the flow reaches approximate equipartition, 
as can be verified by inspecting Figs.~\ref{lor-r} and~\ref{lor-R}. 

%fffffffffffffffffffffffffffffffffffffffffffffffffffffffffffffffffff
%fffffffffffffffffffffffffffffffffffffffffffffffffffffffffffffffffff

%fffffffffffffffffffffffffffffffffffffffffffffffffffffffffffffffffff
%fffffffffffffffffffffffffffffffffffffffffffffffffffffffffffffffffff

%fffffffffffffffffffffffffffffffffffffffffffffffffffffffffffffffffff
%fffffffffffffffffffffffffffffffffffffffffffffffffffffffffffffffffff

\begin{figure*}
\includegraphics[width=65mm]{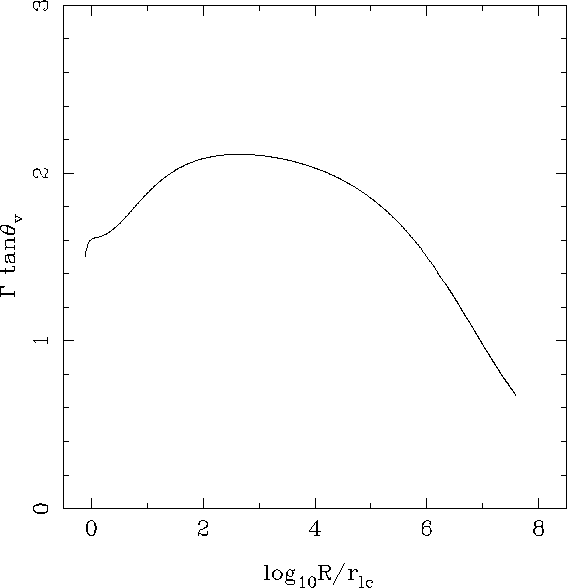}
\includegraphics[width=65mm]{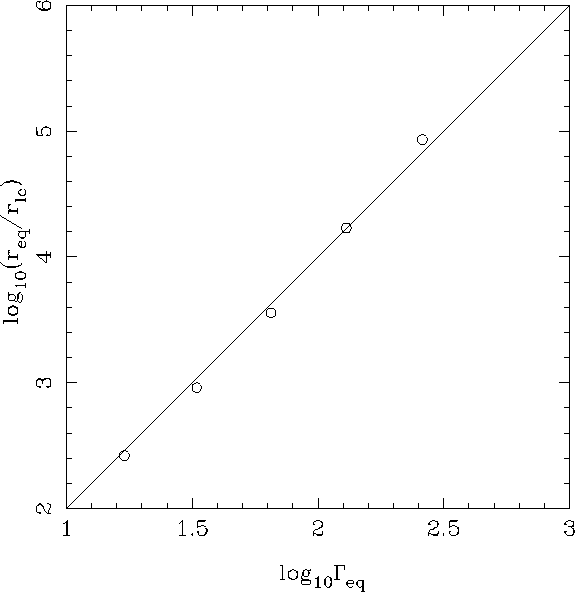}
\caption{
Left panel: variation of $\Gamma\tan\vvartheta$ along 
the flux surface $\Psi=0.8\Psi_{\rm max}$ of model B1. 
Right panel: the diamonds show the equipartition radius (where the 
Poynting and kinetic energy fluxes are equal) along $\Psi=0.8\Psi_{\rm
max}$ as a function of the magnetization parameter $\mu$ for models
B1--B5. The solid line shows the function 
$\log_{10}(r/r_{\rm lc}) = 2\log_{10}(\mu/2)$.     
}
\label{lor-theta-fig}
\end{figure*}

The direct dependence of the flow acceleration on the poloidal curvature
of the magnetic field lines in the regime (iii) leads to an anti-correlation
between the jet Lorentz factor and its opening angle. For a line shape
$z\propto r^b$ ($1<b\le 2$) we find 
\begin{equation}
\Gamma\tan\vvartheta=1/\sqrt{b-1}\,, 
\label{lor-theta-eq}
\end{equation}
where $\vvartheta\equiv \arctan(dr/dz)$ is the local half-opening
angle of the magnetic flux surface. Fig.~\ref{lor-theta-fig} shows the
variation of $\Gamma\tan\vvartheta$ along the flux surface
$\Psi=0.8\Psi_{\rm max}$ of model B1. One can see that this product is
indeed close to $1/\sqrt{b-1}$. It is, however, not exactly a constant,
for the following reasons: the curvature acceleration regime is not
really applicable at small and large spherical radii, the
electromagnetic term in equation~(\ref{transf}) is not exactly equal to
1, and the power-law index $b$ varies along the flow. The figure
nevertheless indicates that equation~(\ref{lor-theta-eq}) provides a
useful estimate of the relationship between $\Gamma$ and $\vvartheta$.

%fffffffffffffffffffffffffffffffffffffffffffffffffffffffffffffffffff
%fffffffffffffffffffffffffffffffffffffffffffffffffffffffffffffffffff

As expected from our discussion in Section~\ref{efficiency} of the close
connection between the acceleration efficiency and the evolution of the
poloidal field-line shape, the trans-field force balance equation, which
determines the variation of the flux-surface shape along the flow,
is seen to provide information on how fast the Lorentz factor increases
with distance from the source. 
For all shape functions $z\propto r^b$ with $1<b\le 2$, the corresponding
power-law dependence of $\Gamma$ leads to a high ($\ga 50\%$)
magnetic-to-kinetic energy conversion efficiency over astrophysically
relevant distances. Using equation~(\ref{G-scaling}) we find that
equipartition between the Poynting and kinetic energy fluxes is attained at 
a cylindrical radius
\begin{equation}
  r_{\rm eq}=r_0 \left(\frac{\mu}{2\Gamma_0}\right)^{1/(b-1)}\,.
\label{eq1}
\end{equation} 
After  substitution $r_0=r_{\rm lc}$ and $\Gamma_0=1$, this equation 
reads
\begin{equation}
  r_{\rm eq}=r_{\rm lc} \left(\frac{\mu}{2}\right)^{1/(b-1)}\,,
\label{equip1}
\end{equation}
which, in fact, agrees very well with our results for models B 
(see Fig.~\ref{lor-theta-fig}). 
In terms of the spherical radius,
assuming again that $R_{\rm lc}\simeq r_{\rm lc}$, we can write this
expression as  
\begin{equation}
  R_{\rm eq}=r_{\rm lc} \left(\frac{\mu}{2}\right)^{b/(b-1)}\,.
\label{equip2}
\end{equation}
For $b>2$ the corresponding relations are (using equation~\ref{Gamma2})
$r_{\rm eq}=(\mu/2) r_{\rm lc} $ 
and $R_{\rm eq}=(\mu/2)^b r_{\rm lc} $.

The derived scaling for the Lorentz factor can be used to find the
behaviour of other quantities. For example, 
for the main part of the flow in which the Poynting flux dominates
the energy flux, one has $\sigma \Gamma \approx \mu$ and 
hence, for $1<b\le 2$,
\begin{equation}
\sigma \approx \mu/\Gamma \propto r/z \propto r^{-(b-1)} \,.
\end{equation}
The predicted behaviour is indeed seen in the left panel 
of Fig.~\ref{sigma-ev}.
This figure further shows that the ``self similar'' structure of the
magnetization curves extends also beyond the equipartition radius, where
they flatten out; in particular, they do not cross each other even in
that regime. Consequently, the magnetization beyond the turning point of the
curve is lower the smaller the inlet value, which goes along with our finding
that the efficiency $\sim 1/(1+\sigma_\infty)$ of magnetic-to-kinetic
energy conversion in cold flows decreases with increasing initial
magnetization.

The high acceleration efficiencies attained by our simulated flows
appear to be inconsistent with the conclusion of \citet{CLB98} that a
transition to a low-$\sigma$ configuration cannot occur gradually in
regions well beyond the light cylinder, where the flow has become
ultra-relativistic. Their analysis was, however, based in part on an
estimate of the change in the angle $\vvartheta$ between the poloidal
flow and the rotation axis as one moves through a length $\Delta \ell$
along the flow (see text after equation~14 in their paper): this
estimate is not generally valid since it assumes that $\Delta \ell \sim
\Delta r$, which only applies to quasi-radial flows. 
If instead we use $\Delta \ell \sim \Delta z$ in equation~(14) of
\citet{CLB98} and concentrate on paraboloidal flows ($z \propto r^b$)
with $b \le 2$, 
we get $\Delta \vvartheta \sim \Delta (z/r) b/ \Gamma^2 (b-1)$, 
which yields the scaling $\Gamma \propto z/r$ found above. On the other
hand, the lower acceleration efficiency exhibited by our model A, in
which the flow morphology is quasi-radial (see Figs.~\ref{model-a},
\ref{geom-effect} and~\ref{S-ev}), appears to be consistent with the
\citet{CLB98} inference of logarithmic collimation and slower
acceleration.  We note in this connection that, beyond the end of the
power-law acceleration phase analysed in this subsection, it is possible
to have an additional, logarithmic acceleration regime in which
potentially up to 100\% of the Poynting flux could be converted into
matter kinetic energy flux (see \citealp{V04} and references therein).
However, this acceleration is too slow to be of astrophysical interest
since it requires exponentially large distances for completion.

\subsection{Dependence on the external pressure distribution}
\label{pressure}

Although we have chosen, for numerical convenience, to prescribe the
shape of the funnels that guide our simulated flows, in reality the
boundary shape of pressure-confined flows will be determined by the
ambient pressure distribution, $p_{\rm ext}$, and we expect a one-to-one correspondence
between the shape of the boundary and the parameters of the confining
medium, enforced through the pressure-balance condition at the
boundary, $p_{\rm int}=p_{\rm ext}$. Here we analyse this issue for 
the asymptotic region of a magnetically accelerated flow, where 
the internal jet pressure, $p_{\rm int}$, is dominated by the contribution 
due to the azimuthal component of magnetic field,
$p_{\rm int}= p+B_{\rm co}^2/8\pi \simeq (B^{\hat\phi})^2/8\pi\Gamma^2$. 
Thus, 
\begin{eqnarray}\nonumber
   \Gamma^{-2} = \frac{8\pi p_{\rm ext}}{(B^{\hat\phi})^2} \,. 
\end{eqnarray}
In the following we assume that the external pressure distribution is a 
power-law 
\begin{eqnarray}\nonumber
p_{\rm ext} = p_{\rm ext,lc} (z/z_{\rm lc})^{-\alphap} \,,
\end{eqnarray}
which is consistent with the funnel shape $z \propto r^a$ adopted in 
our numerical simulations. Moreover, since $\mu_m \propto I \propto r B^{\hat\phi}$
(see equations \ref{I}, \ref{mu_m-def}) is a weak function of distance
we may assume that at the jet 
boundary $B^{\hat\phi}= B^{\hat\phi}_{\rm lc}(r/r_{\rm lc})^{-1}$,
Then we have   
\begin{equation}
\Gamma^{-2}=C x^2 Z^{-\alphap} \,,
\label{C1}
\end{equation}
where $x\equiv r/r_{\rm lc}$ and $Z\equiv z/r_{\rm lc}$ are the dimensionless 
coordinates of the jet boundary and 
\begin{equation}
C=\left( \frac{8\pi p_{\rm ext} }{B_{\hat\phi}^2} \right )_{\rm lc}
\left(\frac{z_{\rm lc}}{r_{\rm lc}}\right)^\alphap
= \frac{(z_{\rm lc}/r_{\rm lc})^\alphap}{\Gamma_{\rm lc}^2} \,.
\label{C2}
\end{equation}
It is easy to see that $C$ is a positive dimensionless constant 
of the order of 1. Provided that $dr/dz\ll 1$ we can approximate the 
curvature radius of the jet boundary via 
\begin{equation}
{\cal R}^{-1} \approx -  \frac{d^2 r}{dz^2} 
= -  \frac{1}{r_{\rm lc}}\frac{d^2 x}{dZ^2}
\label{R-curv1}
\end{equation}
and rewrite equation~(\ref{transf2}) as 
\begin{equation}
x \frac{d^2 x}{dZ^2} +\frac{1}{\Gamma^2} - \frac{1}{x^2} \approx  0 \,.
\label{transf3}
\end{equation}
After the substitution of $\Gamma$ from equation~(\ref{C1}) this  
yields an ordinary differential equation for the jet boundary
\begin{equation}
\frac{d^2 x}{dZ^2} +C \frac{x}{Z^\alphap} - \frac{1}{x^3} =0 \,.
\label{ODE1}
\end{equation}
The first term on the left-hand side of equation~(\ref{ODE1})
represents the effect of poloidal curvature, the second
is the electromagnetic term and the third is the centrifugal term.

Equation~(\ref{ODE1}) can be
solved in closed form in various limits, as discribed in
Appendix~\ref{appA}. Here we simplify the
discussion by looking for almost power-law solutions 
\begin{equation}
x=K^{-1} Z^{1/a}\,,
\label{K}
\end{equation}
with $K$ being positive constants and $a$ varying very slowly. 
Substituting this ansatz into
equation~(\ref{ODE1}) and ignoring all terms including derivatives of $a$, 
we obtain
\begin{equation}
\frac{1}{a} \left( \frac{1}{a} -1 \right) 
+C Z^{2-\alphap} - K^4 Z^{2-4/a} =0 \,.
\label{ODE2}
\end{equation}
We now proceed to analyse this equation for different values of the
exponent $\alphap$.

\subsubsection{$\alphap > 2$}
\label{alphap>2}

In this case the second term on the left-hand side of
equation~(\ref{ODE2}) vanishes as $Z\rightarrow \infty$ and the only 
acceptable asymptotic value of $a$ is unity. Indeed, for $a>2$ the third  
term diverges, for $a=2$ it is constant but negative and so is the first
term, for $a<2$ it vanishes and so must the first one, implying $a\to 1$.  
Thus, asymptotically the boundary adopts conical shape. 
\begin{itemize}
\item When $\alphap <4$ the electromagnetic term of equation~(\ref{ODE2})
dominates over the centrifugal term, and thus $a\to 1^+$ (since
the first term must be negative in order to cancel the second). The
boundary shape is therefore {\em paraboloidal} (with conical
asymptotes). An explicit solution of equation~(\ref{ODE1}) in this limit
is given in Appendix~\ref{appA}.

\item When $\alphap >4$ the centrifugal term dominates over the
electromagnetic term in equation~(\ref{ODE2}) and thus $a\rightarrow
1^-$ (since the first term must be positive in order to cancel the
third). This is case (i) of our analysis of equation~(\ref{transf1}),
which corresponds to a {\em hyperboloidal} shape (with conical
asymptotes), as demonstrated in Appendix~\ref{appA} through an explicit
solution of equation~(\ref{ODE1}) in this limit.

\item
When $\alphap = 4$ one can obtain a solution that is conical ($a=1$) from
the start, with $K^4 = C$. This solution corresponds to our conical
model A during the acceleration phase, when $\Gamma \propto r$ (see
equation~\ref{Gamma2}). Fig.~\ref{p-ext} verifies the predicted scaling
($p_{\rm ext} \propto Z^{-4}$) and also shows that, after the growth of
$\Gamma$ saturates, a conical shape can be maintained only if the
ambient pressure scales as $z^{-2}$, which follows directly from the
scaling $p_{\rm int} \propto \Gamma^{-2} r^{-2}$ discussed at the
beginning of this subsection.
\end{itemize}

In summary, for $\alphap > 2$ the boundary does not simply adjust to the
ambient pressure profile but instead asymptotes to a conical shape. 
This result is consistent with the expectation that in this case the
transverse expansion time of the jet becomes shorter than the
propagation time of magnetosonic waves across the flow, leading to a
loss of causal connectivity and hence to a ``free'' ballistic expansion
in a cone (\citealt{BBR84}; see also Section~\ref{causality}). This is
essentially the behaviour exhibited by our Model E (see Fig.~\ref{geu1}).

\subsubsection{$\alphap = 2$}
\label{alphap=2}

In this case the second term on the left-hand side of
equation~(\ref{ODE2}) is a positive constant. This implies that
$1<a\le 2$. (Indeed, for $a>2$, the third term diverges and hence 
unbalanced. For $a\le 1$ it vanishes but the first term is non-negative
and hence cannot balance the second one.) 
We can distinguish between the following two cases:
\begin{itemize}
\item
$a=2$ --- the power law solution with $K^4=C-1/4$ is exact. 
This implies  $C>1/4$.
\item $1< a<2$ ---  the third term becomes negligible at large $Z$
and balancing of the first two terms requires $a\to2/(1+\sqrt{1-4C})$. 
This implies $C\le1/4$. 
\end{itemize}

In other words, for $C< 1/4$ the centrifugal term is negligible and
the resulting shape is $Z = (z_{\rm lc}/r_{\rm
lc})x^{2/(1+\sqrt{1-4C})}$, whereas for $C>1/4$ the centrifugal term is
comparable to the other two terms and the solution is $Z=\sqrt{C-1/4} \
x^2$. Fig.~\ref{p-ext} verifies that the confining pressure in our
simulated flows scales as $Z^{-2}$ irrespective of the precise
value of $a$ so long as the shape exponent lies in the range $1<a\le
2$. The figure also corroborates the prediction that the $Z^{-2}$
scaling is attained only gradually when $a<2$ (models B and D,
corresponding to $a=3/2$) but that it is present almost from the start 
when $a=2$ (model C).

As shown in Appendix~\ref{appA}, the asymptotic solution for $C=1/4$ is 
$x=Z^{1/2} (C_1+C_2 \ln{Z})$, where $C1$ and $C_2 \neq 0$ are constants. 
(We kept the constant $C_1$
to accommodate the possibility that the solution extends all the way
down to the light-cylinder radius, where $Z\approx 1$.) This
solution is similar to the $C<1/4$ solutions of equation~(\ref{ODE1})
in having a negligible centrifugal contribution.

Although all the funnel shapes whose power-law indices lie in the range
$1<a\le 2$ correspond to a single exponent ($\alphap = 2$) of the
confining pressure distribution, there is nevertheless a one-to-one
match between a given pressure distribution and the resultant funnel
shape. This is because both the power-law index $\alphap$ {\em and} the
magnitude of the confining pressure (as expressed in relation to the
internal magnetic pressure at the light-cylinder radius by the parameter
$C$; see equation~\ref{C2}) play a role in determining the functional
form of the boundary: when $C< 1/4$ the magnitude of $C$ fixes the
exponent of the boundary paraboloid, whereas when $C>1/4$ it fixes the
normalization constant $K$. 
The parameter $C$ is evaluated at the effective base of the asymptotic 
region of the flow and it  
conveys physical properties (e.g. $z_{\rm lc}$ and $\Gamma_{\rm lc}$; see
equation~\ref{C2}) imprinted on the outflow before it reaches the
asymptotic regime. Thus, the asymptotic shape of a jet
propagating through a power-law pressure distribution is determined both
by the exponent of that distribution and by the evolution of the outflow
before entering the asymptotic region. 
 
\subsubsection{$\alphap < 2$}
\label{alphap<2}

\begin{figure}
\includegraphics[width=77mm]{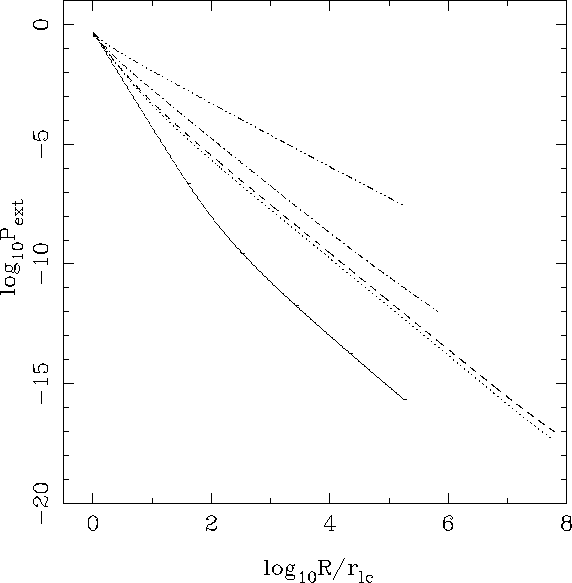}
\caption{Evolution of total pressure along the jet boundary in 
models A (solid line), B1 (dashed line), C (dash-dotted line), D
(dotted line) and F (dash-double-dotted line). 
}
\label{p-ext}
\end{figure}

In this case the second term on the left-hand side of
equation~(\ref{ODE2}) diverges as $Z\rightarrow \infty$. To balance
this term, the third term must also diverge in this limit, which implies
that $a=4/\alphap > 2$ and $C=K^4$. Thus, the jet shape is paraboloidal,  
$Z=C^{1/\alphap} x^{4/\alphap}$.  Like in the $\alphap = 2$ case, 
both the parameters $\alphap$ and $C$ are needed to uniquely fix the 
functional form of the jet shape. For $\alphap=4/3$ we have $a=3$, 
the funnel shape index of our numerical model F. 
Fig.~\ref{p-ext} verifies that the boundary pressure for this model
indeed scales as $Z^{-4/3}$. 

We can collect the results derived in this subsection into a concise
description of the correspondence between the exponent
$\alphap$ of the ambient pressure distribution and the exponent $a$ of
the asymptotic jet shape:
\begin{itemize}
\item $\alphap < 2\ \Leftrightarrow\ a=4/\alphap >2\, ,$
\item $\alphap = 2\  \Leftrightarrow\ 1<a\le2\, ,$
\item $\alphap > 2\ \Leftrightarrow\ a = 1\, .$
\end{itemize}
Similar results for the behaviour of the ambient pressure in a confined 
jet ($\alphap \le 2$) were found by \citet{TMN08} in the
force-free limit, which is consistent with the fact that our expressions
for the spatial profile of $\Gamma$ were obtained in effectively the same
approximation.

As we have seen, $\alphap = 2$ leads to the asymptotic balance between
the electromagnetic and poloidal curvature forces (regime iii) 
whereas $\alphap < 2$  leads to the balance between the electromagnetic and 
centrifugal forces (regime ii; see Section~\ref{power-law}). 
These regimes are characterized by different evolution of many flow 
parameters, which may have observable consequences (see also 
Section~\ref{application}). For example, in regime (ii) 
the product $\Gamma\tan\vvartheta$ is predicted to be a constant
${\cal{O}}(1)$ in the acceleration region, whereas in regime (iii) it is
expected to decrease with distance as $Z^{-(1-2/b)}$, with $b$ being 
slightly larger than $a$ due to the stronger collimation of 
the flow inside the jet. The evolution of the
Lorentz factor in regime (ii) is given by $\Gamma \propto r$
(equation~\ref{Gamma2}) rather than by the $\Gamma \propto r^{b-1}$
scaling of regime (iii). However, in practice this may not translate 
into a significant difference in how fast the jet accelerates 
(for example, $\Gamma \approx z^{1/3}$ for both the 
$\alphap=4/3$ and $\alphap=2$, $b=3/2$ cases).

After the end of the acceleration the internal pressure scales as $r^{-2}$
(since $\Gamma=\Gamma_\infty=$ const).
If the external pressure continues to decline as $z^{-\alphap}$,
the pressure balance implies that
the radial coordinate $r$ increases faster compared to its
variation during the acceleration.
The new flow shape is
$Z=C^{1/\alphap} \Gamma_\infty^{2/\alphap} x^{2/\alphap}$
as a result of equation~(\ref{C1}). For example, in the cases
$\alphap=2$, $1<a<2$, the flow becomes radial
and the opening angle of the jet remains constant.
The quantity $\Gamma \tan \vvartheta$ is also constant and equal to
$C^{-1/2}=a/\sqrt{a-1}$ (using the relation between $C$ and $a$,
see Section~\ref{alphap=2}).
Thus, $\Gamma \tan \vvartheta$ is $a$ times larger compared to
its value during the acceleration phase
(see equation~\ref{lor-theta-eq}).\footnote{
The change of this quantity is smooth and happens as the 
function $\Gamma(Z)$ changes from a power law to a constant.
Equation~(\ref{C1}), written as
$x= C^{-1/2} Z^{\alphap/2} \left[\Gamma(Z) \right]^{-1}$,
gives $\Gamma \tan \vvartheta = \Gamma dx/dZ =
C^{-1/2} (\alphap/2 - d\ln \Gamma / d \ln Z) Z^{\alphap/2-1}$.
In the cases with $\alphap=2$, $1<a<2$
the slope $d\ln \Gamma / d \ln Z$ changes from
$1-1/a$ during the main part of the acceleration phase
(see equation~\ref{G_z_r}) to zero after it ends. As a result,
$\Gamma \tan \vvartheta$ changes from
$1/\sqrt{a-1}$ to $a/\sqrt{a-1}$.
}

%fffffffffffffffffffffffffffffffffffffffffffffffffffffffffffffffffff
%fffffffffffffffffffffffffffffffffffffffffffffffffffffffffffffffffff

%sssssssssssssssssssssssssssssssssssssssssssssss
\subsection{Magnetic acceleration and causality}
%sssssssssssssssssssssssssssssssssssssssssssssss
\label{causality}

We have found that the acceleration efficiency is smaller when the wall
has a conical shape (model A) than in the cases when its shape is
paraboloidal (see Fig.~\ref{geom-effect}). In the conical-wall case the flow
attains equipartition only along field lines that are close to the
rotation axis ($\Psi \le 0.2 \Psi_{\rm max}$). In accordance with our
discussion in Section~\ref{efficiency}, the variation in the
acceleration efficiency is tied to the difference in the degree of the
collimation across the outflow, as seen in Fig.~\ref{par-b}:
Only for small values of $\Psi$ does the exponent $b$ become
significantly larger than 1, corresponding to the innermost field lines
bending toward the rotation axis, which implies that the bunching
function ${\cal S}$ decreases along this portion of the outflow. In
order for collimation to occur, there must, however, exist causal
connectivity across the outflow.
A related discussion of this issue can be found in \citet{ZBB08}. 
However, the simpler flow structure assumed in that paper excludes the 
possibility of magnetic acceleration. In particular, the assumption of zero
azimuthal speed implies that the current $I$ is a constant of motion
(see equation~\ref{angm-def}), which in turn means that $\mu_m$ remains
constant (see equation~\ref{mu_m-def}). 

One can check whether the condition of causal connectivity is
satisfied by comparing the field-line opening angle $\vvartheta$ (defined 
in Section~\ref{power-law}) with the half-angle of the Mach cone of fast
waves, $\thetag$. The latter can be found from the relation
\begin{eqnarray}
  \sin\thetag=\frac{\Gamma_{\rm f} c_{\rm f}}{\Gamma v_p}\,,
\label{mach-angle}
\end{eqnarray}
where $c_{\rm f}$ and $\Gamma_{\rm f}$ are the fast speed and the
corresponding Lorentz factor, respectively.
Since $\Gamma_{\rm f} c_{\rm f} =
B_{\rm co} / \sqrt{4\pi\rho}$, where $B_{\rm co}$ is the magnetic field as measured 
in the fluid frame, and $v_p\approx c$, we have 
\begin{equation}
  \sin\thetag \approx 
  \left ({\frac{B_{\rm co}^2}{4\pi\rho c^2}}\right ) ^{1/2} \frac{1}{\Gamma} =
  \frac{{\sigma}^{1/2}}{\Gamma} \,.
\end{equation}
In the magnetically dominated regime $\sigma \approx \mu/ \Gamma$. 
For highly super-magnetosonic flows  $\thetag\ll 1$. Thus, we may write 
\begin{equation}
\thetag \approx \sqrt{\mu / \Gamma^3}\,.
\end{equation}
In the hydrodynamic limit the fast magnetsonic speed reduces to the sound 
speed and $\Gamma_{\rm f}$, $c_{\rm f}$ in equation~(\ref{mach-angle})
should be replaced by $\Gamma_{\rm s}$, $c_{\rm s}$. 
For the ultra-relativistic equation of state
and $\Gamma\gg 1$  this gives $\theta_{\rm m}\simeq 1/\Gamma$, the value
used for causality analysis  in \citet{ZBB08}. However, in the magnetic
case $\theta_m$ can be much higher because the magnetosonic speed can be
much closer to the speed of light. 

\begin{figure*}
\includegraphics[width=57mm]{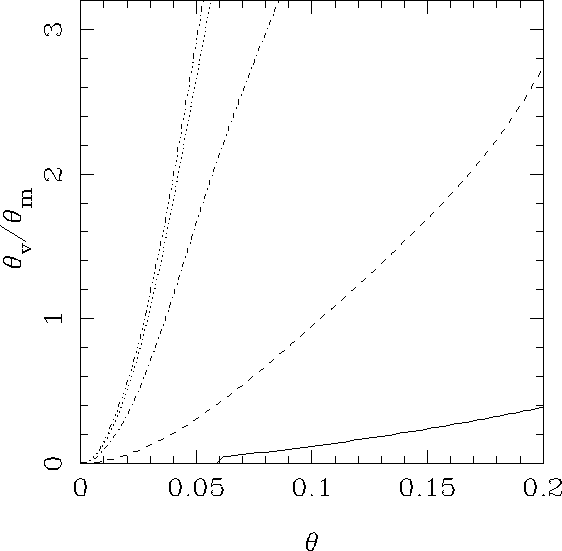}
\includegraphics[width=55mm]{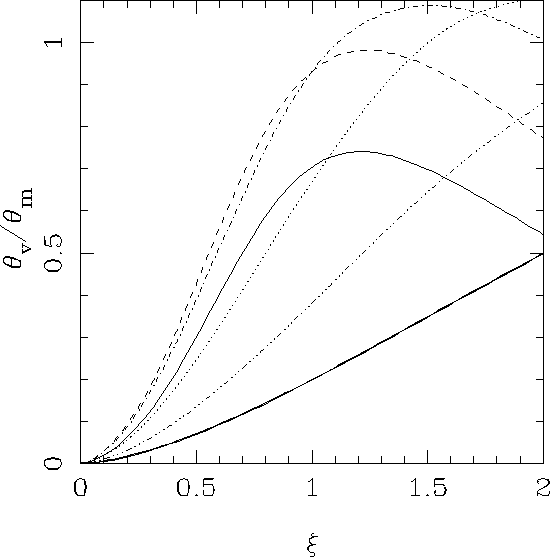}
\includegraphics[width=55mm]{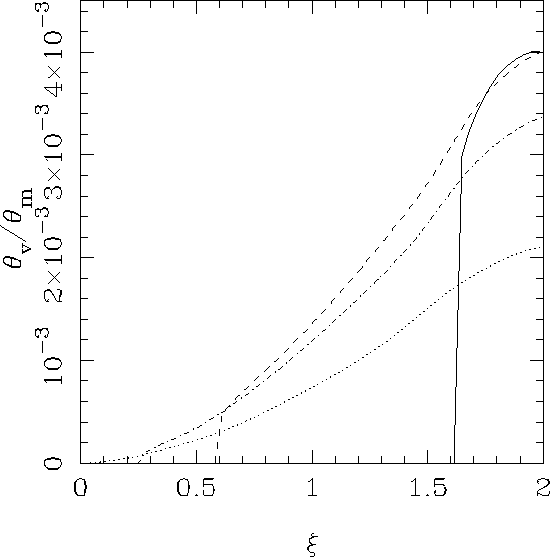}
\caption{The ratio of flow half-angle, $\vvartheta$, to the Mach angle, 
$\thetag$, across the jet for models A (left panel), B3 (middle panel)
and F (right panel). For model A the depicted cross-sections are at 
$R=10$ (solid line),
$R=10^2$ (dashed line),
$R=10^3$ (dash-dotted line),
$R=10^4$ (dotted line) and
$R=10^5$ (dash-triple-dotted line). 
For model B3 the depicted cross-sections are at
$\eta=5\times10^2$ (thin solid line),
$\eta=5\times10^3$ (dashed line),
$\eta=5\times10^4$ (dash-dotted line),
$\eta=5\times10^5$ (dotted line),
$\eta=5\times10^6$ (dash-triple-dotted line) and
$\eta=5\times10^7$ (thick solid line).
For model F the depicted cross-sections are at
$\eta=1.5\times10^3$ (thin solid line),
$\eta=5\times10^3$ (dashed line),
$\eta=1.5\times10^4$ (dash-dotted line), and
$\eta=1.5\times10^5$ (dotted line).
The curves in the right panel dive to zero when the flow becomes 
sub-magnetosonic.
}
\label{mach_angle}
\end{figure*}

In the conical case we have $\Gamma \approx R/r_{\rm lc}$, and 
\begin{equation}
\vvartheta / \thetag \approx (\theta/\sqrt{\mu})
(R/r_{\rm lc})^{3/2}
\label{theta-ratio}
\end{equation}
grows rapidly to a value $> 1$ 
(where it is a good approximation to replace $\vvartheta$ by $\theta$).  
The left panel of Fig.~\ref{mach_angle} 
shows that only the inner part of the jet has $\vvartheta/ \thetag <1$
and thus in causal connection. Collimation (and thus efficient 
acceleration) is possible only in this inner region. In contrast, the
outer parts of the conical jet lack causal connection with the axial
region and the flow there is essentially ballistic.

In the paraboloidal case with $b<2$ (for which $\vvartheta \approx
1/\Gamma$ and $\Gamma \approx (R/r_{\rm lc})^{(b-1)/b}$) 
\begin{equation}
\vvartheta / \thetag \approx 
\left({\Gamma}/{\mu}\right)^{1/2} \approx 
(1/\mu^{1/2}) (R/r_{\rm lc})^{(b-1)/(2b)} \,,
\end{equation}
so this ratio grows much slower compared to the conical case. 
Moreover, the loss of causal contact formally occurs when $\Gamma\simeq\mu$, 
i.e. at the end of the acceleration phase.     
This is confirmed by our simulations. As one can see in the middle and 
right panels of Fig.~\ref{mach_angle}, during the 
power-law acceleration phase $\vvartheta / \thetag$ grows slowly but 
remains less than 1 almost everywhere in our numerical models. It subsequently
decreases again when the growth rate of $\Gamma$ goes down.

In contrast, in the paraboloidal case with $b>2$ (for which $\vvartheta
\approx r/bz$ and $\Gamma \approx r/r_{\rm lc}$),
\begin{eqnarray}
\vvartheta / \thetag& \approx & (1/b \mu^{1/2} C^{b/4}) (r/r_{\rm
lc})^{(5/2)-b} \cr
&=& (1/b \mu^{1/2} C^{5/8}) (R/r_{\rm lc})^{(5/2b)-1}
\end{eqnarray}
(see Section~\ref{alphap<2}), and this ratio actually {\em decreases}
with distance for $b>5/2$! One can also argue quite generally that, even
if $\Gamma$ were to increase all the way up to $\mu$, the value of the
above ratio in that region, which can be estimated to be $\sim 1/b
\mu^{b-2} C^{b/4}$, would likely remain $< 1$ (since $b>2$, $\mu>1$
and $C$ is of the order of 1; see equation~\ref{C2}). 
Thus, the necessary (but not sufficient) conditon for acceleration
is satisfied in this case. This suggests that the acceleration efficiency
may be comparable to the $1\le b< 2$ cases. 

The behaviour of an unconfined wind is similar to that of an outflow in
a conical funnel, which is not surprising given the fact that 
the former is a limiting case of the latter. 
As seen in Fig.~\ref{gpw},
the acceleration in model AW is $\ga 50 \%$ efficient only along field
lines that are close to the rotation axis ($\Psi \le 0.1 \Psi_{\rm
max}$), similarly to the situation in model~A.

%fffffffffffffffffffffffffffffffffffffffffffffffffffffffffffffffffff
%fffffffffffffffffffffffffffffffffffffffffffffffffffffffffffffffffff

%sssssssssssssssssssssssssssssssssssssssssssss
\subsection{Hot flows}
%sssssssssssssssssssssssssssssssssssssssssssss
\label{hot}

When $w/\rho c^2$ is significantly larger than 1 at the inlet there is
an additional reservoir of energy for the flow acceleration --- the
thermal energy of particles. As the flow expands the enthalpy per unit
rest mass $w/\rho =c^2+[s/(s-1)] (p/\rho)$ (equation~\ref{w-def})
decreases until it reaches its minimum value $(=c^2)$, and beyond that
point the flow can be regarded as cold. In the pure hydrodynamic
case the thermal energy is directly transferred to the bulk
kinetic energy of the fluid.  In the magnetic case there is an additional
possibility --- the thermal energy can also be transferred to the
Poynting flux.  Indeed, since $\mu c^2 = (w/\rho) \Gamma + \mu_m c^2$,
it is possible to have both $\Gamma$ and $\mu_m c^2$ increasing when
$w/\rho$ decreases, and this in fact is what we observe in model B2H
(Fig.~\ref{gbut}).

%fffffffffffffffffffffffffffffffffffffffffffffffffffffffffffffffffff
\begin{figure*}
\includegraphics[width=77mm]{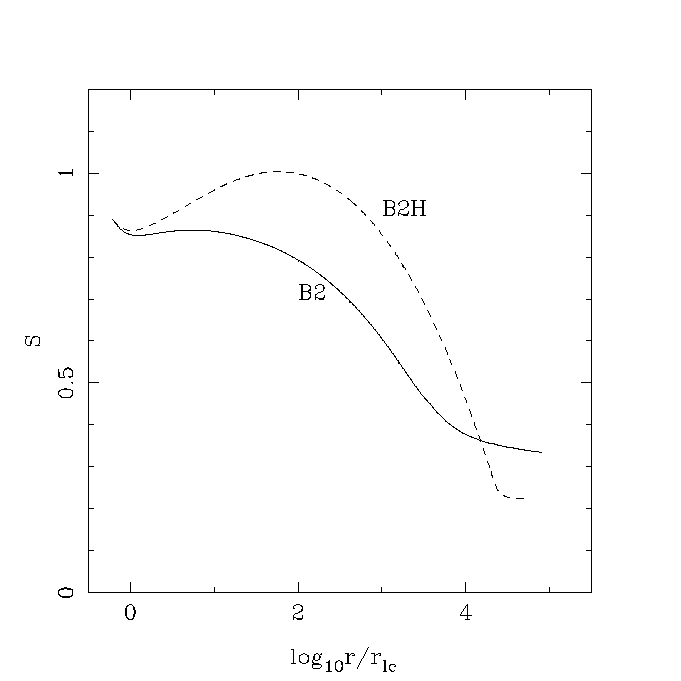}
\includegraphics[width=77mm]{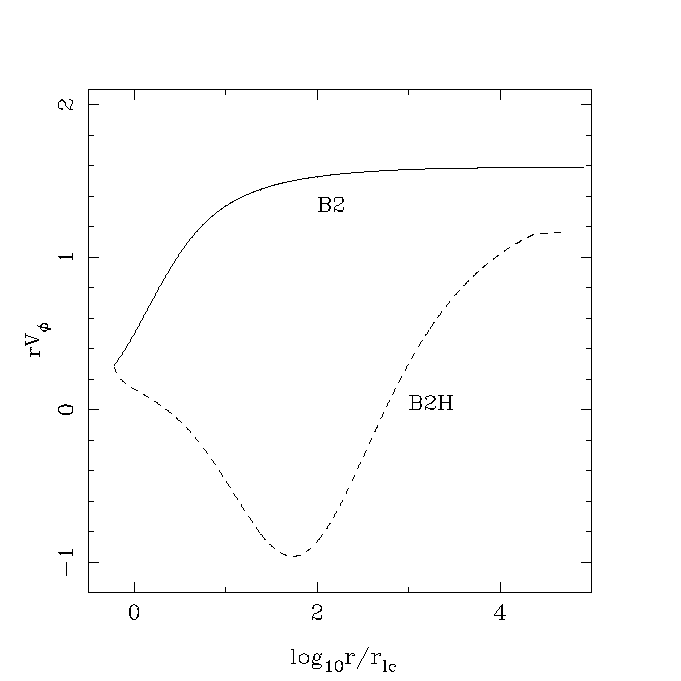}
\caption{Effects of thermal acceleration. 
Left panel: the bunching function ${\cal S}$ along the magnetic 
field line with $\Psi=0.5\Psi_{\rm max}$.
Right panel: $rv^{\hat{\phi}}$ along the magnetic field line with 
$\Psi=0.5\Psi_{\rm max}$. Solid lines: model B2; dashed lines: model B2H.
}
\label{gbut-extra}
\end{figure*}
%fffffffffffffffffffffffffffffffffffffffffffffffffffffffffffffffffff

We have already noted in Section~\ref{efficiency} that in
Poynting flux-dominated flows $\mu_m$ is proportional to the bunching
function ${\cal S}$ (see equation~\ref{S}). In agreement with this result, 
the left panel of Fig.~\ref{gbut-extra} shows that in model B2H
${\cal S}$ exhibits the same evolution as $\mu_m$ 
(which is shown in the middle panel of Fig.~\ref{gbut}). 

In the super-Alfv\'enic regime the trans-field force balance for 
hot flows is described by equation~(\ref{transf}) even for hot outflows
provided that $p$ remains  $\ll B_{\rm co}^2/8\pi$. 
Therefore we still have ${\cal R} \sim \Gamma^2 r$
and hence $\Gamma \propto r^{b-1}$ along magnetic field lines of
paraboloidal jets with exponents in the range $1<b\le 2$. Combining the
mass conservation relation~(\ref{kappa}) and equation~(\ref{calS}) we obtain 
\begin{eqnarray}
\Gamma \rho =\frac{\massint \Psi \cal S}{\pi r^2} \propto r^{-2}\,,
\nonumber
\end{eqnarray}
where we took account of the fact that ${\cal S}$ is a weak function of
$r$. This enables us to write  the variation of the thermodynamic
parameters as
\begin{eqnarray}
\rho \propto r^{-b-1}, \qquad p\propto r^{-s(b+1)}\,.
\nonumber
\end{eqnarray}
In the limit $w\gg \rho c^2$ equation~(\ref{w-def}) gives
$w \propto p \propto r^{-s(b+1)}$, and therefore $\mu_h = (w/\rho c^2)
\Gamma$ scales as
\begin{equation}
\mu_h \propto r^\delta, \qquad \delta= b(2-s) - s\,.
\label{mu_h}
\end{equation}
For model B2H with $b \approx 3/2$ and $s=4/3$ 
this yields  $\mu_h \propto r^{-1/3}$. Hence $\mu_h$ is expected 
to decrease and $\mu_m=\mu-\mu_h$ to increase along the field lines,
in agreement with what is observed in the simulation.

Similar behaviour has been found in the self-similar solutions of
\cite{VK03b}, but only in cases where the flow is super-Alfv\'enic
from the start (see also \citealp{VPK03}).
In their trans-Alfv\'enic, hot-flow solutions \citep{VK03a}, $\mu_m$
remained constant throughout the thermal acceleration phase. This could
be understood from the fact that these solutions corresponded to
$b\approx 2$ and therefore to $\delta \approx 0$ in
equation~(\ref{mu_h}), resulting in constant $\mu_h$ and $\mu_m$ in the
thermal acceleration region.\footnote{As was shown analytically in the
magnetodynamic self-similar solutions of \cite{NMcKF07}, the field-line
shape is $z\propto r^{2/(2-F)}$, where $F$ is a constant parameter entering the
self-similarity expression of the magnetic flux function, $\Psi = r^F
{\cal F}(r/z)$. The MHD self-similar solutions follow the same
scaling in their force-free regime. The trans-Alfv\'enic solutions
presented in \citet{VK03a} were characterized by $F\approx 1$, which
implies $b \approx 2$. Note in this connection that the $F=1$
magnetodynamic solution is exactly the paraboloidal force-free solution
presented by \citet{Bland76}.
\label{footnote_bland}}
In contrast, the super-Alfv\'enic solutions presented in \citet{VK03b}
corresponded to $b \approx 3/2$ and hence to $\delta \approx -1/3$ (the
same values as in our models B and D), and therefore they exhibited the
same behaviour in the thermal acceleration zone as our simulated flows.

The increase of the Poynting-to-mass flux ratio $\mu_m$ in the thermal
acceleration regime leads to a rather unusual behaviour of the azimuthal
velocity.  The right panel of Fig.~\ref{gbut-extra} shows the variation
of $r v^{\hat{\phi}}$ along the same magnetic surface in models B2 and
B2H. For the cold jet it always grows with cylindrical radius and hence
with the distance from the jet origin. This reflects the fact that the
plasma is being spun up by the rotating magnetic field:
in this case $|B_p / B^{\hat\phi}| \gg |v_p/v^{\hat\phi} |$ in
equation~(\ref{omega-def}) and $v^{\hat\phi} \approx r \Omega \propto r$.
However, in the hot jet $r v^{\hat{\phi}}$ (and therefore also
$v^{\hat{\phi}}$) initially decreases with increasing $r$ and even 
attains negative values, indicating counter rotation of the plasma. 
Eventually the cold-jet behaviour is restored, with the switch
taking place at the turning point of $\mu_m$. The decrease in $r
v^{\hat\phi}$ when $\mu_m$ increases along a field line follows from the
following relation,
\begin{equation}\label{rvphi}
\frac{r \Omega v^{\hat\phi}}{c^2}= 
1- \frac{1-{l \Omega}/{\mu c^2}} {1 - {\mu_m}/{\mu}} \,,
\end{equation}
obtained by combining equations~(\ref{angm-def})
and~(\ref{kap-def}).\footnote{The inequality ${l \Omega}/{\mu c^2} < 1$
always holds in trans-Alfv\'enic flows, since $({l \Omega}/{\mu
c^2})^{1/2}$ equals the value of $r/r_{\rm lc}$ at the Alfv\'en surface
(e.g. \citealp{VK03a}), and the Alfv\'en surface is located closer to
the source than the light cylinder (with the two surfaces almost
coinciding for highly magnetized flows).} 
Physically, the increase in $\mu_m$ implies that the magnetic
contribution to the total angular momentum per unit rest mass goes
up (see equations~\ref{mu_m-def} and~\ref{angm-def}), which, by the
conservation of $l$ along a field line (and taking account of energy
conservation) implies that the specific material angular momentum  $r
v^{\hat\phi}$ must decline.

The efficiency of the acceleration in model B2H is higher than in the 
cold models, as can be seen in Fig.~\ref{gbut}.
This is connected to the behaviour of the function ${\cal S}$.
The increase of ${\cal S}$ during the 
thermal acceleration phase results in a higher ${\cal S}_{\rm f}$, the value
of the function ${\cal S}$ at the fast magnetosonic surface. In addition,
the asymptotic value ${\cal S}_\infty$ is smaller than in cold models
(see Fig.~\ref{gbut-extra}). Both effects result in a higher
value of $\Gamma_\infty / \mu$ (see equation~\ref{G_infty}).

%sssssssssssssssssssssssssssssssssssssssssssssss
\subsection{Comparison with semi-analytic solutions}
%sssssssssssssssssssssssssssssssssssssssssssssss

As discussed in Section~\ref{introduction}, it is possible to find exact
solutions of the relativistic MHD equations by assuming radial
self-similarity \citep{LCB92,Con94,VK03a,VK03b,VK04}.  Due to the
mathematical complexity of the equations, these are the only possible
exact semi-analytic solutions describing cold or polytropic flows
\citep{VK03a}.  Similarly to their non-relativistic counterparts (the
Blandford-Payne--type models), they successfully capture the physics of
magnetically driven jets and yield the general characteristics of the
flow acceleration and collimation.\footnote{The self-similar solutions
of \cite{VK03a} have a line-shape $z\propto r^2$ (see footnote
\ref{footnote_bland}) and thus most closely resemble our model C.}  In
particular, the results of \cite{VK03a,VK03b} for ultra-relativistic GRB
jets follow the general scaling relationships derived here.
In fact, the scaling $\Gamma \propto r^{b-1}$, corresponding to a
streamline shape $z\propto r^b$ (for $1<b\le 2$; the regime (iii) of
equation~\ref{transf2}), was first presented in
\citet{VK03b}. Note in this connection that both the $\Gamma \simeq
r/r_{\rm lc}$ (equation~\ref{Gamma2}) and the $\Gamma \simeq
z/r$ (equation~\ref{G_z_r}) scalings exhibited by our solutions could be
captured through the basic radial-self-similarity ansatz $\Gamma=\Gamma(\theta)$ 
because both $r/r_{\rm lc}$ and $z/r_{\rm lc}$ are functions of the
polar angle $\theta$ in the self-similar solutions.
The semi-analytic solutions exhibit as high an acceleration efficiency ($\ga
50\%$) as the simulated $b\le 2$ solutions, and, correspondingly, have a
similar value for the asymptotic shape function (${\cal S}_\infty \sim 1/2$;
\citealp{V04dogl}). Self-collimation also acts in a similar way in both
types of solution, with the inner field lines at any given height $z$
being better aligned with the rotation axis than the poloidal field at
larger values of~$r$.

Despite their qualitative similarity in regard to the acceleration and
collimation processes, the semi-analytic and numerical solutions do of
course differ in their details, reflecting the fact that in the
self-similar model the angular velocity at the base necessarily scales
as $1/r$ and that only one current-flow regime is allowed. In
particular, the spatial distributions of the integrals of motion is not
the same in these two cases. For example, the energy integral, which is
constant in the self-similar model, is roughly proportional to the
magnetic flux function in the simulated uniform-rotation jets, and the
adiabat $Q$, which is given as a power of the magnetic flux function in
the self-similar model, is a global constant in the simulations. We also
note that, while the far-asymptotic (beyond the acceleration region)
flow shape in the self-similar models is either cylindrical or conical,
only the innermost field lines become cylindrical in the simulated jets,
whereas further out the streamlines remain paraboloidal. However, this
is evidently related to the imposed boundary shape, and we can expect
that, if the flow were followed to still larger distances, even more of the
interior field lines would tend to cylinders \citep[see][]{CLB91} or
(in the case of an initially ``hot'' flow) to cones.

The high acceleration efficiency inferred from the self-similar and
numerical solutions for non-radial, relativistic MHD outflows
was also deduced by \cite{BN06} on the basis of a perturbative analysis around 
a parabolic ($z \propto r^2$) flow. These authors found that the Lorentz factor
increases with distance from the origin as $\Gamma \propto z^{1/2}$, 
in agreement with our general result for paraboloidal jets of this type,
$\Gamma \sim z/r$.

%sssssssssssssssssssssssssssssssssssssssssssssss
\section{Application to GRB Jets}
%sssssssssssssssssssssssssssssssssssssssssssssss
\label{application}

The observational study of GRBs has not yet reached the stage where the
basic parameters of the flows producing prompt $\gamma$-ray emission and
afterglows have become well established. There is no general consensus
yet on the angular structure, degree of collimation, distance from the
central source or composition of GRB jets. These parameters may vary
significantly from burst to burst. The anisotropy of $\gamma$-ray emission
due to relativistic beaming further complicates the problem as the same
burst could have a very different appearance when observed from
different viewing angles.  In this section we test our theory against
the current, not yet very stringent, observational constraints and
provide a guide for future observations.

The maximum terminal Lorentz factors in our numerical models of
parabolic jets, $\sim 100 - 300$, are close to those inferred for
long/soft GRB jets and also high enough to ensure that we have captured
the properties of magnetic acceleration in the ultra-relativistic
regime. Although real GRB jets may be even faster \citep[e.g.][]{LS01},
the analytic results verified by our numerical study can be applied to
such jets with a high degree of confidence.

To make detailed comparisons between our theory and the observations 
we need to determine the characteristic light-cylinder radius  
at the source of the jets. In the case of a millisecond magnetar 
\begin{eqnarray}
  r_{\rm lc}=\frac{cT}{2\pi} \simeq 5\times10^6
  \left(\frac{T}{1\,\mbox{ms}}\right)\,\mbox{cm}\,,
\nonumber
\end{eqnarray}
and for a maximally rotating black hole 
\begin{eqnarray}
   r_{\rm lc}=4r_g \approx 6\times10^5
   \left(\frac{M}{M_\odot}\right)\,\mbox{cm}\,.
\nonumber
\end{eqnarray}
Thus, $L=10^6$cm is a suitable reference length-scale for this application. 

Given the extended nature of magnetic acceleration,
the first question that one has to address is whether the Lorentz 
factors deduced from observations can be reached in our model 
on the inferred scale of the $\gamma$-ray emission region. According to
equations~(\ref{Gamma2}) and~(\ref{G-scaling}),  
\begin{eqnarray}
R\simeq10^{12}\left(\frac{\Gamma}{100}\right)^3\mbox{cm}
\nonumber
\end{eqnarray}
for paraboloidal jets with $b=3/2$ and $b=3$, and 
\begin{eqnarray}
R\simeq10^{10}\left(\frac{\Gamma}{100}\right)^2\mbox{cm}
\nonumber
\end{eqnarray}
for paraboloidal jets with $b=2$. These estimates are lower than the distance 
to the $\gamma$-ray production region inferred from the burst variability in 
the internal-shocks model of GRBs,
\begin{eqnarray}
R_{\gamma} \sim \Gamma^2 c \delta t = 3\times 10^{13} 
  \left(\frac{\Gamma}{100}\right)^2 
  \left(\frac{\delta t}{0.1\; {\rm s}}\right)\; {\rm cm}\, ,
\nonumber
\end{eqnarray}
where $\delta t$ is the internal variability time-scale \citep[e.g.][]{Pir05}.
In fact, recent {\it Swift}\/ observations indicate even
larger distances ($\sim 10^{15}-10^{16}\; {\rm cm}$;
e.g. \citealt{Lyu06a,Kum07}). The theory thus appears to be
consistent with the observations in this respect.

We emphasize that the above results have been derived in the context
of ideal and axisymmetric MHD. In reality, various instabilities, and
in particular non-axisymmetric, current-driven ones occurring near the
jet axis, may result in magnetic reconnection and dissipation. It is
interesting to note in this connection that the dissipation of
Poynting flux would naturally generate a negative magnetic pressure
gradient (associated with the azimuthal field component) along the flow
and that this process was argued to be capable, on its own, to
accelerate the flow to a high Lorentz factor
\citep[e.g.][]{DS02,Dre02}. In this respect our ideal-MHD simulations
may be yielding only lower limits on the terminal Lorentz factor in
the modelled jets.

A related issue is whether there is an adequate confining 
medium, as required for the establishment of the ``power-law''
acceleration regime described by equation~(\ref{G-scaling}). If the 
confinement of a long/soft GRB jet is provided only by the envelope 
of the progenitor massive star, as proposed by \citet{TMN08}, the
acceleration would need to take place on a scale smaller than the
stellar radius, $\sim 10^{11}-10^{12}\;$cm.  
Downstream of the stellar surface the jet is expected to enter the regime 
of ``free'' (ballistic) expansion, as in our model E, which is
characterized by a less efficient magnetic acceleration.\footnote{
It has been suggested that matter-dominated GRB jets could
remain confined by the expanding cocoon of relativistically hot 
shocked jet material after they break out through the stellar surface
\citep[e.g.][]{RCR02} and could continue to accelerate during that phase
\citep[e.g.][]{LB05}. In contrast, Poynting-dominated jets do not inflate 
large cocoons but instead create the so-called ``nose cones''
\citep[e.g.][]{Kom99}. In fact, given the low compression ratio of a fast
shock in a magnetically dominated plasma, a jet termination shock is
unlikely to form before the jet emerges from the star --- instead, the
jet would have the form of a super-Alfv\'enic but sub--fast-magnetosonic
outflow, as has been observed in recent computer simulations
\citep[e.g.][]{KB07,BK08}.} But even this rather restrictive constraint 
on the size of the acceleration region, and hence on $\Gamma_\infty$, 
is in principle consistent with the theory. 
An alternative possibility is that the GRB outflow is confined by a wind
launched from the surface of a disc that surrounds the central object
\citep[e.g.][]{LE00}. This mechanism is a prime candidate for the
confinement of short/hard GRB outflows, which evidently do not originate
inside a star. In this case the collimation might be attained smoothly, with
the disc-driven and central object-driven components constituting parts of a
coherent outflow configuration \citep[e.g.][]{TMN08}.  However, the outflow 
may also involve shocks formed at the interface of these two
components \citep[e.g.][]{BL07}. If the GRB jet and disc outflow
commence at the same time, the spatial extent of the confining medium in
this picture can be estimated as 
\begin{eqnarray}
R_{\rm wind} \approx 3 \times 10^9 \left (\frac{v_{\rm wind}}{0.1\,
c}\right ) \left ( \frac{\Delta t}{1\, {\rm s}}\right )\; {\rm cm}\,,
\nonumber
\end{eqnarray}
where $v_{\rm wind}$ is the mean wind speed over this distance and
$\Delta t$ is the GRB duration (normalized here to a fiducial value
appropriate for a short/hard burst). This should be compared with the
above theoretical relationships between $R$ and $\Gamma$, which for
$\Gamma = 30$ (a fiducial value for the lower limit on $\Gamma_\infty$
in short/hard GRBs; e.g. \citealt{Nak07}) yields $R \approx 3 \times
10^{10}\; {\rm cm}$ for $b=3/2$ or $b=3$ and $R \approx 9 \times 10^{8}\; {\rm
cm}$ for $b=2$. This comparison indicates that, over the time $\Delta
t$, a moderately relativistic disc outflow could form a sheath around
the jet acceleration region. Given that the size of a disc that forms
during a binary (NS-NS or NS-BH) merger that gives rise to a short/hard
GRB event is not expected to exceed a few times $10^6\; {\rm cm}$
(i.e. significantly less than than the expected cylindrical radius of
the jet in the main acceleration region), meaningful confinement would
be attained only if the wind had sufficiently large inertia, which would
require the wind-to-jet total energy ratio to be $\gg 1$
\citep[cf.][]{LE00}. 
If the initial magnetizations of short/hard and
long/soft GRB outflows are comparable, this scenario provides a
plausible explanation of the finding (from the best available current
data) that short-GRB jets are on average less relativistic than their
long-duration counterparts. A concomitant prediction, which could be
tested when more afterglow data for short/hard GRBs become available, is
that short/hard GRB outflows should also be less well collimated, on
average, than long/soft ones.
 
The internal-shocks model envisions the prompt GRB emission to be
powered by the collision of successively ejected relativistic ``shells''
\citep[e.g.][]{Pir05}. This scenario requires the jet to be kinetic

energy-dominated on the scale of 
the emission region; otherwise, the flow deceleration 
and dissipation at fast shocks is too weak 
(or else, if the flow is inhomogeneous, the energy
requirements are strongly increased). 
The numerical solutions presented in this paper have demonstrated the 
possibility of efficient conversion of Poynting flux into bulk
kinetic energy, with $\ga 50\%$ efficiency attained by the end of 
the power-law--like acceleration regime. However, the distance
$R_\gamma$ of the prompt 
emission region from the central source imposes a constraint on the initial 
magnetization of GRB jets in this model. Using 
equation~(\ref{equip1}), we obtain  
\begin{equation}
  \mu 
\approx 2 \Gamma_\infty < \left\{
\begin{array}{lcc}
2(r_\gamma/r_{\rm lc})^{b-1} & \mbox{if} & b\le 2 \\
2(r_\gamma/r_{\rm lc}) & \mbox{if} & b\ge2 \\
\end{array}\right. .
\nonumber
\end{equation}
For paraboloidal jets with
$b=3/2$ or $b=3$ this gives (setting $R_{\rm lc} \approx r_{\rm lc}$) 
\begin{eqnarray}
  \mu < 430 \left(\frac{R_\gamma}{10^{13}\mbox{cm}}\right)^{1/3}\,,
\nonumber
\end{eqnarray}
whereas for $b=2$ we obtain 
\begin{eqnarray}
   \mu <6\times10^3\left(\frac{R_\gamma}{10^{13}\mbox{cm}}\right)^{1/2}\,.
\nonumber
\end{eqnarray}
By approximating $\Gamma_\infty \dot M_j c^2 \approx {\cal{E}}/\Delta
t$, where ${\cal{E}}$ the outflow kinetic energy as inferred from
afterglow observations and $\Delta t$ is the burst duration, we
estimate the mass outflow rate in the jet to be
\begin{eqnarray}
  \dot{M}_j \approx 5.6\times 10^{-8} \; 
\left(\frac{\cal{E}}{10^{51}\mbox{erg}}\right) 
\left(\frac{\Delta t}{10\mbox{s}}\right)^{-1}   
   \left(\frac{\Gamma}{10^3}\right)^{-1} M_\odot\;\mbox{s}^{-1}\,,
\nonumber
\end{eqnarray}
where we normalized by values appropriate to long/soft bursts.  This is
very much lower than the expected mass accretion rate onto the central
black hole in the collapsar model ($\sim 0.05 - 1\;
M_\odot\,\mbox{s}^{-1}$; e.g.  \citealt{PWF99}) and constitutes the
so-called ``baryon loading problem'' in GRB source models. Such a
comparatively low mass outflow rate might be produced if the
GRB-emitting outflow originates on magnetic field lines that thread the
horizon of a spinning black hole and tap its rotational energy via the
Blandford-Znajek mechanism \citep[e.g.][]{LE93}; in this case the flow
would initially be baryon-free and would require a baryon-injection
mechanism as it propagates outward.  Alternatively, jets launched from
an accretion disc may experience such a low mass loading if they are
initially thermally driven along magnetic field lines inclined at a
small ($\la 15^\circ$) angle to the rotation axis
\citep{BL08}.\footnote{It was also proposed that the problem could be
alleviated in a magnetically driven disc outflow that is initially
neutron rich and hot
if the neutrons decouple from the protons well
before the latter attain their terminal Lorentz factor 
(see \citealt{VPK03} and \citealt{FPA00}). 
There are indications from studies
of discs around non-rotating black holes that this might not work in
practice because outflows may be required to be comparatively massive to
remain neutron rich \citep[e.g.][]{Lev06,BL08}, but this conclusion
still needs to be verified in the case of discs around rapidly rotating
black holes. }

The internal-shocks model of GRBs has been questioned on account of
the relatively high emission efficiency that it requires, and these
challenges have become significantly stronger following observations
made by {\it Swift}\/ \citep[e.g.][]{GKP06,Kum07}. Various suggestions
have been made (and continue to be made) in the literature for
reconciling this scenario with the observations \citep[e.g.][]{KZ07} or
else for modifying or replacing it. Perhaps the main alternative
picture proposed to date is based on the assumption that 
the prompt high-energy emission is produced directly from
the dissipation of magnetic energy without requiring it to be
converted into kinetic energy first \citep[e.g.][]{Kum07}, which
circumvents the efficiency problem that has troubled the
internal-shocks model. Although magnetic dissipation could in principle
occur also in the context of the MHD model \citep[e.g.][]{DS02}, perhaps
the most extreme realization of this idea occurs within the framework of
the magnetodynamics scenario, in which GRB outflows are regarded as
remaining Poynting flux-dominated (and sub--fast-magnetosonic)
in the $\gamma$-ray emission region \citep[e.g.][]{Bla02,Lyu06b}.
In this scenario, neither the internal nor the reverse shocks
of the standard model would develop, which could be the basis for an 
observational test.\footnote{Note in this connection that, in some of 
the proposed interpretations of the {\it Swift}\/ data 
\citep[e.g.][]{UB07,GDM07}, the entire afterglow emission is 
attributed to a reverse shock that is driven into the ejecta.}

As we discussed in Section~\ref{power-law}, a key prediction of the
magnetic acceleration model is the approximate inverse proportionality
between the Lorentz factor along a poloidal magnetic surface and
$\tan\vvartheta$ for that surface for paraboloidal jets with $1<b \le 2$
(see equation~\ref{lor-theta-eq}). For a small opening angle and $b$ not
very close to 1 this result
can be approximated as $\Gamma \vvartheta \approx 1$. This implies that
GRB outflows with $b\le 2$ that attain $\Gamma \sim 100$, 
the approximate inferred lower limit for long/soft
GRBs, must have $\vvartheta \sim 0.6^\circ$, essentially independent of
the details of the acceleration process.
When $b>2$, $\Gamma\vvartheta \approx b^{-1} (R/r_{\rm lc})^{-(1-2/b)}$
{\em decreases} with $R$ in the magnetic acceleration region, implying
an even smaller value of $\vvartheta$ at the end of this zone.  The
relation $\Gamma\vvartheta \sim 1$ may be useful for differentiating
between magnetic and fireball models of GRB flows.  Indeed, this
property is generic to the magnetic acceleration mechanism, whereas for
the thermal acceleration the terminal bulk Lorentz factor is essentially
given by the thermal Lorentz factor at the base of the flow and is
fairly independent on the flow collimation, which means that the product
$\Gamma \vvartheta$ can in principle become $\gg 1$. Interestingly, one
of the proposals made for interpreting the apparent GRB ``tails''
observed by {\it Swift}\/ invokes a GRB-emitting outflow component whose
opening half-angle must be $< 1^\circ$ \citep{Pan07}. While the currently
available data are not sufficient for favouring this interpretation over
other suggested explanations of the ``tails,'' it is noteworthy that the
requirement arrived at by \citet{Pan07} on strictly phenomenological
grounds is consistent with a distinguishing property of the magnetic
acceleration model. It is also noteworthy that there is already at least
one source (GRB 070401) in which such a small opening half-angle has
been inferred directly from a measurement of an early break in the X-ray
afterglow light curve \citep{Kamb08}.
Such small asymptotic opening angles and even 
$\Gamma\vvartheta\la 1$ could in principle be 
attained also in purely hydrodynamical jet models, although this would
require a very high efficiency of collimation and acceleration within
the stellar interior. Specifically, the jets would need to emerge from
the star with $\theta_v<1^\circ$ and $\Gamma\ga
1/\theta_v \simeq 60$, which, in view of recent analytic and numerical studies 
\citep[e.g.][]{LB05,MLB07}, is unlikely to be achieved in practice. 

The original fireball model for GRB jets envisions a uniform conical
outflow that becomes accelerated to Lorentz factors $\Gamma \gg
1/\vvartheta$ and predicts that during the afterglow phase the Lorentz
factor of the forward shock driven by the jet into the ambient medium
will decrease to values $< 1/\vvartheta$. The observational consequence
of this transition is a panchromatic break in the afterglow light curve
(referred to as the ``jet break'') occurring when $\vvartheta \Gamma$ becomes
$\sim 1$ \citep[e.g.][]{Rho99,Sari99}. In view of the results presented
in this paper, the predictions of the MHD model for GRB outflows that are
efficiently accelerated --- and therefore necessarily confined (by
either thermal, magnetic or ram pressure) during the acceleration phase
--- are radically different.
Specifically, the MHD model predicts that the afterglow light curve would
exhibit either a very early jet break (in cases where
$\Gamma\vvartheta\approx 1$ at the end of the acceleration phase, as
expected in jets with $b \le 2$) or no jet break at all (if
$\Gamma\vvartheta < 1$ at the end of the magnetic acceleration region,
as expected in jets with $b>2$).\footnote{If the low current detection
rate of jet breaks in the early afterglow light curves of GRB sources
would prove to be more than just the result of observational
difficulties, this could be an indication, when interpreted in the
context of the magnetic acceleration model, that these jets are
characterized by effective shape-function exponents $b>2$.} This prediction 
is seemingly at odds with the inference from a number of pre-{\it Swift}\/
GRB sources of breaks of this type occurring on a time-scale of days
(see e.g. \citealt{LZ05} for a compilation). The paucity of ``textbook''
jet breaks in {\it Swift}\/ GRB sources \citep[e.g.][]{Lia08}, which has
even cast doubts on the interpretation of the alleged pre-{\it Swift}\/
jet breaks, points to one way out of this dilemma: it may be that indeed
there are no bona fide jet breaks at later times. We recall, however,
that the jet-break interpretation lies at the basis of the
identification of GRB outflows as collimated jets, which has
significantly reduced the otherwise prohibitive energy requirements in
some sources. Alternatively, it could be that the difficulties in
finding late-time jet breaks in {\it Swift}\/ sources are to a large
extent observational \citep[e.g.][]{Zhang07}, in which case other
explanations for late-break candidates must be sought.

One natural possibility is that the outflow possesses more than one
kinematic component. In its simplest incarnation, this is the ``two
component'' model, which envisions the prompt emission to originate in
an ultra-relativistic, highly collimated jet and the afterglow emission
to be dominated by a less relativistic, wider outflow component. The
suggestion in \citet{Pan07} and in \citet{Kamb08} that the $\gamma$-ray
emitting jet is very narrow was made in the context of this model, and a
similar picture was used by \citet{GKP06} to explain other aspects of
the early GRB X-ray emission measured by {\it Swift} \citep[see
also][]{Zhang07}. In fact, a two-component outflow configuration had
already been proposed in the pre-{\it Swift}\/ era to account for
certain observations \citep[e.g.][]{Ber03} and as a means of alleviating
the efficiency requirements on the internal-shocks model
\citep{PKG05}. The separation into two components could arise either
from an interaction of the outflow with the envelope of a massive
progenitor star or represent an intrinsic property of the central engine
(see \citealt{PKG05} for a summary of some specific proposals). In the
context of the magnetically driven outflow model, there are at least two
possibilities for an intrinsic origin. First, neutron-rich, hot outflow
may split into two components when the neutrons and protons decouple
before the protons have attained their terminal Lorentz factor
\citep{VPK03}. Second, a baryon-poor ultra-relativistic outflow launched
from the black hole can be surrounded by a magnetically driven,
relativistic outflow from the accretion disc itself
\citep[see][]{GKP06}.\footnote{In the latter scenario, the disc wind
could provide a ready source for seeding the central funnel with baryons
\citep[e.g.][]{LE03} and could also help collimate the interior outflow
\citep{LE00}.}  
We stress that, in reality, the outflow may be more complex than in the
schematic ``two component'' picture sketched above. For example,
inhomogeneities in the accretion flow may result in several distinct
outflow components emerging from the disc, associated, perhaps, with
isolated magnetic flux tubes that thread the disc at different
locations. Phenomenologically, this situation might resemble the
``patchy shell'' scenario considered by \citet{KP00}.

%fffffffffffffffffffffffffffffffffffffffffffffffffffffffffffffffffff
\begin{figure*}
\includegraphics[width=60mm]{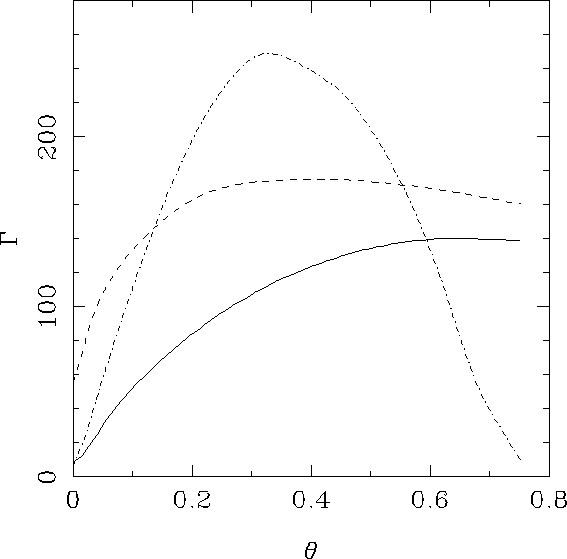}
\includegraphics[width=60mm]{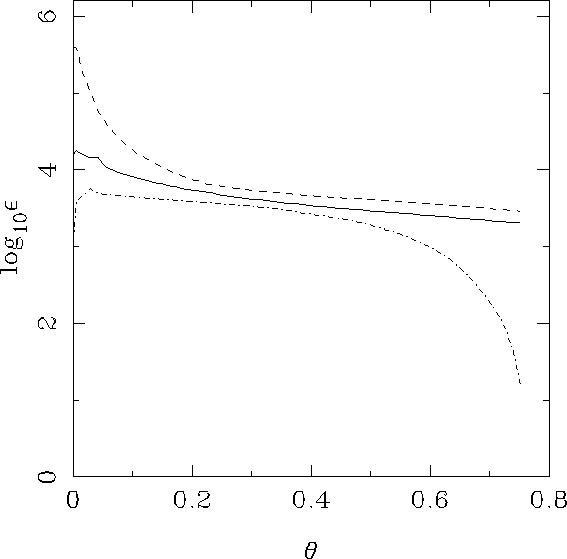}
\includegraphics[width=60mm]{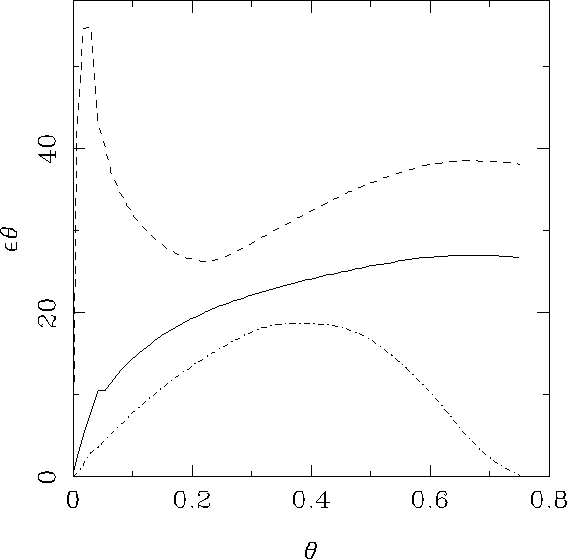}
\includegraphics[width=60mm]{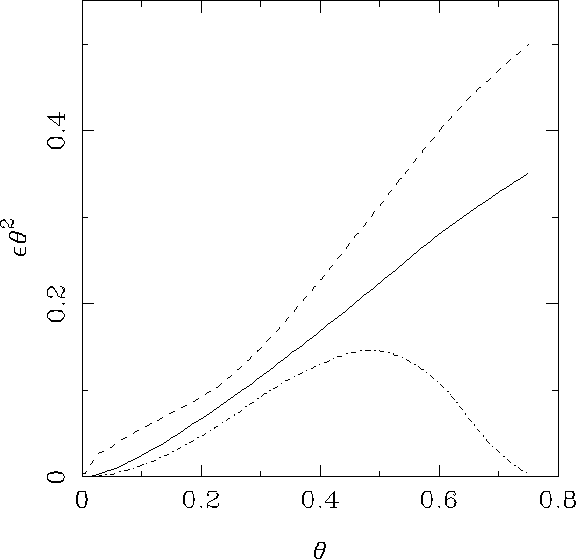}
\caption{Angular distributions of the Lorentz factor (top left panel), 
the kinetic power per unit solid angle in the local direction of 
the flow ($\epsilon$, top right panel), the kinetic power per annulus of 
unit angular size ($\epsilon\theta$, bottom left panel)
and $\epsilon\theta^2$ (bottom right panel) in the asymptotic regime,
plotted as functions of polar angle.  
The variable $\epsilon$ is given in units of
$c B_0^2 L^2 / 4 \pi$, and when it is multiplied by $\theta$ or $\theta^2$ 
the polar angle is measured in radians. Note, however, that the polar 
angle along the horizontal axis is given in degrees.
The solid lines show model B2, the dashed lines
model B2H, and the dash-dotted lines model D. 
}
\label{grb-profiles}
\end{figure*}
%fffffffffffffffffffffffffffffffffffffffffffffffffffffffffffffffffff
%
The distribution of the terminal Lorentz factor and of the kinetic power
across the jet directly affects the evolution of the light
curve of the GRB afterglow \citep[e.g.][]{Gra05} as well as the
statistical properties of a GRB sample \citep[e.g.][]{NGG04} and the
detectability of ``orphan'' afterglows (afterglows detected without an
associated GRB; e.g. \citealt{NP03}). One could in turn attempt to use
such observations to probe the jet structure and to test the
underlying acceleration and collimation models.
With this in mind, we present in Fig.~\ref{grb-profiles} illustrative
asymptotic distributions of the Lorentz factor and of the kinetic power 
from our simulations. 

We consider first the $\Gamma_\infty$ distribution.  The top left panel 
of Fig.~\ref{grb-profiles} shows that in all of the cases the Lorentz
factor decreases toward the axis --- this is a generic feature of the
axisymmetric, ideal-MHD acceleration mechanism as the azimuthal
magnetic field and hence the Poynting flux vanish along the symmetry
axis.  This feature may not, however, be as pronounced when
non-axisymmetric instabilities and resistive dissipation of magnetic
energy (which are not incorporated into our study) are taken into
account. In fact, we find that even in our solutions $\Gamma\not=1$ at
$\theta=0$ because of numerical dissipation. In the case of an
initially hot outflow $\Gamma(\theta=0)>1$ is due to the 
thermal acceleration. In initially cold outflows that have uniform
rotation and mass density distribution 
at the base $\Gamma$ peaks at the jet
boundary. It is seen, however, that if the flow is initially hot the
anisotropy of the Lorentz factor distribution within the jet is
reduced.  Uniform rotation is a robust prediction of models with a
magnetar or a magnetized black hole as a central rotator. 
The assumption 
a uniform mass-flux distribution at the jet base is more of an approximation: 
for example, when the central source is a black hole the degree of baryon 
loading is likely to be higher near the jet boundary due to various 
boundary interactions with the jet surroundings \citep[e.g.][]{LE03}. 
Such a mass distribution would lead to lower terminal Lorentz factors near the
boundary compared to that found in our simulations.  
If the inner regions of an accretion
disc contribute to the magnetic driving of the GRB-emitting outflow
component then a model with differential rotation, with $\Omega$
decreasing away from the centre, is more suitable. As seen from the
figure, in this case the terminal Lorentz factor peaks at intermediate
angles. 
In practice it may, however, be difficult to distinguish this case from
that of uniform rotation with nonuniform mass loading.

Turning now to the distribution of energy flux across the jets in the
asymptotic regime, we recall that the observational consequences of this
energy are strongly influenced by relativistic beaming --- whenever a
fraction of this energy is dissipated and converted into radiation, this
radiation will be beamed in the direction of motion of the corresponding
fluid element, given by $\vvartheta$.  Most phenomenological 
models of GRBs have assumed that the jet is conical 
and has radial streamlines. Thus, the streamline angle, $\vvartheta$,
is equal to $\theta$, the polar angle of
the fluid element.  In our model the streamlines are curved and
asymptotically their shape is close to that of the boundary (with the
exception of the cylindrical core).  Hence we have
$\vvartheta\simeq\theta/a$. Consider a surface element normal to the
$\eta$ coordinate lines (streamlines) 
$d\Sigma_\eta=\sqrt{g_{\phi \phi} g_{\xi \xi}} d\phi d\xi$, 
where $g_{\phi \phi}$ and $g_{\xi \xi}$ are components of the metric tensor. Since in
the asymptotic regime $\theta,\vvartheta\ll 1$, we can write $d\xi=a
z^{1-1/a}d\vvartheta$, $g_{\phi \phi}=r^2$ and $g_{\xi \xi}=z^{2/a}$ 
(see Appendix A of Paper I),
and hence
\begin{eqnarray}\nonumber
d\Sigma_\eta=a^2z^2 d\omega,
\end{eqnarray}
where $d\omega=\vvartheta d\vvartheta d\phi$ is the solid angle defined by the
tangents to the streamlines passing through the surface element. The
power per unit solid angle is then given by 
\begin{eqnarray}\nonumber
d{\cal L}/d\omega \equiv \epsilon = S^{\eta} a^2 R^2, 
\end{eqnarray}
where $S^{\eta}$ is the component of the energy flux density in the $\eta$
direction.

The top right panel of Fig.~\ref{grb-profiles} shows the distribution of
kinetic power per unit solid angle, $\epsilon$ 
(the total power has a very similar distribution). 
One can see that in all models it peaks at, or very close to, $\theta=0$.  
The reason for this behaviour, which seemingly 
conflicts with the Lorentz factor distribution shown in the left
panel of the figure, is that the density distribution across the jet
is highly nonuniform, with the mass density strongly peaking near the symmetry
axis on account of the enhanced collimation of the flow in that region
(see Figs.~\ref{model-a}--\ref{model-b2h}). The bottom left panel of
Fig.~\ref{grb-profiles} shows the distribution of $\epsilon\theta$: this
quantity tells us how the jet power is distributed between annuli of
equal size in $\theta$. 
One can see that in model D (differential rotation) more power comes 
from the intermediate annuli, in model B (uniform rotation at
the base) from the outer annuli, and that
a significant core component emerges in model B2H (initially hot jet).
Note, however, that the distributions of the Lorentz factor and the power
depend on the choices of the density, magnetic flux and angular velocity
distributions at the inlet boundary, so different profiles may be possible.

The derived distributions of $\Gamma(\theta)$ and $\epsilon(\theta)$ are
markedly different from those commonly adopted in phenomenological 
GRB jet models, which either take
them to be uniform within the jet half-opening angle
$\theta_j$ or else assume that the flow has a universal structure, with
$\epsilon$ being a Gaussian or a power-law in $\theta$ (in particular,
$\epsilon \propto \theta^{-2}$; 
\citealt{RLR02} ---
compare with the bottom right panel of Fig.~\ref{grb-profiles}) outside
a uniform-core region.\footnote{
In the force-free electromagnetic model for GRBs it is envisioned that the
current flows along the axis of rotation and returns through the
equatorial plane; this yields an energy distribution $\propto
\theta^{-2}$ in the associated electromagnetic shell
\citep[e.g.][]{Bla02,Lyu06b}. A universal structured outflow with
$\epsilon \propto \theta^{-2}$ could
potentially also be produced when a relativistic GRB jet with possibly a
different initial energy distribution breaks out through the surface of
a massive progenitor star \citep{LB05}.}  The structure exhibited by our
model jets is also different from that of a ``hollow cone,'' where the
flow occupies the region $\theta\in[\theta_j-\Delta\theta,\theta_j]$
\citep[e.g.][]{EL04,LB05}.  Although the distribution of Lorentz factors
is reminiscent of such a cone, the distribution of kinetic power
actually peaks near the symmetry axis. Moreover, in contrast with the
phenomenological hollow-cone models considered in the literature, in
which $\Delta \theta \ll \theta_j$, our solutions yield configurations
with $\Delta \theta \sim \theta_j$.  The detailed observational
implications of these structures remain to be explored.

%%%%%%%%%%%%%%%%%%%%%%%%%%%%%%%%%%%%%%%%%%%%%%%%%%%%%%%%%%%%%%%%%
\section{Conclusion}
%%%%%%%%%%%%%%%%%%%%%%%%%%%%%%%%%%%%%%%%%%%%%%%%%%%%%%%%%%%%%%%%%
\label{conclusion}

In this paper we extend our previous numerical study of magnetically 
accelerated relativistic jets (Paper I) from the case of terminal Lorentz 
factors $\Gamma_\infty \sim 10$, appropriate to AGN jets, 
to $\Gamma_\infty \ga 10^2$,
appropriate to GRB jets. The larger values of $\Gamma_\infty$
reached in the present study enable us to compare results of our simulations,
carried out using the equations of special-relativistic ideal MHD, with the
asymptotic analytic formulae that we obtain from the constituent
equations in the limit $\Gamma \gg 1$. Our analysis of the results also
benefits from a comparison with semi-analytic solutions that were derived
under the assumption of radial self-similarity. We can summarize our
conclusions regarding the magnetic acceleration of ultra-relativistic
outflows as follows.

\begin{enumerate} 

\item Our simulations verify that the MHD
acceleration mechanism remains robust even when the terminal Lorentz
factors reach the ultra-relativistic regime
($\Gamma_\infty \ga 10^2$). The simulated flows rapidly settle into
quasi-steady and seemingly stable configurations. A complete model would
need to incorporate non-axisymmetric effects, which we have not
considered.

\item A key property of magnetically driven relativistic flows in the
ideal-MHD regime is the spatially extended nature of their
acceleration. This property, which was first revealed by the
self-similar solutions and subsequently confirmed in the moderately
relativistic regime by the simulations reported in Paper I, is also a
distinguishing characteristic of jets accelerated to ultra-relativistic
speeds. For initially Poynting flux-dominated jets whose magnetic flux
surfaces can be approximated by paraboloids of the form $z\propto r^b$
(with $b\ge 1$), the Lorentz factor during the main magnetic
acceleration phase increases as $\Gamma \simeq (b/\sqrt{b-1}) z/r$ when
$1<b\le 2$ and as $\Gamma \simeq r/r_{\rm lc}$ when $b=1$ or $b>2$. After the
(increasing) kinetic energy flux becomes comparable to the (decreasing)
Poynting flux the growth of $\Gamma$ saturates, and thereafter it
increases at a much slower rate. (We have not been able to reach this 
phase in models with $b>2$ due to the limitations of our numerical method.)

\item The conversion efficiency $\Gamma_\infty/\mu$ of
total injected energy to kinetic energy at the end of the
power-law acceleration phase lies in the range $55-75\%$ for the
initially cold simulated paraboloidal flows whose effective exponents
lie in the range $1<b \le 2$; the efficiency is smaller
the larger the initial magnetization (or, equivalently, the higher the
value of $\Gamma_\infty$).
A higher efficiency is attained in jets with $b<2$ that are initially
relativistically hot than in the corresponding initially-cold outflows:
in this case a measurable fraction
($>50\%$ in the example that we show) of the thermal energy flux is at
first converted into Poynting flux, thereby reducing the initial
thermal acceleration of the flow and enhancing the subsequent magnetic
acceleration.

\item In our simulations the flow is confined by a rigid wall whose
shape is described by $z\propto r^a$, with $a$ ranging from 2/3 to
3. We have conducted a
detailed analytic investigation of the relationship between a confining
pressure distribution of the form $p_{\rm ext}\propto z^{-\alphap}$ and
the shape of the jet boundary in the asymptotic regime of the magnetic
acceleration zone. We found that there is a one-to-one correspondence
between the functional forms of the pressure distribution and of the
boundary shape. Except for one special case (for which $a$ remains close
to 2), the jet becomes an exact paraboloid of the form given above,
with $a=4/\alphap>2$ for $\alphap < 2$ and $1<a\le 2$ for $\alphap = 2$. When
$\alphap > 2$ the jet cannot maintain pressure equilibrium with the
ambient medium and asymptotes to a conical shape. This situation is
reproduced in our simulations by unconfined flows as well as by flows 
with $a \le 1$. In this case the outer regions of the jet become causally
disconnected (the local opening half-angle of the field lines becomes
larger than the local half-angle of the Mach cone of fast-magnetosonic
waves), and only the innermost regions continue to collimate and
accelerate.

\item We find that for all current-carrying jets (irrespective of whether
the return current flows inside or outside the jet) the innermost
field lines are more strongly collimated than the exterior ones,
indicating ``self collimation'' by the magnetic hoop stress (see also
Paper I). This redistribution of the poloidal field lines within the
jet is directly responsible for the high acceleration efficiency of the
flow.

\end{enumerate}

We have applied our results to GRB sources, taking into account the
constraints imposed by the detected prompt and afterglow emission on the
properties of the ultrarelativistic jets that evidently give rise to the
GRB phenomenon. Our main conclusions are:

\begin{enumerate}      
\item 

Initially Poynting flux-dominated outflows can be magnetically
accelerated to a Lorentz factor exceeding the minimum ($\Gamma\sim
10^2$) inferred in long/soft GRBs within a distance of $\sim
10^{11}-10^{12}\; {\rm cm}$ from a rapidly rotating stellar-mass black
hole or a millisecond magnetar.  Thus, most of the acceleration of
long/soft GRB jets can be achieved inside a typical progenitor star
in the collapsar model, whose envelope provides a natural confining
environment for the jets. Lack of confinement outside of the star may
result in a radial outflow characterized by loss of causal
connectivity across the jet and inefficient acceleration. 
An alternative confinement mechanism that is of particular relevance to
short/hard GRBs, which likely form through a
merger of compact stars rather than in the collapse of a massive star,
is a disc wind.
The MHD acceleration mechanism implies that the minimum bulk Lorentz
factor inferred in short/hard GRBs ($\Gamma\sim 30$) could be attained
within the distance that such a wind covers over the burst duration if
the disc outflow (which might also be driven magnetically) has at least
a moderately relativistic speed ($\sim 0.1-1\, c$).

\item 
The MHD acceleration model entails a high ($\ga 50\%$) asymptotic 
conversion efficiency of injected magnetic and thermal energy into 
bulk kinetic energy for effectively confined flows. 
If the initial magnetization is of the same 
order as that of the inferred Lorentz factor of a GRB jet, 
$\sigma_0\sim 10^2-10^3$, the energy conversion can be attained on a 
spatial scale that is smaller than the indicated size of the prompt
emission region. The model is then compatible with the internal-shocks
scenario for GRBs. For a much higher initial magnetization the jet
remains Poynting flux-dominated on these scales and the prompt
emission has to be attributed to direct magnetic energy dissipation,
as in the magnetodynamics scenario. A full treatment of the dynamics
of such jets in the context of MHD would require taking account of
the acceleration induced by the field-dissipation process and the use of
a non-ideal, relativistic-MHD code.

\item
We have found that the MHD jet model places a strong
constraint on the product of the Lorentz factor and 
the half-opening angle of the streamline in the asymptotic regime of the
main acceleration region: $\Gamma\vvartheta\simeq 1$ 
along paraboloidal streamlines $z \propto r^b$ when $b\le 2$ (but $b$ not
too close to 1), and $\Gamma\vvartheta\propto z^{-(1-2/b)}$ (and thus
attaining even smaller values at the end of the main acceleration phase)
when $b>2$.
This feature is unique to the ideal MHD
mechanism and could potentially serve to distinguish it from alternative
models, notably the classical fireball scenario (in which
$\Gamma\vvartheta$ is envisioned to be $\gg 1$ at the end of the
acceleration region). In particular, this property implies that, if
long/soft GRB jets with $\Gamma \ga 100$ are magnetically accelerated, 
they must be collimated to $\vvartheta \la 1^\circ$. 
This result is consistent with one of the interpretations of the
prompt emission ``tails'' discovered by {\it Swift},
although this is not the only possible explanation of a very small
collimation angle. This relationship also indicates that
the $\gamma$-ray emitting outflow component might exhibit a
panchromatic jet break (corresponding to $\Gamma\vvartheta\simeq 1$
decreasing from a value $>1$ to a valure $<1$)
soon after it enters the afterglow phase, although in principle no such
break need to occur (corresponding to cases where this product is $<1$ at
the end of the acceleration zone). A later jet break could
potentially be seen if the outflow has more than one kinematic component.

\item
The magnetic acceleration model also makes specific predictions about
the angular distributions of the terminal Lorentz factor and of the
kinetic and total energy per unit solid angle across the jet, which can
be probed by a variety of observations. These distributions depend on
the magnetization profile and the thermal energy content of the jet at
the inlet boundary, which could in principle be constrained by the
observations. A general characteristic of this model is that
$\Gamma_\infty(\theta)$ {\it decreases}\/ with decreasing polar angle
$\theta$ near the symmetry axis.

\end{enumerate}      

Although our analytic scalings have been derived in the limit where the
jet is in the force-free regime, we emphasize that key parameters of
interest for astrophysical applications --- including the jet velocity
and the magnetic-to-kinetic energy conversion efficiency --- could have
only been obtained within the magneto{\em hydro}dynamics formalism that we
adopted and not in the magnetodynamics (or force-free electrodynamics)
approximation adopted in other recent semi-analytic and numerical
investigations. Another point worth emphasizing is that the acceleration
mechanism investigated in this paper is identical to that considered in
paper I. Our results are consistent with the view that the main
difference between ``superluminal'' AGN jets and GRB jets is that the
latter outflows have a higher initial magnetization (and possibly also a
higher initial enthalpy), which leads to their correspondingly higher
terminal Lorentz factors. If this picture is correct, one could use
observations of AGN and GRB sources to deduce complementary aspects of
the same basic phenomenon. For example, one could take advantage of the
fact that the acceleration region in AGN jets is potentially resolvable
by radio interferometry to probe the details of the acceleration
process; one could then consider the implications to GRB jets, which are
not directly accessible to such observations.

%%%%%%%%%%%%%%%%%%%%%%%%%%%%%%%%%%%%%%%%%%%%%%%%%%%%%%%%%%%%%%%%%
\section*{Acknowledgments}
%%%%%%%%%%%%%%%%%%%%%%%%%%%%%%%%%%%%%%%%%%%%%%%%%%%%%%%%%%%%%%%%%
This research was funded by PPARC under the rolling grant
``Theoretical Astrophysics in Leeds'' (SSK and MVB).
NV acknowledges partial support by the Special Account 
for Research Grants of the National and Kapodistrian University of Athens.
AK was partially supported by a NASA Theoretical Astrophysics Program grant.
We thank Vasily Beskin for many helpful comments on the magnetic acceleration 
mechanism and Jonathan Granot for useful discussions of GRB issues. 
%%%%%%%%%%%%%%%%%%%%%%%%%%%%%%%%%%%%%%%%%%%%%%%%%%%%%%%%%%%%%%%%%

\appendix
\section{Solutions of equation~(\ref{ODE1})}
\label{appA}

In Section~\ref{pressure} we considered the dependence of the jet
boundary shape on the external pressure distribution, and we derived a
second-order ordinary differential equation (equation~\ref{ODE1}) that
expresses this dependence in the asymptotic regime of the main magnetic
acceleration region for the case where the external pressure scales as
$p_{\rm ext}\propto z^{-\alphap}$. For convenience, we reproduce this
equation here, keeping the original notation:
\begin{equation}
\frac{d^2 x}{dZ^2} +C \frac{x}{Z^\alphap} - \frac{1}{x^3} =0 \,,
\label{ODE}
\end{equation}
where $C$ is a constant of the order of 1 (equation~\ref{C2}).
In Section~\ref{pressure} we obtained solutions for this equation after
making a power-law ansatz for $x(Z)$ (equation~\ref{K}). In this
appendix we consider general solutions of this equation without assuming
from the start that they have a power-law form.

\subsection{$\alphap > 2$}
\label{Aalphap>2}

One can identify three different regimes in this case.

\subsubsection{$2<\alphap<4$}

When $\alphap < 4$ the ratio of the second (electromagnetic) to the third
(centrifugal) terms on the left-hand side of equation~(\ref{ODE})
diverges as $Z\rightarrow \infty$, and one can therefore neglect the
centrifugal term in the asymptotic regime (as we also inferred in
Section~\ref{alphap>2}). Changing variables to
\begin{equation}
\label{var_def}
y\equiv \frac{\sqrt{C}}{|1-\alphap/2|} Z^{1-\alphap/2} \,,\quad
f(y)\equiv \frac{x}{Z^{1/2}} \, ,
\end{equation}
this equation can then be written as
\begin{equation}
\label{ODE3}
y^2 \frac{d^2 f}{dy^2} + y \frac{d f}{dy} + 
\left[y^2 - \left(\frac{1}{|2-\alphap|}\right)^2 \right] f =0 \,,
\end{equation}
whose solution is
\begin{equation}
\label{f(Y)}
f(y)=C_1 J_{1/|2-\alphap|}(y) + C_2 Y_{1/|2-\alphap|} (y) \,.
\end{equation}
For $\alphap > 2$ we have $1-\alphap/2 < 0$, so the limit $Z
\rightarrow \infty$ corresponds to $y \rightarrow 0$, in which case
$J_\nu (y) \approx y^\nu$ and $Y_\nu (y) \approx 1/y^\nu$. Thus the term
involving the Neumann function $Y_\nu (y)$ dominates, implying an
asymptotic solution $f(y)\approx C_2 / y^{1/(\alphap-2)}$, or, using the
definition of $y$ (equation~\ref{var_def}), 
\begin{equation}
r\approx C_2 z \,.
\end{equation}
Thus, the solution is essentially paraboloidal (concave) with conical
asymptotes.

\subsubsection{$\alphap=4$}

Changing the variable $x(Z)$ to $g(Z)\equiv x/Z$, equation~(\ref{ODE})
becomes 
\begin{equation}
Z^2 \frac{d^2g}{dZ^2} +2 Z \frac{dg}{dZ}  + \frac{Cg-1/g^3}{Z^2} =0 \,.
\end{equation}
This equation has an exact solution, $g= {\rm const} = C^{-1/4}$,
representing a flow that is conical from the start, $Z=C^{1/4}x$ (as we
already found in Section~\ref{alphap>2}), and in which the
electromagnetic and centrifugal forces have comparable contributions.

\subsubsection{$\alphap>4$}
\label{A1.3}

When $\alphap > 4$ the ratio of the third (centrifugal) to the second
(electromagnetic) terms on the left-hand side of equation~(\ref{ODE})
diverges as $Z\rightarrow \infty$, and one can therefore neglect the
electromagnetic term in the asymptotic regime (as we also inferred in
Section~\ref{alphap>2}). Without this term, equation~(\ref{ODE}) becomes
\begin{equation}
\label{ODE4}
\frac{dx^2}{dZ^2}=\frac{1}{x^3} \,.
\end{equation}
Multiplying by $2dx/dZ$, this equation can be integrated to give
\begin{equation}
\label{ODE4_1}
\left ( \frac{dx}{dZ} \right )^2 + \frac{1}{x^2}=\frac{1}{D}\, ,
\end{equation}
where $D$ is a constant of integration. Equation~(\ref{ODE4_1})
can be further integrated to yield 
\begin{equation}
\label{hyper}
\frac{x^2}{D}-\frac{(Z-Z_0)^2}{D^2}=1 \,,
\end{equation}
where $Z_0$ is another constant of integration. Equation~(\ref{hyper})
explicitly shows that the jet assumes a hyperboloidal shape in this
case, with the asymptotes again being conical.

\subsection{$\alphap = 2$}
\label{Aalphap=2}

Changing variables to $f \equiv x/Z^{1/2}$ (as in
equation~\ref{var_def}) and $q \equiv \ln{Z}$, equation~(\ref{ODE})
becomes in this case
\begin{equation}
\label{ODE5}
\frac{d^2 f}{dq^2} = \left(\frac{1}{4}-C\right) f + \frac{1}{f^3} \,.
\end{equation}

One obvious solution is $f=$const, with $f^{-1/4}=C-1/4$,
or $Z=\sqrt{C-1/4} \  x^2$.
This solution is real only for $C>1/4$.

For $df/dq \neq 0$, equation~(\ref{ODE5}) can be multiplied by $2 df/dq$
and rewritten as
\begin{equation}
\frac{d}{dq} \left[\left(\frac{df}{dq}\right)^2
+ \left(C-\frac{1}{4}\right)f^2 + \frac{1}{f^2}\right]=0 \,,
\end{equation}
or, with $u\equiv f^2$,
\begin{equation}
\left(\frac{du}{dq}\right)^2=(1-4C) u^2 + 4 E u -4   \,,
\end{equation}
where $E$ is a constant of integration. This can be integrated to give
\begin{eqnarray}
\pm 2 \int dq = \int \frac{du}{\sqrt{\left(\frac{1}{4}-C\right)u^2 + E
u -1}}=
\nonumber
\end{eqnarray}
\begin{eqnarray}
\label{soln_5}
\left\{
\begin{array}{l}
\displaystyle
\frac{
\displaystyle
 \ln \left(\frac{{E}/{2}}{\frac{1}{4}-C}+ u
+ \sqrt{ u^2 + \frac{E u -1 }{\frac{1}{4}-C} }  \right)
}{\sqrt{\frac{1}{4}-C}} 
\,,\, C<\frac{1}{4}
\\
\displaystyle 
2\frac{\displaystyle \sqrt{E u -1}}{ \displaystyle E} \,,\, C= \frac{1}{4}
\\
\displaystyle \frac{-1}{\sqrt{C-\frac{1}{4}}}
\arctan \frac{
\displaystyle
\frac{E/2}{C-\frac{1}{4}}-u}
{\sqrt{
\displaystyle 
-u^2 + \frac{E u -1}{C-\frac{1}{4}} }} \,,\, C>\frac{1}{4} \,.
\end{array} \right.
\end{eqnarray}
The quantity inside the square root in the integrand must be positive
in the asymptotic regime. This implies that $u$ cannot tend to zero as
$Z\rightarrow \infty$, and so it either tends to
a constant or to $\infty$. The first option is unacceptable as it gives
$q= {\rm const} \Leftrightarrow Z= {\rm const}$. 
Thus, we must have $u\rightarrow \infty$ as $Z\rightarrow \infty$. 
Since $u=f^2= x^2/Z$, this means that the shape is 
$Z \propto x^b$ with $b<2$.
The dominant term inside the square root in the integrand when
$Z\rightarrow \infty$ is the first one, 
which implies $C\le 1/4$ (and hence that the $C>1/4$ solution branch of
equation~\ref{soln_5} is not physical).

For $C=1/4$, equation~(\ref{soln_5}) gives $x=Z^{1/2} \sqrt{1/E+ E
\left[\ln{(Z/Z_ 0)}\right]^2}$, or, keeping only the dominant terms,
$x=Z^{1/2} (C_1+C_2 \ln{Z})$. In view of the requirement $u = x^2/Z
\rightarrow \infty$ as $Z\rightarrow \infty$ derived above, one must
keep the logarithmic term in this solution (i.e. $C_2 \neq 0$). We also
keep the other term in this solution, which may be needed to match to
the conditions at the base of the flow (see Section~\ref{alphap=2}). Note
that this solution could not be derived from the ansatz employed in
Section~\ref{pressure} since it does not have a pure power-law form.

For $C<1/4$, equation~(\ref{soln_5}) gives (keeping only the dominant terms)
$x =C_1 Z^{\frac{1}{2} \pm\frac{1}{2} \sqrt{1-4C} }$. Since, as we found
above, the exponent $b$ in $Z \propto x^b$ must be $<2$ the only
acceptable solution is
$x =C_1 Z^{\frac{1}{2} +\frac{1}{2} \sqrt{1-4C} }$.

\noindent
Summarizing, 
\\ for $C>1/4$, $Z=\sqrt{C-1/4} \  x^2$,
\\ for $C=1/4$, $x=Z^{1/2} (C_1+C_2 \ln Z)$ with $C_2 \neq 0$, and
\\ for $C<1/4$, $x =C_1 Z^{\frac{1}{2} +\frac{1}{2} \sqrt{1-4C} }$.
\\
The first case ($C>1/4$) represents a balance between the poloidal
curvature, electromagnetic and centrifugal terms in
equation~(\ref{ODE}), whereas in the last two cases ($C\le 1/4$) only the
poloidal curvature and electromagnetic terms play a role.

\subsection{$\alphap < 2$}
\label{Aalphap<2}

As discussed in Section~\ref{alphap<2}, in this case the poloidal
curvature term --- the first term on the left-hand side of
equation~(\ref{ODE}) --- can be neglected. The power-law form for $x(Z)$
given in the main text is then an exact solution of this equation.


\begin{thebibliography}{}

\bibitem[Barkov \& Komissarov(2008)]{BK08}
Barkov M. V., Komissarov S. S., 2008, MNRAS, 385, L28 

\bibitem[Barzilay \& Levinson(2008)]{BL08}
Barzilay Y., Levinson A., 2008, New Astr., 13, 386

\bibitem[\protect\citeauthoryear{Begelman \& Li}{Begelman \&
Li}{1994}]{BL94} Begelman M.~C., Li Z.-Y., 1994, ApJ, 326, 269

\bibitem[Berger et al.(2003a)]{BKF03}
Berger E., Kulkarni S.~R., Frail D.~A., 2003, ApJ, 590, 379

\bibitem[Berger et al.(2003b)]{Ber03}
Berger E., et al., 2003, Nature, 426, 154

\bibitem[\protect\citeauthoryear{{Beskin}, {Kuznetsova} 
\& {Rafikov}}{{Beskin} et~al.}{1998}]{BKR98}
{Beskin} V.~S.,  {Kuznetsova} I.~V.,    {Rafikov} R.~R.,  1998, \mnras, 299,
  341

\bibitem[\protect\citeauthoryear{{Beskin} \& {Malyshkin}}{{Beskin} \&
  {Malyshkin}}{2000}]{2000AstL...26..208B}
{Beskin} V.~S.,  {Malyshkin} L.~M.,  2000, Astronomy Letters, 26, 208

\bibitem[\protect\citeauthoryear{{Beskin} \& {Nokhrina}}{{Beskin} \&
  {Nokhrina}}{2006}]{BN06}
{Beskin} V.~S.,  {Nokhrina} E.~E.,  2006, \mnras, 367, 375

\bibitem[Begelman et al.(1984)]{BBR84}
Begelman M.~C., Blandford R.~D., Rees M.~J., 1984, Rev. Mod. Phys., 56,
255

\bibitem[\protect\citeauthoryear{{Blandford}}{{Blandford}}{1976}]{Bland76}
{Blandford} R.~D., 1976, \mnras, 176, 465

\bibitem[\protect\citeauthoryear{{Blandford}}{{Blandford}}{2002}]{Bla02}
{Blandford} R.~D., 2002, in {Gilfanov} M., et al.,
% {Sunyeav} R., {Churazov} E.,
eds, Lighthouses of the Universe.
%: The Most Luminous Celestial Objects and Their Use for Cosmology. 
Springer-Verlag, Berlin, p. 381

\bibitem[Blandford \& Payne(1982)]{BP82} 
Blandford R.~D., Payne D.~G., 1982, \mnras, 199, 883

%\bibitem[\protect\citeauthoryear{{Blandford} \& {Rees}}{{Blandford} \&
%  {Rees}}{1974}]{BR74}
%{Blandford} R.~D.,  {Rees} M.~J.,  1974, \mnras, 169, 395

\bibitem[\protect\citeauthoryear{{Bogovalov}}{{Bogovalov}}{1995}]{Bog95}
{Bogovalov} S.~V.,  1995, Astronomy Letters, 21, 565

\bibitem[Bromberg \& Levinson(2007)]{BL07}
Bromberg O., Levinson A., 2007, ApJ, 671, 678

\bibitem[\protect\citeauthoryear{Bucciantini et al. }{2008}]{B08}
 Bucciantini N., Quataert E., Arons J., Metzger B. B., Thompson T. A.,
 2008, MNRAS, 383, 25

\bibitem[\protect\citeauthoryear{{Chiueh}, {Li} \& {Begelman}}{{Chiueh}
  et~al.}{1991}]{CLB91}
{Chiueh} T.,  {Li} Z.-Y.,    {Begelman} M.~C.,  1991, \apj, 377, 462

\bibitem[\protect\citeauthoryear{{Chiueh}, {Li} \& {Begelman}}{{Chiueh}
  et~al.}{1998}]{CLB98}
{Chiueh} T.,  {Li} Z.-Y.,    {Begelman} M.~C.,  1998, \apj, 505, 835

\bibitem[\protect\citeauthoryear{{Contopoulos}}{{Contopoulos}}{1994}]{Con94}
{Contopoulos} J.,  1994, \apj, 432, 508

\bibitem[\protect\citeauthoryear{{Contopoulos} \& {Kazanas}}{{Contopoulos} \&
  {Kazanas}}{2002}]{2002ApJ...566..336C}
{Contopoulos} I.,  {Kazanas} D.,  2002, \apj, 566, 336

%\bibitem[Contopoulos \& Lovelace(1994)]{CL94}
%Contopoulos J., Lovelace R.~V.~E., 1994, ApJ, 429, 139

\bibitem[Drenkhahn(2002)]{Dre02}
Drenkhahn G., 2002, \aap, 387, 714

\bibitem[\protect\citeauthoryear{{Drenkhahn} \& {Spruit}}
{{Drenkhahn} \& {Spruit}}2002]{DS02}
{Drenkhahn} G., {Spruit} H.~C., 2002, \aap, 391, 1141

\bibitem[Di Matteo et al.(2002)]{DPN02}
Di Matteo T., Perna R., Narayan R., 2002, ApJ, 579, 706

\bibitem[Eichler \& Levinson(2004)]{EL04}
Eichler D., Levinson A., 2004, ApJ, 614, L13

\bibitem[Frail et al.(2005)]{Fra05}
Frail D.~A., Soderberg A.~M., Kulkarni S.~R., Berger E., Yost S., Fox
D.~W., Harrison F.~A., 2005, ApJ, 619, 994

\bibitem[Fuller et al.(2000)]{FPA00}
Fuller G.~M., Pruet J., Abazajian K., 2000, Phys. Rev. Lett., 85, 2673

\bibitem[Genet et al. (2007)]{GDM07}
Genet F., Daigne F., Mochkovitch R., 2007, MNRAS, 381, 732

\bibitem[Granot(2005)]{Gra05}
Granot J., 2005, ApJ, 631, 1022

\bibitem[Granot et al.(2006)]{GKP06}
Granot J., K\"onigl A., Piran T., 2006, MNRAS, 370, 1946

\bibitem[Kamble et al.(2008)]{Kamb08}
Kamble A., Misra K., Bhattacharya D., Sagar R., 2008, MNRAS in press 
(arXiv:0806.4270)

\bibitem[Katz(1997)]{K97} Katz J.~I., 1997, \apj, 490, 633

\bibitem[Klu\'zniak \& Ruderman(1998)]{KR98}
Klu\'zniak W., Ruderman M., 1998, \apj, 505, L113

\bibitem[Kobayashi \& Zhang(2007)]{KZ07}
Kobayashi S., Zhang B., 2007, ApJ, 655, 973

%\bibitem[\protect\citeauthoryear{{K\"onigl}}{{K\"onigl}}{1982}]{K82}
%{K\"onigl} A.,  1982, \apj, 261, 115

\bibitem[\protect\citeauthoryear{{Komissarov}}{{Komissarov}}{1999a}]{K99}
{Komissarov} S.~S.,  1999a, \mnras, 303, 343

\bibitem[\protect\citeauthoryear{{Komissarov}}{{Komissarov}}{1999b}]{Kom99}
{Komissarov} S.~S.,  1999b, \mnras, 308, 1069

\bibitem[\protect\citeauthoryear{{Komissarov}}{{Komissarov}}{2004}]{K04}
{Komissarov} S.~S.,  2004, \mnras, 350, 1431

\bibitem[\protect\citeauthoryear{{Komissarov}, {Barkov}, {Vlahakis} \&
{K{\"o}nigl}}{{Komissarov} et~al.}{2007}]{KBVK07}
{Komissarov} S.~S., {Barkov} M.~V., {Vlahakis} N., {K{\"o}nigl} A., 2007,
\mnras, 380, 51 
(Paper~I)

\bibitem[\protect\citeauthoryear{{Komissarov} \& {Barkov}}
{{Komissarov} \& {Barkov}}{2007}]{KB07}
{Komissarov} S.~S., {Barkov} M.~V., 2007,\mnras, 382, 1089

\bibitem[\protect\citeauthoryear{{Komissarov} \& {Lyubarsky}}{{Komissarov} \&
{Lyubarsky}}{2004}]{KL04}
{Komissarov} S.~S.,  {Lyubarsky} Y.~E.,  2004, \mnras, 349, 779

\bibitem[Kumar et al.(2007)]{Kum07}
Kumar P., et al., 2007, MNRAS, 376, L57

\bibitem[Kumar \& Piran(2000)]{KP00}
Kumar P., Piran T., ApJ, 535, 152

\bibitem[Lazzati \& Begelman(2005)]{LB05}
Lazzati D., Begelman, M.~C., 2005, ApJ, 629, 903

\bibitem[Levinson(2006)]{Lev06}
Levinson A., 2006, ApJ, 648, 510

\bibitem[Levinson \& Eichler(1993)]{LE93}
Levinson A., Eichler D., 1993, ApJ, 418, 386

\bibitem[Levinson \& Eichler(2000)]{LE00}
Levinson A., Eichler D., 2000, Phys. Rev. Lett., 85, 236

\bibitem[Levinson \& Eichler(2003)]{LE03}
Levinson A., Eichler D., 2003, ApJ, 594, L19

\bibitem[Li et al.(1992)]{LCB92} 
Li Z.-Y., Chiueh T., Begelman M.~C., 1992, \apj, 394, 459


\bibitem[Liang et al.(2008)]{Lia08}
Liang E.-W., Racusin J.~L., Zhang B., Zhang B.-B., Burrows D.~N., 2008,
ApJ, 675, 528

\bibitem[Liang \& Zhang(2005)]{LZ05}
Liang E., Zhang B., 2005, ApJ, 633, 611
  
\bibitem[Lithwick \& Sari(2001)]{LS01}
Lithwick Y., Sari, R., 2001, ApJ, 555, 540

\bibitem[\protect\citeauthoryear{{Lyutikov}}
{{Lyutikov}}{2006a}]{Lyu06a}
{Lyutikov} M., 2006a, MNRAS, 369, L5

\bibitem[\protect\citeauthoryear{{Lyutikov}}
{{Lyutikov}}{2006b}]{Lyu06b}
{Lyutikov} M., 2006b, New J. Phys., 8, 119

\bibitem[\protect\citeauthoryear{{Lyubarsky} \& {Eichler}}{{Lyubarsky} \&
  {Eichler}}{2001}]{2001ApJ...562..494L}
{Lyubarsky} Y.,  {Eichler} D.,  2001, \apj, 562, 494

\bibitem[\protect\citeauthoryear{McKinney }{2006}]{M06}
 McKinney J.~C., 2006, MNRAS,368,1561

\bibitem[M\'esz\'aros(2006)]{Mes06}
M\'esz\'aros P., 2006, Rep. Prog. Phys., 69, 2259

\bibitem[M\'{e}sz\'{a}ros \& Rees(1997)]{MR97}
M\'{e}sz\'{a}ros P., Rees M.~J., 1997, \apj, 482, L29

\bibitem[Morsony et al. (2007)]{MLB07}
Morsony B. J., Lazzati D., Begelman M. C., 2007, \apj, 665, 569

\bibitem[Nakar(2007)]{Nak07}
Nakar E., 2007, Phys. Rep., 442, 166

\bibitem[Nakar et al.(2004)]{NGG04}
Nakar E., Granot J., Guetta D., 2004, ApJ, 606, L37

\bibitem[Nakar \& Piran(2003)]{NP03}
Nakar E., Piran T., 2003, New Astr., 8, 141

\bibitem[\protect\citeauthoryear
{{Narayan}, {McKinney} \& {Farmer}}{{Narayan} et~al.}{2007}]{NMcKF07}
Narayan R., McKinney J.~C., Farmer A.~J., 2007, \mnras, 375, 548

\bibitem[\protect\citeauthoryear{{Okamoto}}{{Okamoto}}{2002}]
{2002ApJ...573L..31O}
{Okamoto} I.,  2002, \apj, 573, L31

\bibitem[Paczy\'nsky \& Wiita(1980)]{PW80}
Paczy\'nski B., Wiita P., 1980, \aa, 88, 23

\bibitem[Panaitescu(2007)]{Pan07}
Panaitescu A., 2007, MNRAS, 379, 331

%\bibitem[Panaitescu \& M\'esz\'aros(1999)]{PM99}
%Panaitescu A., M\'esz\'aros P., 1999, ApJ, 526, 707

\bibitem[Peng et al.(2005)]{PKG05}
Peng F., K\"onigl A., Granot J., 2005, ApJ, 626, 966

%\bibitem[Piran(1994)]{Pir94}
%Piran T., 1994, in AIP Conf. Proc. 307, Gamma-Ray Bursts: Second
%Huntsville Workshop, ed. G.J.Fishman, J.J.Brainerd, \& 
%K.Hurley (New York: AIP), p. 495

\bibitem[Piran(2005)]{Pir05}
Piran T., 2005, Rev. Mod. Phys., 76, 1143

\bibitem[Popham et al.(1999)]{PWF99}
Popham R., Woosley S.~E., Fryer C., 1999, ApJ, 518, 356

\bibitem[\protect\citeauthoryear{{Proga} et al. }{2003}]{P03}
 Proga D., MacFadyen A.~I., Armitage P.~J., Begelman M.~C., 
 2003,ApJ,629,397

\bibitem[Ramirez-Ruiz et al.(2002)]{RCR02}
Ramirez-Ruiz E., Celotti A., Rees M.~J., 2002, MNRAS, 337, 1349

\bibitem[Rhoads(1999)]{Rho99}
Rhoads J.~E., 1999, ApJ, 525, 737

\bibitem[Rossi et al.(2002)]{RLR02}
Rossi E., Lazzati D., Rees M.~J., 2002, MNRAS, 332, 945

\bibitem[Sari et al.(1999)]{Sari99}
Sari R., Piran T., Halpern J. P., 1999, ApJ, 519, L17

\bibitem[Tchekhovskoy et al.(2008)]{TMN08}
Tchekhovskoy A., McKinney J.~C., Narayan R., 2008, MNRAS, 388, 551

\bibitem[Thompson(1994)]{T94}
Thompson C., 1994, \mnras, 270, 480

\bibitem[\protect\citeauthoryear{
{Tomimatsu} \& {Takahashi}}{{Tomimatsu} \&
  {Takahashi}}{2003}]{2003ApJ...592..321T}
{Tomimatsu} A.,  {Takahashi} M.,  2003, \apj, 592, 321

\bibitem[Uhm \& Beloborodov(2007)]{UB07}
Uhm Z.~L., Beloborodov A.~M., 2007, ApJ, 665, L93

\bibitem[Usov(1992)]{U92} Usov V.~V., 1992, Nature, 357, 472

\bibitem[\protect\citeauthoryear{{Vlahakis}}{{Vlahakis}}{2004a}]{V04}
{Vlahakis} N.,  2004a, \apj, 600, 324

\bibitem[\protect\citeauthoryear{{Vlahakis}}{{Vlahakis}}{2004b}]{V04dogl}
{Vlahakis} N.,  2004b, \apss, 293, 67

\bibitem[\protect\citeauthoryear{{Vlahakis} \& {K{\"o}nigl}}{{Vlahakis} \&
  {K{\"o}nigl}}{2001}]{VK01}
{Vlahakis} N.,  {K{\"o}nigl} A.,  2001, \apj, 563, L129

\bibitem[\protect\citeauthoryear{{Vlahakis} \& {K{\"o}nigl}}{{Vlahakis} \&
  {K{\"o}nigl}}{2003a}]{VK03a}
{Vlahakis} N.,  {K{\"o}nigl} A.,  2003a, \apj, 596, 1080

\bibitem[\protect\citeauthoryear{{Vlahakis} \& {K{\"o}nigl}}{{Vlahakis} \&
  {K{\"o}nigl}}{2003b}]{VK03b}
{Vlahakis} N.,  {K{\"o}nigl} A.,  2003b, \apj, 596, 1104

\bibitem[\protect\citeauthoryear{{Vlahakis} \& {K{\"o}nigl}}{{Vlahakis} \&
  {K{\"o}nigl}}{2004}]{VK04}
{Vlahakis} N.,  {K{\"o}nigl} A.,  2004, \apj, 605, 656

\bibitem[Vlahakis at al.(2003)]{VPK03}
Vlahakis N., Peng F., K\"onigl A., 2003, ApJ, 594, L23

\bibitem[Vlahakis et al.(2000)]{V00}
Vlahakis N., Tsinganos K., Sauty C., Trussoni E., 2000, \mnras, 318, 417

\bibitem[\protect\citeauthoryear{{Zakamska}, {Begelman} \& {Blandford}}
{{Zakamska} et~al.}{2008}]{ZBB08}
Zakamska N.~L., Begelman M.~C., Blandford R.~D., 2008, \apj, 679, 990

\bibitem[Zhang(2007)]{Zhang07}
Zhang B., 2007, Chin. J. Astron. Astrophys., 7, 1
\end{thebibliography}
\end{document}